\documentclass[fleqn,usenatbib]{mnras}

\usepackage{newtxtext,newtxmath}

\usepackage[T1]{fontenc}

\DeclareRobustCommand{\VAN}[3]{#2}
\let\VANthebibliography\thebibliography
\def\thebibliography{\DeclareRobustCommand{\VAN}[3]{##3}\VANthebibliography}

\usepackage{natbib, twoopt}
\usepackage{float}
\usepackage{graphicx,float}
\usepackage{epstopdf}
\usepackage{enumitem}
\usepackage[caption=false]{subfig}
\usepackage{color}\definecolor{AliceBlue}{rgb}{0.94,0.97,1.00}
\definecolor{AntiqueWhite1}{rgb}{1.00,0.94,0.86}
\definecolor{AntiqueWhite2}{rgb}{0.93,0.87,0.80}
\definecolor{AntiqueWhite3}{rgb}{0.80,0.75,0.69}
\definecolor{AntiqueWhite4}{rgb}{0.55,0.51,0.47}
\definecolor{AntiqueWhite}{rgb}{0.98,0.92,0.84}
\definecolor{BlanchedAlmond}{rgb}{1.00,0.92,0.80}
\definecolor{BlueViolet}{rgb}{0.54,0.17,0.89}
\definecolor{CadetBlue1}{rgb}{0.60,0.96,1.00}
\definecolor{CadetBlue2}{rgb}{0.56,0.90,0.93}
\definecolor{CadetBlue3}{rgb}{0.48,0.77,0.80}
\definecolor{CadetBlue4}{rgb}{0.33,0.53,0.55}
\definecolor{CadetBlue}{rgb}{0.37,0.62,0.63}
\definecolor{CornflowerBlue}{rgb}{0.39,0.58,0.93}
\definecolor{DarkBlue}{rgb}{0.00,0.00,0.55}
\definecolor{DarkCyan}{rgb}{0.00,0.55,0.55}
\definecolor{DarkGoldenrod1}{rgb}{1.00,0.73,0.06}
\definecolor{DarkGoldenrod2}{rgb}{0.93,0.68,0.05}
\definecolor{DarkGoldenrod3}{rgb}{0.80,0.58,0.05}
\definecolor{DarkGoldenrod4}{rgb}{0.55,0.40,0.03}
\definecolor{DarkGoldenrod}{rgb}{0.72,0.53,0.04}
\definecolor{DarkGray}{rgb}{0.66,0.66,0.66}
\definecolor{DarkGreen}{rgb}{0.00,0.39,0.00}
\definecolor{DarkGrey}{rgb}{0.66,0.66,0.66}
\definecolor{DarkKhaki}{rgb}{0.74,0.72,0.42}
\definecolor{DarkMagenta}{rgb}{0.55,0.00,0.55}
\definecolor{DarkOliveGreen1}{rgb}{0.79,1.00,0.44}
\definecolor{DarkOliveGreen2}{rgb}{0.74,0.93,0.41}
\definecolor{DarkOliveGreen3}{rgb}{0.64,0.80,0.35}
\definecolor{DarkOliveGreen4}{rgb}{0.43,0.55,0.24}
\definecolor{DarkOliveGreen}{rgb}{0.33,0.42,0.18}
\definecolor{DarkOrange1}{rgb}{1.00,0.50,0.00}
\definecolor{DarkOrange2}{rgb}{0.93,0.46,0.00}
\definecolor{DarkOrange3}{rgb}{0.80,0.40,0.00}
\definecolor{DarkOrange4}{rgb}{0.55,0.27,0.00}
\definecolor{DarkOrange}{rgb}{1.00,0.55,0.00}
\definecolor{DarkOrchid1}{rgb}{0.75,0.24,1.00}
\definecolor{DarkOrchid2}{rgb}{0.70,0.23,0.93}
\definecolor{DarkOrchid3}{rgb}{0.60,0.20,0.80}
\definecolor{DarkOrchid4}{rgb}{0.41,0.13,0.55}
\definecolor{DarkOrchid}{rgb}{0.60,0.20,0.80}
\definecolor{DarkRed}{rgb}{0.55,0.00,0.00}
\definecolor{DarkSalmon}{rgb}{0.91,0.59,0.48}
\definecolor{DarkSeaGreen1}{rgb}{0.76,1.00,0.76}
\definecolor{DarkSeaGreen2}{rgb}{0.71,0.93,0.71}
\definecolor{DarkSeaGreen3}{rgb}{0.61,0.80,0.61}
\definecolor{DarkSeaGreen4}{rgb}{0.41,0.55,0.41}
\definecolor{DarkSeaGreen}{rgb}{0.56,0.74,0.56}
\definecolor{DarkSlateBlue}{rgb}{0.28,0.24,0.55}
\definecolor{DarkSlateGray1}{rgb}{0.59,1.00,1.00}
\definecolor{DarkSlateGray2}{rgb}{0.55,0.93,0.93}
\definecolor{DarkSlateGray3}{rgb}{0.47,0.80,0.80}
\definecolor{DarkSlateGray4}{rgb}{0.32,0.55,0.55}
\definecolor{DarkSlateGray}{rgb}{0.18,0.31,0.31}
\definecolor{DarkSlateGrey}{rgb}{0.18,0.31,0.31}
\definecolor{DarkTurquoise}{rgb}{0.00,0.81,0.82}
\definecolor{DarkViolet}{rgb}{0.58,0.00,0.83}
\definecolor{DeepPink1}{rgb}{1.00,0.08,0.58}
\definecolor{DeepPink2}{rgb}{0.93,0.07,0.54}
\definecolor{DeepPink3}{rgb}{0.80,0.06,0.46}
\definecolor{DeepPink4}{rgb}{0.55,0.04,0.31}
\definecolor{DeepPink}{rgb}{1.00,0.08,0.58}
\definecolor{DeepSkyBlue1}{rgb}{0.00,0.75,1.00}
\definecolor{DeepSkyBlue2}{rgb}{0.00,0.70,0.93}
\definecolor{DeepSkyBlue3}{rgb}{0.00,0.60,0.80}
\definecolor{DeepSkyBlue4}{rgb}{0.00,0.41,0.55}
\definecolor{DeepSkyBlue}{rgb}{0.00,0.75,1.00}
\definecolor{DimGray}{rgb}{0.41,0.41,0.41}
\definecolor{DimGrey}{rgb}{0.41,0.41,0.41}
\definecolor{DodgerBlue1}{rgb}{0.12,0.56,1.00}
\definecolor{DodgerBlue2}{rgb}{0.11,0.53,0.93}
\definecolor{DodgerBlue3}{rgb}{0.09,0.45,0.80}
\definecolor{DodgerBlue4}{rgb}{0.06,0.31,0.55}
\definecolor{DodgerBlue}{rgb}{0.12,0.56,1.00}
\definecolor{FloralWhite}{rgb}{1.00,0.98,0.94}
\definecolor{ForestGreen}{rgb}{0.13,0.55,0.13}
\definecolor{GhostWhite}{rgb}{0.97,0.97,1.00}
\definecolor{GreenYellow}{rgb}{0.68,1.00,0.18}
\definecolor{HotPink1}{rgb}{1.00,0.43,0.71}
\definecolor{HotPink2}{rgb}{0.93,0.42,0.65}
\definecolor{HotPink3}{rgb}{0.80,0.38,0.56}
\definecolor{HotPink4}{rgb}{0.55,0.23,0.38}
\definecolor{HotPink}{rgb}{1.00,0.41,0.71}
\definecolor{IndianRed1}{rgb}{1.00,0.42,0.42}
\definecolor{IndianRed2}{rgb}{0.93,0.39,0.39}
\definecolor{IndianRed3}{rgb}{0.80,0.33,0.33}
\definecolor{IndianRed4}{rgb}{0.55,0.23,0.23}
\definecolor{IndianRed}{rgb}{0.80,0.36,0.36}
\definecolor{LavenderBlush1}{rgb}{1.00,0.94,0.96}
\definecolor{LavenderBlush2}{rgb}{0.93,0.88,0.90}
\definecolor{LavenderBlush3}{rgb}{0.80,0.76,0.77}
\definecolor{LavenderBlush4}{rgb}{0.55,0.51,0.53}
\definecolor{LavenderBlush}{rgb}{1.00,0.94,0.96}
\definecolor{LawnGreen}{rgb}{0.49,0.99,0.00}
\definecolor{LemonChiffon1}{rgb}{1.00,0.98,0.80}
\definecolor{LemonChiffon2}{rgb}{0.93,0.91,0.75}
\definecolor{LemonChiffon3}{rgb}{0.80,0.79,0.65}
\definecolor{LemonChiffon4}{rgb}{0.55,0.54,0.44}
\definecolor{LemonChiffon}{rgb}{1.00,0.98,0.80}
\definecolor{LightBlue1}{rgb}{0.75,0.94,1.00}
\definecolor{LightBlue2}{rgb}{0.70,0.87,0.93}
\definecolor{LightBlue3}{rgb}{0.60,0.75,0.80}
\definecolor{LightBlue4}{rgb}{0.41,0.51,0.55}
\definecolor{LightBlue}{rgb}{0.68,0.85,0.90}
\definecolor{LightCoral}{rgb}{0.94,0.50,0.50}
\definecolor{LightCyan1}{rgb}{0.88,1.00,1.00}
\definecolor{LightCyan2}{rgb}{0.82,0.93,0.93}
\definecolor{LightCyan3}{rgb}{0.71,0.80,0.80}
\definecolor{LightCyan4}{rgb}{0.48,0.55,0.55}
\definecolor{LightCyan}{rgb}{0.88,1.00,1.00}
\definecolor{LightGoldenrod1}{rgb}{1.00,0.93,0.55}
\definecolor{LightGoldenrod2}{rgb}{0.93,0.86,0.51}
\definecolor{LightGoldenrod3}{rgb}{0.80,0.75,0.44}
\definecolor{LightGoldenrod4}{rgb}{0.55,0.51,0.30}
\definecolor{LightGoldenrodYellow}{rgb}{0.98,0.98,0.82}
\definecolor{LightGoldenrod}{rgb}{0.93,0.87,0.51}
\definecolor{LightGray}{rgb}{0.83,0.83,0.83}
\definecolor{LightGreen}{rgb}{0.56,0.93,0.56}
\definecolor{LightGrey}{rgb}{0.83,0.83,0.83}
\definecolor{LightPink1}{rgb}{1.00,0.68,0.73}
\definecolor{LightPink2}{rgb}{0.93,0.64,0.68}
\definecolor{LightPink3}{rgb}{0.80,0.55,0.58}
\definecolor{LightPink4}{rgb}{0.55,0.37,0.40}
\definecolor{LightPink}{rgb}{1.00,0.71,0.76}
\definecolor{LightSalmon1}{rgb}{1.00,0.63,0.48}
\definecolor{LightSalmon2}{rgb}{0.93,0.58,0.45}
\definecolor{LightSalmon3}{rgb}{0.80,0.51,0.38}
\definecolor{LightSalmon4}{rgb}{0.55,0.34,0.26}
\definecolor{LightSalmon}{rgb}{1.00,0.63,0.48}
\definecolor{LightSeaGreen}{rgb}{0.13,0.70,0.67}
\definecolor{LightSkyBlue1}{rgb}{0.69,0.89,1.00}
\definecolor{LightSkyBlue2}{rgb}{0.64,0.83,0.93}
\definecolor{LightSkyBlue3}{rgb}{0.55,0.71,0.80}
\definecolor{LightSkyBlue4}{rgb}{0.38,0.48,0.55}
\definecolor{LightSkyBlue}{rgb}{0.53,0.81,0.98}
\definecolor{LightSlateBlue}{rgb}{0.52,0.44,1.00}
\definecolor{LightSlateGray}{rgb}{0.47,0.53,0.60}
\definecolor{LightSlateGrey}{rgb}{0.47,0.53,0.60}
\definecolor{LightSteelBlue1}{rgb}{0.79,0.88,1.00}
\definecolor{LightSteelBlue2}{rgb}{0.74,0.82,0.93}
\definecolor{LightSteelBlue3}{rgb}{0.64,0.71,0.80}
\definecolor{LightSteelBlue4}{rgb}{0.43,0.48,0.55}
\definecolor{LightSteelBlue}{rgb}{0.69,0.77,0.87}
\definecolor{LightYellow1}{rgb}{1.00,1.00,0.88}
\definecolor{LightYellow2}{rgb}{0.93,0.93,0.82}
\definecolor{LightYellow3}{rgb}{0.80,0.80,0.71}
\definecolor{LightYellow4}{rgb}{0.55,0.55,0.48}
\definecolor{LightYellow}{rgb}{1.00,1.00,0.88}
\definecolor{LimeGreen}{rgb}{0.20,0.80,0.20}
\definecolor{MediumAquamarine}{rgb}{0.40,0.80,0.67}
\definecolor{MediumBlue}{rgb}{0.00,0.00,0.80}
\definecolor{MediumOrchid1}{rgb}{0.88,0.40,1.00}
\definecolor{MediumOrchid2}{rgb}{0.82,0.37,0.93}
\definecolor{MediumOrchid3}{rgb}{0.71,0.32,0.80}
\definecolor{MediumOrchid4}{rgb}{0.48,0.22,0.55}
\definecolor{MediumOrchid}{rgb}{0.73,0.33,0.83}
\definecolor{MediumPurple1}{rgb}{0.67,0.51,1.00}
\definecolor{MediumPurple2}{rgb}{0.62,0.47,0.93}
\definecolor{MediumPurple3}{rgb}{0.54,0.41,0.80}
\definecolor{MediumPurple4}{rgb}{0.36,0.28,0.55}
\definecolor{MediumPurple}{rgb}{0.58,0.44,0.86}
\definecolor{MediumSeaGreen}{rgb}{0.24,0.70,0.44}
\definecolor{MediumSlateBlue}{rgb}{0.48,0.41,0.93}
\definecolor{MediumSpringGreen}{rgb}{0.00,0.98,0.60}
\definecolor{MediumTurquoise}{rgb}{0.28,0.82,0.80}
\definecolor{MediumVioletRed}{rgb}{0.78,0.08,0.52}
\definecolor{MidnightBlue}{rgb}{0.10,0.10,0.44}
\definecolor{MintCream}{rgb}{0.96,1.00,0.98}
\definecolor{MistyRose1}{rgb}{1.00,0.89,0.88}
\definecolor{MistyRose2}{rgb}{0.93,0.84,0.82}
\definecolor{MistyRose3}{rgb}{0.80,0.72,0.71}
\definecolor{MistyRose4}{rgb}{0.55,0.49,0.48}
\definecolor{MistyRose}{rgb}{1.00,0.89,0.88}
\definecolor{NavajoWhite1}{rgb}{1.00,0.87,0.68}
\definecolor{NavajoWhite2}{rgb}{0.93,0.81,0.63}
\definecolor{NavajoWhite3}{rgb}{0.80,0.70,0.55}
\definecolor{NavajoWhite4}{rgb}{0.55,0.47,0.37}
\definecolor{NavajoWhite}{rgb}{1.00,0.87,0.68}
\definecolor{NavyBlue}{rgb}{0.00,0.00,0.50}
\definecolor{OldLace}{rgb}{0.99,0.96,0.90}
\definecolor{OliveDrab1}{rgb}{0.75,1.00,0.24}
\definecolor{OliveDrab2}{rgb}{0.70,0.93,0.23}
\definecolor{OliveDrab3}{rgb}{0.60,0.80,0.20}
\definecolor{OliveDrab4}{rgb}{0.41,0.55,0.13}
\definecolor{OliveDrab}{rgb}{0.42,0.56,0.14}
\definecolor{OrangeRed1}{rgb}{1.00,0.27,0.00}
\definecolor{OrangeRed2}{rgb}{0.93,0.25,0.00}
\definecolor{OrangeRed3}{rgb}{0.80,0.22,0.00}
\definecolor{OrangeRed4}{rgb}{0.55,0.15,0.00}
\definecolor{OrangeRed}{rgb}{1.00,0.27,0.00}
\definecolor{PaleGoldenrod}{rgb}{0.93,0.91,0.67}
\definecolor{PaleGreen1}{rgb}{0.60,1.00,0.60}
\definecolor{PaleGreen2}{rgb}{0.56,0.93,0.56}
\definecolor{PaleGreen3}{rgb}{0.49,0.80,0.49}
\definecolor{PaleGreen4}{rgb}{0.33,0.55,0.33}
\definecolor{PaleGreen}{rgb}{0.60,0.98,0.60}
\definecolor{PaleTurquoise1}{rgb}{0.73,1.00,1.00}
\definecolor{PaleTurquoise2}{rgb}{0.68,0.93,0.93}
\definecolor{PaleTurquoise3}{rgb}{0.59,0.80,0.80}
\definecolor{PaleTurquoise4}{rgb}{0.40,0.55,0.55}
\definecolor{PaleTurquoise}{rgb}{0.69,0.93,0.93}
\definecolor{PaleVioletRed1}{rgb}{1.00,0.51,0.67}
\definecolor{PaleVioletRed2}{rgb}{0.93,0.47,0.62}
\definecolor{PaleVioletRed3}{rgb}{0.80,0.41,0.54}
\definecolor{PaleVioletRed4}{rgb}{0.55,0.28,0.36}
\definecolor{PaleVioletRed}{rgb}{0.86,0.44,0.58}
\definecolor{PapayaWhip}{rgb}{1.00,0.94,0.84}
\definecolor{PeachPuff1}{rgb}{1.00,0.85,0.73}
\definecolor{PeachPuff2}{rgb}{0.93,0.80,0.68}
\definecolor{PeachPuff3}{rgb}{0.80,0.69,0.58}
\definecolor{PeachPuff4}{rgb}{0.55,0.47,0.40}
\definecolor{PeachPuff}{rgb}{1.00,0.85,0.73}
\definecolor{PowderBlue}{rgb}{0.69,0.88,0.90}
\definecolor{RosyBrown1}{rgb}{1.00,0.76,0.76}
\definecolor{RosyBrown2}{rgb}{0.93,0.71,0.71}
\definecolor{RosyBrown3}{rgb}{0.80,0.61,0.61}
\definecolor{RosyBrown4}{rgb}{0.55,0.41,0.41}
\definecolor{RosyBrown}{rgb}{0.74,0.56,0.56}
\definecolor{RoyalBlue1}{rgb}{0.28,0.46,1.00}
\definecolor{RoyalBlue2}{rgb}{0.26,0.43,0.93}
\definecolor{RoyalBlue3}{rgb}{0.23,0.37,0.80}
\definecolor{RoyalBlue4}{rgb}{0.15,0.25,0.55}
\definecolor{RoyalBlue}{rgb}{0.25,0.41,0.88}
\definecolor{SaddleBrown}{rgb}{0.55,0.27,0.07}
\definecolor{SandyBrown}{rgb}{0.96,0.64,0.38}
\definecolor{SeaGreen1}{rgb}{0.33,1.00,0.62}
\definecolor{SeaGreen2}{rgb}{0.31,0.93,0.58}
\definecolor{SeaGreen3}{rgb}{0.26,0.80,0.50}
\definecolor{SeaGreen4}{rgb}{0.18,0.55,0.34}
\definecolor{SeaGreen}{rgb}{0.18,0.55,0.34}
\definecolor{SkyBlue1}{rgb}{0.53,0.81,1.00}
\definecolor{SkyBlue2}{rgb}{0.49,0.75,0.93}
\definecolor{SkyBlue3}{rgb}{0.42,0.65,0.80}
\definecolor{SkyBlue4}{rgb}{0.29,0.44,0.55}
\definecolor{SkyBlue}{rgb}{0.53,0.81,0.92}
\definecolor{SlateBlue1}{rgb}{0.51,0.44,1.00}
\definecolor{SlateBlue2}{rgb}{0.48,0.40,0.93}
\definecolor{SlateBlue3}{rgb}{0.41,0.35,0.80}
\definecolor{SlateBlue4}{rgb}{0.28,0.24,0.55}
\definecolor{SlateBlue}{rgb}{0.42,0.35,0.80}
\definecolor{SlateGray1}{rgb}{0.78,0.89,1.00}
\definecolor{SlateGray2}{rgb}{0.73,0.83,0.93}
\definecolor{SlateGray3}{rgb}{0.62,0.71,0.80}
\definecolor{SlateGray4}{rgb}{0.42,0.48,0.55}
\definecolor{SlateGray}{rgb}{0.44,0.50,0.56}
\definecolor{SlateGrey}{rgb}{0.44,0.50,0.56}
\definecolor{SpringGreen1}{rgb}{0.00,1.00,0.50}
\definecolor{SpringGreen2}{rgb}{0.00,0.93,0.46}
\definecolor{SpringGreen3}{rgb}{0.00,0.80,0.40}
\definecolor{SpringGreen4}{rgb}{0.00,0.55,0.27}
\definecolor{SpringGreen}{rgb}{0.00,1.00,0.50}
\definecolor{SteelBlue1}{rgb}{0.39,0.72,1.00}
\definecolor{SteelBlue2}{rgb}{0.36,0.67,0.93}
\definecolor{SteelBlue3}{rgb}{0.31,0.58,0.80}
\definecolor{SteelBlue4}{rgb}{0.21,0.39,0.55}
\definecolor{SteelBlue}{rgb}{0.27,0.51,0.71}
\definecolor{VioletRed1}{rgb}{1.00,0.24,0.59}
\definecolor{VioletRed2}{rgb}{0.93,0.23,0.55}
\definecolor{VioletRed3}{rgb}{0.80,0.20,0.47}
\definecolor{VioletRed4}{rgb}{0.55,0.13,0.32}
\definecolor{VioletRed}{rgb}{0.82,0.13,0.56}
\definecolor{WhiteSmoke}{rgb}{0.96,0.96,0.96}
\definecolor{YellowGreen}{rgb}{0.60,0.80,0.20}
\definecolor{aliceblue}{rgb}{0.94,0.97,1.00}
\definecolor{antiquewhite}{rgb}{0.98,0.92,0.84}
\definecolor{aquamarine1}{rgb}{0.50,1.00,0.83}
\definecolor{aquamarine2}{rgb}{0.46,0.93,0.78}
\definecolor{aquamarine3}{rgb}{0.40,0.80,0.67}
\definecolor{aquamarine4}{rgb}{0.27,0.55,0.45}
\definecolor{aquamarine}{rgb}{0.50,1.00,0.83}
\definecolor{azure1}{rgb}{0.94,1.00,1.00}
\definecolor{azure2}{rgb}{0.88,0.93,0.93}
\definecolor{azure3}{rgb}{0.76,0.80,0.80}
\definecolor{azure4}{rgb}{0.51,0.55,0.55}
\definecolor{azure}{rgb}{0.94,1.00,1.00}
\definecolor{beige}{rgb}{0.96,0.96,0.86}
\definecolor{bisque1}{rgb}{1.00,0.89,0.77}
\definecolor{bisque2}{rgb}{0.93,0.84,0.72}
\definecolor{bisque3}{rgb}{0.80,0.72,0.62}
\definecolor{bisque4}{rgb}{0.55,0.49,0.42}
\definecolor{bisque}{rgb}{1.00,0.89,0.77}
\definecolor{black}{rgb}{0.00,0.00,0.00}
\definecolor{blanchedalmond}{rgb}{1.00,0.92,0.80}
\definecolor{blue1}{rgb}{0.00,0.00,1.00}
\definecolor{blue2}{rgb}{0.00,0.00,0.93}
\definecolor{blue3}{rgb}{0.00,0.00,0.80}
\definecolor{blue4}{rgb}{0.00,0.00,0.55}
\definecolor{blueviolet}{rgb}{0.54,0.17,0.89}
\definecolor{blue}{rgb}{0.00,0.00,1.00}
\definecolor{brown1}{rgb}{1.00,0.25,0.25}
\definecolor{brown2}{rgb}{0.93,0.23,0.23}
\definecolor{brown3}{rgb}{0.80,0.20,0.20}
\definecolor{brown4}{rgb}{0.55,0.14,0.14}
\definecolor{brown}{rgb}{0.65,0.16,0.16}
\definecolor{burlywood1}{rgb}{1.00,0.83,0.61}
\definecolor{burlywood2}{rgb}{0.93,0.77,0.57}
\definecolor{burlywood3}{rgb}{0.80,0.67,0.49}
\definecolor{burlywood4}{rgb}{0.55,0.45,0.33}
\definecolor{burlywood}{rgb}{0.87,0.72,0.53}
\definecolor{cadetblue}{rgb}{0.37,0.62,0.63}
\definecolor{chartreuse1}{rgb}{0.50,1.00,0.00}
\definecolor{chartreuse2}{rgb}{0.46,0.93,0.00}
\definecolor{chartreuse3}{rgb}{0.40,0.80,0.00}
\definecolor{chartreuse4}{rgb}{0.27,0.55,0.00}
\definecolor{chartreuse}{rgb}{0.50,1.00,0.00}
\definecolor{chocolate1}{rgb}{1.00,0.50,0.14}
\definecolor{chocolate2}{rgb}{0.93,0.46,0.13}
\definecolor{chocolate3}{rgb}{0.80,0.40,0.11}
\definecolor{chocolate4}{rgb}{0.55,0.27,0.07}
\definecolor{chocolate}{rgb}{0.82,0.41,0.12}
\definecolor{coral1}{rgb}{1.00,0.45,0.34}
\definecolor{coral2}{rgb}{0.93,0.42,0.31}
\definecolor{coral3}{rgb}{0.80,0.36,0.27}
\definecolor{coral4}{rgb}{0.55,0.24,0.18}
\definecolor{coral}{rgb}{1.00,0.50,0.31}
\definecolor{cornflowerblue}{rgb}{0.39,0.58,0.93}
\definecolor{cornsilk1}{rgb}{1.00,0.97,0.86}
\definecolor{cornsilk2}{rgb}{0.93,0.91,0.80}
\definecolor{cornsilk3}{rgb}{0.80,0.78,0.69}
\definecolor{cornsilk4}{rgb}{0.55,0.53,0.47}
\definecolor{cornsilk}{rgb}{1.00,0.97,0.86}
\definecolor{cyan1}{rgb}{0.00,1.00,1.00}
\definecolor{cyan2}{rgb}{0.00,0.93,0.93}
\definecolor{cyan3}{rgb}{0.00,0.80,0.80}
\definecolor{cyan4}{rgb}{0.00,0.55,0.55}
\definecolor{cyan}{rgb}{0.00,1.00,1.00}
\definecolor{darkblue}{rgb}{0.00,0.00,0.55}
\definecolor{darkcyan}{rgb}{0.00,0.55,0.55}
\definecolor{darkgoldenrod}{rgb}{0.72,0.53,0.04}
\definecolor{darkgray}{rgb}{0.66,0.66,0.66}
\definecolor{darkgreen}{rgb}{0.00,0.39,0.00}
\definecolor{darkgrey}{rgb}{0.66,0.66,0.66}
\definecolor{darkkhaki}{rgb}{0.74,0.72,0.42}
\definecolor{darkmagenta}{rgb}{0.55,0.00,0.55}
\definecolor{darkolive}{rgb}{0.33,0.42,0.18}
\definecolor{darkorange}{rgb}{1.00,0.55,0.00}
\definecolor{darkorchid}{rgb}{0.60,0.20,0.80}
\definecolor{darkred}{rgb}{0.55,0.00,0.00}
\definecolor{darksalmon}{rgb}{0.91,0.59,0.48}
\definecolor{darksea}{rgb}{0.56,0.74,0.56}
\definecolor{darkslate}{rgb}{0.18,0.31,0.31}
\definecolor{darkslate}{rgb}{0.18,0.31,0.31}
\definecolor{darkslate}{rgb}{0.28,0.24,0.55}
\definecolor{darkturquoise}{rgb}{0.00,0.81,0.82}
\definecolor{darkviolet}{rgb}{0.58,0.00,0.83}
\definecolor{deeppink}{rgb}{1.00,0.08,0.58}
\definecolor{deepsky}{rgb}{0.00,0.75,1.00}
\definecolor{dimgray}{rgb}{0.41,0.41,0.41}
\definecolor{dimgrey}{rgb}{0.41,0.41,0.41}
\definecolor{dodgerblue}{rgb}{0.12,0.56,1.00}
\definecolor{firebrick1}{rgb}{1.00,0.19,0.19}
\definecolor{firebrick2}{rgb}{0.93,0.17,0.17}
\definecolor{firebrick3}{rgb}{0.80,0.15,0.15}
\definecolor{firebrick4}{rgb}{0.55,0.10,0.10}
\definecolor{firebrick}{rgb}{0.70,0.13,0.13}
\definecolor{floralwhite}{rgb}{1.00,0.98,0.94}
\definecolor{forestgreen}{rgb}{0.13,0.55,0.13}
\definecolor{gainsboro}{rgb}{0.86,0.86,0.86}
\definecolor{ghostwhite}{rgb}{0.97,0.97,1.00}
\definecolor{gold1}{rgb}{1.00,0.84,0.00}
\definecolor{gold2}{rgb}{0.93,0.79,0.00}
\definecolor{gold3}{rgb}{0.80,0.68,0.00}
\definecolor{gold4}{rgb}{0.55,0.46,0.00}
\definecolor{goldenrod1}{rgb}{1.00,0.76,0.15}
\definecolor{goldenrod2}{rgb}{0.93,0.71,0.13}
\definecolor{goldenrod3}{rgb}{0.80,0.61,0.11}
\definecolor{goldenrod4}{rgb}{0.55,0.41,0.08}
\definecolor{goldenrod}{rgb}{0.85,0.65,0.13}
\definecolor{gold}{rgb}{1.00,0.84,0.00}
\definecolor{gray0}{rgb}{0.00,0.00,0.00}
\definecolor{gray100}{rgb}{1.00,1.00,1.00}
\definecolor{gray10}{rgb}{0.10,0.10,0.10}
\definecolor{gray11}{rgb}{0.11,0.11,0.11}
\definecolor{gray12}{rgb}{0.12,0.12,0.12}
\definecolor{gray13}{rgb}{0.13,0.13,0.13}
\definecolor{gray14}{rgb}{0.14,0.14,0.14}
\definecolor{gray15}{rgb}{0.15,0.15,0.15}
\definecolor{gray16}{rgb}{0.16,0.16,0.16}
\definecolor{gray17}{rgb}{0.17,0.17,0.17}
\definecolor{gray18}{rgb}{0.18,0.18,0.18}
\definecolor{gray19}{rgb}{0.19,0.19,0.19}
\definecolor{gray1}{rgb}{0.01,0.01,0.01}
\definecolor{gray20}{rgb}{0.20,0.20,0.20}
\definecolor{gray21}{rgb}{0.21,0.21,0.21}
\definecolor{gray22}{rgb}{0.22,0.22,0.22}
\definecolor{gray23}{rgb}{0.23,0.23,0.23}
\definecolor{gray24}{rgb}{0.24,0.24,0.24}
\definecolor{gray25}{rgb}{0.25,0.25,0.25}
\definecolor{gray26}{rgb}{0.26,0.26,0.26}
\definecolor{gray27}{rgb}{0.27,0.27,0.27}
\definecolor{gray28}{rgb}{0.28,0.28,0.28}
\definecolor{gray29}{rgb}{0.29,0.29,0.29}
\definecolor{gray2}{rgb}{0.02,0.02,0.02}
\definecolor{gray30}{rgb}{0.30,0.30,0.30}
\definecolor{gray31}{rgb}{0.31,0.31,0.31}
\definecolor{gray32}{rgb}{0.32,0.32,0.32}
\definecolor{gray33}{rgb}{0.33,0.33,0.33}
\definecolor{gray34}{rgb}{0.34,0.34,0.34}
\definecolor{gray35}{rgb}{0.35,0.35,0.35}
\definecolor{gray36}{rgb}{0.36,0.36,0.36}
\definecolor{gray37}{rgb}{0.37,0.37,0.37}
\definecolor{gray38}{rgb}{0.38,0.38,0.38}
\definecolor{gray39}{rgb}{0.39,0.39,0.39}
\definecolor{gray3}{rgb}{0.03,0.03,0.03}
\definecolor{gray40}{rgb}{0.40,0.40,0.40}
\definecolor{gray41}{rgb}{0.41,0.41,0.41}
\definecolor{gray42}{rgb}{0.42,0.42,0.42}
\definecolor{gray43}{rgb}{0.43,0.43,0.43}
\definecolor{gray44}{rgb}{0.44,0.44,0.44}
\definecolor{gray45}{rgb}{0.45,0.45,0.45}
\definecolor{gray46}{rgb}{0.46,0.46,0.46}
\definecolor{gray47}{rgb}{0.47,0.47,0.47}
\definecolor{gray48}{rgb}{0.48,0.48,0.48}
\definecolor{gray49}{rgb}{0.49,0.49,0.49}
\definecolor{gray4}{rgb}{0.04,0.04,0.04}
\definecolor{gray50}{rgb}{0.50,0.50,0.50}
\definecolor{gray51}{rgb}{0.51,0.51,0.51}
\definecolor{gray52}{rgb}{0.52,0.52,0.52}
\definecolor{gray53}{rgb}{0.53,0.53,0.53}
\definecolor{gray54}{rgb}{0.54,0.54,0.54}
\definecolor{gray55}{rgb}{0.55,0.55,0.55}
\definecolor{gray56}{rgb}{0.56,0.56,0.56}
\definecolor{gray57}{rgb}{0.57,0.57,0.57}
\definecolor{gray58}{rgb}{0.58,0.58,0.58}
\definecolor{gray59}{rgb}{0.59,0.59,0.59}
\definecolor{gray5}{rgb}{0.05,0.05,0.05}
\definecolor{gray60}{rgb}{0.60,0.60,0.60}
\definecolor{gray61}{rgb}{0.61,0.61,0.61}
\definecolor{gray62}{rgb}{0.62,0.62,0.62}
\definecolor{gray63}{rgb}{0.63,0.63,0.63}
\definecolor{gray64}{rgb}{0.64,0.64,0.64}
\definecolor{gray65}{rgb}{0.65,0.65,0.65}
\definecolor{gray66}{rgb}{0.66,0.66,0.66}
\definecolor{gray67}{rgb}{0.67,0.67,0.67}
\definecolor{gray68}{rgb}{0.68,0.68,0.68}
\definecolor{gray69}{rgb}{0.69,0.69,0.69}
\definecolor{gray6}{rgb}{0.06,0.06,0.06}
\definecolor{gray70}{rgb}{0.70,0.70,0.70}
\definecolor{gray71}{rgb}{0.71,0.71,0.71}
\definecolor{gray72}{rgb}{0.72,0.72,0.72}
\definecolor{gray73}{rgb}{0.73,0.73,0.73}
\definecolor{gray74}{rgb}{0.74,0.74,0.74}
\definecolor{gray75}{rgb}{0.75,0.75,0.75}
\definecolor{gray76}{rgb}{0.76,0.76,0.76}
\definecolor{gray77}{rgb}{0.77,0.77,0.77}
\definecolor{gray78}{rgb}{0.78,0.78,0.78}
\definecolor{gray79}{rgb}{0.79,0.79,0.79}
\definecolor{gray7}{rgb}{0.07,0.07,0.07}
\definecolor{gray80}{rgb}{0.80,0.80,0.80}
\definecolor{gray81}{rgb}{0.81,0.81,0.81}
\definecolor{gray82}{rgb}{0.82,0.82,0.82}
\definecolor{gray83}{rgb}{0.83,0.83,0.83}
\definecolor{gray84}{rgb}{0.84,0.84,0.84}
\definecolor{gray85}{rgb}{0.85,0.85,0.85}
\definecolor{gray86}{rgb}{0.86,0.86,0.86}
\definecolor{gray87}{rgb}{0.87,0.87,0.87}
\definecolor{gray88}{rgb}{0.88,0.88,0.88}
\definecolor{gray89}{rgb}{0.89,0.89,0.89}
\definecolor{gray8}{rgb}{0.08,0.08,0.08}
\definecolor{gray90}{rgb}{0.90,0.90,0.90}
\definecolor{gray91}{rgb}{0.91,0.91,0.91}
\definecolor{gray92}{rgb}{0.92,0.92,0.92}
\definecolor{gray93}{rgb}{0.93,0.93,0.93}
\definecolor{gray94}{rgb}{0.94,0.94,0.94}
\definecolor{gray95}{rgb}{0.95,0.95,0.95}
\definecolor{gray96}{rgb}{0.96,0.96,0.96}
\definecolor{gray97}{rgb}{0.97,0.97,0.97}
\definecolor{gray98}{rgb}{0.98,0.98,0.98}
\definecolor{gray99}{rgb}{0.99,0.99,0.99}
\definecolor{gray9}{rgb}{0.09,0.09,0.09}
\definecolor{gray}{rgb}{0.75,0.75,0.75}
\definecolor{green1}{rgb}{0.00,1.00,0.00}
\definecolor{green2}{rgb}{0.00,0.93,0.00}
\definecolor{green3}{rgb}{0.00,0.80,0.00}
\definecolor{green4}{rgb}{0.00,0.55,0.00}
\definecolor{greenyellow}{rgb}{0.68,1.00,0.18}
\definecolor{green}{rgb}{0.00,1.00,0.00}
\definecolor{grey0}{rgb}{0.00,0.00,0.00}
\definecolor{grey100}{rgb}{1.00,1.00,1.00}
\definecolor{grey10}{rgb}{0.10,0.10,0.10}
\definecolor{grey11}{rgb}{0.11,0.11,0.11}
\definecolor{grey12}{rgb}{0.12,0.12,0.12}
\definecolor{grey13}{rgb}{0.13,0.13,0.13}
\definecolor{grey14}{rgb}{0.14,0.14,0.14}
\definecolor{grey15}{rgb}{0.15,0.15,0.15}
\definecolor{grey16}{rgb}{0.16,0.16,0.16}
\definecolor{grey17}{rgb}{0.17,0.17,0.17}
\definecolor{grey18}{rgb}{0.18,0.18,0.18}
\definecolor{grey19}{rgb}{0.19,0.19,0.19}
\definecolor{grey1}{rgb}{0.01,0.01,0.01}
\definecolor{grey20}{rgb}{0.20,0.20,0.20}
\definecolor{grey21}{rgb}{0.21,0.21,0.21}
\definecolor{grey22}{rgb}{0.22,0.22,0.22}
\definecolor{grey23}{rgb}{0.23,0.23,0.23}
\definecolor{grey24}{rgb}{0.24,0.24,0.24}
\definecolor{grey25}{rgb}{0.25,0.25,0.25}
\definecolor{grey26}{rgb}{0.26,0.26,0.26}
\definecolor{grey27}{rgb}{0.27,0.27,0.27}
\definecolor{grey28}{rgb}{0.28,0.28,0.28}
\definecolor{grey29}{rgb}{0.29,0.29,0.29}
\definecolor{grey2}{rgb}{0.02,0.02,0.02}
\definecolor{grey30}{rgb}{0.30,0.30,0.30}
\definecolor{grey31}{rgb}{0.31,0.31,0.31}
\definecolor{grey32}{rgb}{0.32,0.32,0.32}
\definecolor{grey33}{rgb}{0.33,0.33,0.33}
\definecolor{grey34}{rgb}{0.34,0.34,0.34}
\definecolor{grey35}{rgb}{0.35,0.35,0.35}
\definecolor{grey36}{rgb}{0.36,0.36,0.36}
\definecolor{grey37}{rgb}{0.37,0.37,0.37}
\definecolor{grey38}{rgb}{0.38,0.38,0.38}
\definecolor{grey39}{rgb}{0.39,0.39,0.39}
\definecolor{grey3}{rgb}{0.03,0.03,0.03}
\definecolor{grey40}{rgb}{0.40,0.40,0.40}
\definecolor{grey41}{rgb}{0.41,0.41,0.41}
\definecolor{grey42}{rgb}{0.42,0.42,0.42}
\definecolor{grey43}{rgb}{0.43,0.43,0.43}
\definecolor{grey44}{rgb}{0.44,0.44,0.44}
\definecolor{grey45}{rgb}{0.45,0.45,0.45}
\definecolor{grey46}{rgb}{0.46,0.46,0.46}
\definecolor{grey47}{rgb}{0.47,0.47,0.47}
\definecolor{grey48}{rgb}{0.48,0.48,0.48}
\definecolor{grey49}{rgb}{0.49,0.49,0.49}
\definecolor{grey4}{rgb}{0.04,0.04,0.04}
\definecolor{grey50}{rgb}{0.50,0.50,0.50}
\definecolor{grey51}{rgb}{0.51,0.51,0.51}
\definecolor{grey52}{rgb}{0.52,0.52,0.52}
\definecolor{grey53}{rgb}{0.53,0.53,0.53}
\definecolor{grey54}{rgb}{0.54,0.54,0.54}
\definecolor{grey55}{rgb}{0.55,0.55,0.55}
\definecolor{grey56}{rgb}{0.56,0.56,0.56}
\definecolor{grey57}{rgb}{0.57,0.57,0.57}
\definecolor{grey58}{rgb}{0.58,0.58,0.58}
\definecolor{grey59}{rgb}{0.59,0.59,0.59}
\definecolor{grey5}{rgb}{0.05,0.05,0.05}
\definecolor{grey60}{rgb}{0.60,0.60,0.60}
\definecolor{grey61}{rgb}{0.61,0.61,0.61}
\definecolor{grey62}{rgb}{0.62,0.62,0.62}
\definecolor{grey63}{rgb}{0.63,0.63,0.63}
\definecolor{grey64}{rgb}{0.64,0.64,0.64}
\definecolor{grey65}{rgb}{0.65,0.65,0.65}
\definecolor{grey66}{rgb}{0.66,0.66,0.66}
\definecolor{grey67}{rgb}{0.67,0.67,0.67}
\definecolor{grey68}{rgb}{0.68,0.68,0.68}
\definecolor{grey69}{rgb}{0.69,0.69,0.69}
\definecolor{grey6}{rgb}{0.06,0.06,0.06}
\definecolor{grey70}{rgb}{0.70,0.70,0.70}
\definecolor{grey71}{rgb}{0.71,0.71,0.71}
\definecolor{grey72}{rgb}{0.72,0.72,0.72}
\definecolor{grey73}{rgb}{0.73,0.73,0.73}
\definecolor{grey74}{rgb}{0.74,0.74,0.74}
\definecolor{grey75}{rgb}{0.75,0.75,0.75}
\definecolor{grey76}{rgb}{0.76,0.76,0.76}
\definecolor{grey77}{rgb}{0.77,0.77,0.77}
\definecolor{grey78}{rgb}{0.78,0.78,0.78}
\definecolor{grey79}{rgb}{0.79,0.79,0.79}
\definecolor{grey7}{rgb}{0.07,0.07,0.07}
\definecolor{grey80}{rgb}{0.80,0.80,0.80}
\definecolor{grey81}{rgb}{0.81,0.81,0.81}
\definecolor{grey82}{rgb}{0.82,0.82,0.82}
\definecolor{grey83}{rgb}{0.83,0.83,0.83}
\definecolor{grey84}{rgb}{0.84,0.84,0.84}
\definecolor{grey85}{rgb}{0.85,0.85,0.85}
\definecolor{grey86}{rgb}{0.86,0.86,0.86}
\definecolor{grey87}{rgb}{0.87,0.87,0.87}
\definecolor{grey88}{rgb}{0.88,0.88,0.88}
\definecolor{grey89}{rgb}{0.89,0.89,0.89}
\definecolor{grey8}{rgb}{0.08,0.08,0.08}
\definecolor{grey90}{rgb}{0.90,0.90,0.90}
\definecolor{grey91}{rgb}{0.91,0.91,0.91}
\definecolor{grey92}{rgb}{0.92,0.92,0.92}
\definecolor{grey93}{rgb}{0.93,0.93,0.93}
\definecolor{grey94}{rgb}{0.94,0.94,0.94}
\definecolor{grey95}{rgb}{0.95,0.95,0.95}
\definecolor{grey96}{rgb}{0.96,0.96,0.96}
\definecolor{grey97}{rgb}{0.97,0.97,0.97}
\definecolor{grey98}{rgb}{0.98,0.98,0.98}
\definecolor{grey99}{rgb}{0.99,0.99,0.99}
\definecolor{grey9}{rgb}{0.09,0.09,0.09}
\definecolor{grey}{rgb}{0.75,0.75,0.75}
\definecolor{honeydew1}{rgb}{0.94,1.00,0.94}
\definecolor{honeydew2}{rgb}{0.88,0.93,0.88}
\definecolor{honeydew3}{rgb}{0.76,0.80,0.76}
\definecolor{honeydew4}{rgb}{0.51,0.55,0.51}
\definecolor{honeydew}{rgb}{0.94,1.00,0.94}
\definecolor{hotpink}{rgb}{1.00,0.41,0.71}
\definecolor{indianred}{rgb}{0.80,0.36,0.36}
\definecolor{ivory1}{rgb}{1.00,1.00,0.94}
\definecolor{ivory2}{rgb}{0.93,0.93,0.88}
\definecolor{ivory3}{rgb}{0.80,0.80,0.76}
\definecolor{ivory4}{rgb}{0.55,0.55,0.51}
\definecolor{ivory}{rgb}{1.00,1.00,0.94}
\definecolor{khaki1}{rgb}{1.00,0.96,0.56}
\definecolor{khaki2}{rgb}{0.93,0.90,0.52}
\definecolor{khaki3}{rgb}{0.80,0.78,0.45}
\definecolor{khaki4}{rgb}{0.55,0.53,0.31}
\definecolor{khaki}{rgb}{0.94,0.90,0.55}
\definecolor{lavenderblush}{rgb}{1.00,0.94,0.96}
\definecolor{lavender}{rgb}{0.90,0.90,0.98}
\definecolor{lawngreen}{rgb}{0.49,0.99,0.00}
\definecolor{lemonchiffon}{rgb}{1.00,0.98,0.80}
\definecolor{lightblue}{rgb}{0.68,0.85,0.90}
\definecolor{lightcoral}{rgb}{0.94,0.50,0.50}
\definecolor{lightcyan}{rgb}{0.88,1.00,1.00}
\definecolor{lightgoldenrod}{rgb}{0.93,0.87,0.51}
\definecolor{lightgoldenrod}{rgb}{0.98,0.98,0.82}
\definecolor{lightgray}{rgb}{0.83,0.83,0.83}
\definecolor{lightgreen}{rgb}{0.56,0.93,0.56}
\definecolor{lightgrey}{rgb}{0.83,0.83,0.83}
\definecolor{lightpink}{rgb}{1.00,0.71,0.76}
\definecolor{lightsalmon}{rgb}{1.00,0.63,0.48}
\definecolor{lightsea}{rgb}{0.13,0.70,0.67}
\definecolor{lightsky}{rgb}{0.53,0.81,0.98}
\definecolor{lightslate}{rgb}{0.47,0.53,0.60}
\definecolor{lightslate}{rgb}{0.47,0.53,0.60}
\definecolor{lightslate}{rgb}{0.52,0.44,1.00}
\definecolor{lightsteel}{rgb}{0.69,0.77,0.87}
\definecolor{lightyellow}{rgb}{1.00,1.00,0.88}
\definecolor{limegreen}{rgb}{0.20,0.80,0.20}
\definecolor{linen}{rgb}{0.98,0.94,0.90}
\definecolor{magenta1}{rgb}{1.00,0.00,1.00}
\definecolor{magenta2}{rgb}{0.93,0.00,0.93}
\definecolor{magenta3}{rgb}{0.80,0.00,0.80}
\definecolor{magenta4}{rgb}{0.55,0.00,0.55}
\definecolor{magenta}{rgb}{1.00,0.00,1.00}
\definecolor{maroon1}{rgb}{1.00,0.20,0.70}
\definecolor{maroon2}{rgb}{0.93,0.19,0.65}
\definecolor{maroon3}{rgb}{0.80,0.16,0.56}
\definecolor{maroon4}{rgb}{0.55,0.11,0.38}
\definecolor{maroon}{rgb}{0.69,0.19,0.38}
\definecolor{mediumaquamarine}{rgb}{0.40,0.80,0.67}
\definecolor{mediumblue}{rgb}{0.00,0.00,0.80}
\definecolor{mediumorchid}{rgb}{0.73,0.33,0.83}
\definecolor{mediumpurple}{rgb}{0.58,0.44,0.86}
\definecolor{mediumsea}{rgb}{0.24,0.70,0.44}
\definecolor{mediumslate}{rgb}{0.48,0.41,0.93}
\definecolor{mediumspring}{rgb}{0.00,0.98,0.60}
\definecolor{mediumturquoise}{rgb}{0.28,0.82,0.80}
\definecolor{mediumviolet}{rgb}{0.78,0.08,0.52}
\definecolor{midnightblue}{rgb}{0.10,0.10,0.44}
\definecolor{mintcream}{rgb}{0.96,1.00,0.98}
\definecolor{mistyrose}{rgb}{1.00,0.89,0.88}
\definecolor{moccasin}{rgb}{1.00,0.89,0.71}
\definecolor{navajowhite}{rgb}{1.00,0.87,0.68}
\definecolor{navyblue}{rgb}{0.00,0.00,0.50}
\definecolor{navy}{rgb}{0.00,0.00,0.50}
\definecolor{oldlace}{rgb}{0.99,0.96,0.90}
\definecolor{olivedrab}{rgb}{0.42,0.56,0.14}
\definecolor{orange1}{rgb}{1.00,0.65,0.00}
\definecolor{orange2}{rgb}{0.93,0.60,0.00}
\definecolor{orange3}{rgb}{0.80,0.52,0.00}
\definecolor{orange4}{rgb}{0.55,0.35,0.00}
\definecolor{orangered}{rgb}{1.00,0.27,0.00}
\definecolor{orange}{rgb}{1.00,0.65,0.00}
\definecolor{orchid1}{rgb}{1.00,0.51,0.98}
\definecolor{orchid2}{rgb}{0.93,0.48,0.91}
\definecolor{orchid3}{rgb}{0.80,0.41,0.79}
\definecolor{orchid4}{rgb}{0.55,0.28,0.54}
\definecolor{orchid}{rgb}{0.85,0.44,0.84}
\definecolor{palegoldenrod}{rgb}{0.93,0.91,0.67}
\definecolor{palegreen}{rgb}{0.60,0.98,0.60}
\definecolor{paleturquoise}{rgb}{0.69,0.93,0.93}
\definecolor{paleviolet}{rgb}{0.86,0.44,0.58}
\definecolor{papayawhip}{rgb}{1.00,0.94,0.84}
\definecolor{peachpuff}{rgb}{1.00,0.85,0.73}
\definecolor{peru}{rgb}{0.80,0.52,0.25}
\definecolor{pink1}{rgb}{1.00,0.71,0.77}
\definecolor{pink2}{rgb}{0.93,0.66,0.72}
\definecolor{pink3}{rgb}{0.80,0.57,0.62}
\definecolor{pink4}{rgb}{0.55,0.39,0.42}
\definecolor{pink}{rgb}{1.00,0.75,0.80}
\definecolor{plum1}{rgb}{1.00,0.73,1.00}
\definecolor{plum2}{rgb}{0.93,0.68,0.93}
\definecolor{plum3}{rgb}{0.80,0.59,0.80}
\definecolor{plum4}{rgb}{0.55,0.40,0.55}
\definecolor{plum}{rgb}{0.87,0.63,0.87}
\definecolor{powderblue}{rgb}{0.69,0.88,0.90}
\definecolor{purple1}{rgb}{0.61,0.19,1.00}
\definecolor{purple2}{rgb}{0.57,0.17,0.93}
\definecolor{purple3}{rgb}{0.49,0.15,0.80}
\definecolor{purple4}{rgb}{0.33,0.10,0.55}
\definecolor{purple}{rgb}{0.63,0.13,0.94}
\definecolor{red1}{rgb}{1.00,0.00,0.00}
\definecolor{red2}{rgb}{0.93,0.00,0.00}
\definecolor{red3}{rgb}{0.80,0.00,0.00}
\definecolor{red4}{rgb}{0.55,0.00,0.00}
\definecolor{red}{rgb}{1.00,0.00,0.00}
\definecolor{rosybrown}{rgb}{0.74,0.56,0.56}
\definecolor{royalblue}{rgb}{0.25,0.41,0.88}
\definecolor{saddlebrown}{rgb}{0.55,0.27,0.07}
\definecolor{salmon1}{rgb}{1.00,0.55,0.41}
\definecolor{salmon2}{rgb}{0.93,0.51,0.38}
\definecolor{salmon3}{rgb}{0.80,0.44,0.33}
\definecolor{salmon4}{rgb}{0.55,0.30,0.22}
\definecolor{salmon}{rgb}{0.98,0.50,0.45}
\definecolor{sandybrown}{rgb}{0.96,0.64,0.38}
\definecolor{seagreen}{rgb}{0.18,0.55,0.34}
\definecolor{seashell1}{rgb}{1.00,0.96,0.93}
\definecolor{seashell2}{rgb}{0.93,0.90,0.87}
\definecolor{seashell3}{rgb}{0.80,0.77,0.75}
\definecolor{seashell4}{rgb}{0.55,0.53,0.51}
\definecolor{seashell}{rgb}{1.00,0.96,0.93}
\definecolor{sienna1}{rgb}{1.00,0.51,0.28}
\definecolor{sienna2}{rgb}{0.93,0.47,0.26}
\definecolor{sienna3}{rgb}{0.80,0.41,0.22}
\definecolor{sienna4}{rgb}{0.55,0.28,0.15}
\definecolor{sienna}{rgb}{0.63,0.32,0.18}
\definecolor{skyblue}{rgb}{0.53,0.81,0.92}
\definecolor{slateblue}{rgb}{0.42,0.35,0.80}
\definecolor{slategray}{rgb}{0.44,0.50,0.56}
\definecolor{slategrey}{rgb}{0.44,0.50,0.56}
\definecolor{snow1}{rgb}{1.00,0.98,0.98}
\definecolor{snow2}{rgb}{0.93,0.91,0.91}
\definecolor{snow3}{rgb}{0.80,0.79,0.79}
\definecolor{snow4}{rgb}{0.55,0.54,0.54}
\definecolor{snow}{rgb}{1.00,0.98,0.98}
\definecolor{springgreen}{rgb}{0.00,1.00,0.50}
\definecolor{steelblue}{rgb}{0.27,0.51,0.71}
\definecolor{tan1}{rgb}{1.00,0.65,0.31}
\definecolor{tan2}{rgb}{0.93,0.60,0.29}
\definecolor{tan3}{rgb}{0.80,0.52,0.25}
\definecolor{tan4}{rgb}{0.55,0.35,0.17}
\definecolor{tan}{rgb}{0.82,0.71,0.55}
\definecolor{thistle1}{rgb}{1.00,0.88,1.00}
\definecolor{thistle2}{rgb}{0.93,0.82,0.93}
\definecolor{thistle3}{rgb}{0.80,0.71,0.80}
\definecolor{thistle4}{rgb}{0.55,0.48,0.55}
\definecolor{thistle}{rgb}{0.85,0.75,0.85}
\definecolor{tomato1}{rgb}{1.00,0.39,0.28}
\definecolor{tomato2}{rgb}{0.93,0.36,0.26}
\definecolor{tomato3}{rgb}{0.80,0.31,0.22}
\definecolor{tomato4}{rgb}{0.55,0.21,0.15}
\definecolor{tomato}{rgb}{1.00,0.39,0.28}
\definecolor{turquoise1}{rgb}{0.00,0.96,1.00}
\definecolor{turquoise2}{rgb}{0.00,0.90,0.93}
\definecolor{turquoise3}{rgb}{0.00,0.77,0.80}
\definecolor{turquoise4}{rgb}{0.00,0.53,0.55}
\definecolor{turquoise}{rgb}{0.25,0.88,0.82}
\definecolor{violetred}{rgb}{0.82,0.13,0.56}
\definecolor{violet}{rgb}{0.93,0.51,0.93}
\definecolor{wheat1}{rgb}{1.00,0.91,0.73}
\definecolor{wheat2}{rgb}{0.93,0.85,0.68}
\definecolor{wheat3}{rgb}{0.80,0.73,0.59}
\definecolor{wheat4}{rgb}{0.55,0.49,0.40}
\definecolor{wheat}{rgb}{0.96,0.87,0.70}
\definecolor{whitesmoke}{rgb}{0.96,0.96,0.96}
\definecolor{white}{rgb}{1.00,1.00,1.00}
\definecolor{yellow1}{rgb}{1.00,1.00,0.00}
\definecolor{yellow2}{rgb}{0.93,0.93,0.00}
\definecolor{yellow3}{rgb}{0.80,0.80,0.00}
\definecolor{yellow4}{rgb}{0.55,0.55,0.00}
\definecolor{yellowgreen}{rgb}{0.60,0.80,0.20}
\definecolor{yellow}{rgb}{1.00,1.00,0.00}
\usepackage{amsfonts}
\usepackage{bm}
\usepackage{pdflscape}
\usepackage{longtable}
\usepackage{multirow}
\usepackage{multicol}        
\usepackage{pdflscape}	
\usepackage{pifont}
\usepackage{gensymb}
\usepackage{amsmath}

\usepackage{booktabs}

\usepackage{graphicx}	
\usepackage{amsmath}	

\newcommand{\gskfont}{
  \bfseries 
  \color{red}
}
\newcommand{\revfont}{
  \bfseries 
  \color{blue}
}

\DeclareTextFontCommand{\gsk}{\gskfont}
\DeclareTextFontCommand{\rev}{\gskfont}
\DeclareTextFontCommand{\revm}{\revfont}

\newcommand{\radyn}{\texttt{RADYN}}
\newcommand{\radynfp}{\texttt{RADYN+FP}}
\newcommand{\fpc}{\texttt{FP}}
\newcommand{\rhpar}{\texttt{RH15D}}
\newcommand{\rh}{\texttt{RH}}

\title[\ion{O}{i} Nonthermal widths in a Solar Flare]{An Optically Thin View of the Flaring Chromosphere: Nonthermal widths in a chromospheric condensation during an X-class solar flare}

\author[G.S. Kerr et al.]{
Graham S. Kerr,$^{1,2}$\thanks{E-mail: kerrg@cua.edu}
Adam F. Kowalski,$^{3,4}$
Joel C. Allred,$^{1}$
Adrian N. Daw,$^{1}$
Melissa R. Kane,$^{3}$
\\
$^{1}$NASA Goddard Space Flight Center, Heliophysics Science Division, Code 671, 8800 Greenbelt Rd., Greenbelt, MD 20771, USA\\
$^{2}$Department of Physics, Catholic University of America, 620 Michigan Avenue, Northeast, Washington, DC 20064, USA\\
$^{3}$Department of Astrophysical and Planetary Sciences, University of Colorado, Boulder 2000 Colorado Ave, CO 80305, USA\\
$^{4}$National Solar Observatory, University of Colorado Boulder, 3665 Discovery Drive, Boulder CO 80303, USA
}

\date{Accepted XXX. Received YYY; in original form ZZZ}

\pubyear{2023}

\begin{document}
\label{firstpage}
\pagerange{\pageref{firstpage}--\pageref{lastpage}}
\maketitle

\begin{abstract}
The bulk of solar flare energy is deposited in the chromosphere. Flare ribbons and footpoints in the chromosphere therefore offer great diagnostic potential of flare energy release and transport processes. High quality observations from the IRIS spacecraft have transformed our view of the Sun's atmospheric response to flares. Since most of the chromospheric lines observed by IRIS are optically thick, forward modelling is required to fully appreciate and extract the information they carry. Reproducing certain aspects of the \ion{Mg}{ii} lines remain frustratingly out of reach in state-of-the-art flare models, which are unable to satisfactorily reproduce the very broad line profiles. A commonly proposed resolution to this is to assert that very large values of `microturbulence' is present. We asses the validity of that approach by analysing optically thin lines in the flare chromosphere from the X-class flare SOL2014-10-25T17:08:00, using the derived value of nonthermal width as a constraint to our numerical models. A nonthermal width of the order 10~km~s$^{-1}$ was found within the short-lived red wing components of three spectral lines, with relatively narrow stationary components. Simulations of this flare were produced, and in the post-processing spectral synthesis we include within the downflows a microturbulence of 10~km~s$^{-1}$. While we can reproduce the \ion{O}{i} 1355.598~\AA\ line rather well, and we can capture the general shape and properties of the \ion{Mg}{ii} line widths, the synthetic lines are still too narrow.
\end{abstract}



\section{Introduction}

The broadband transient, yet intense, enhancement to the Sun's radiative output is what characterises a solar flare. Although the bulk of the flare radiative output originates from the chromosphere, where dramatic ribbon-like features, alongside more compact footpoint-like sources emit strongly in the UV, optical and infrared \citep[][]{2011SSRv..159...19F}, the energy release site is generally agreed to located in the solar corona. Following magnetic reconnection in the corona a tremendous amount of energy is released, manifesting in many forms, including plasma heating, the production of magnetohydrodynamic waves, and particle acceleration. This energy is subsequently transported to the chromosphere, where it is dissipated, heating and ionising the plasma there. Thus, the radiation produced in the flaring chromosphere is an important, yet not fully understood, source of diagnostic potential of these various processes. It is the lower atmosphere's window on the flare's energy release, and energy transport processes, including the dynamics of the current sheet since newly reconnected magnetic field lines are rooted in the chromosphere.

In the standard flare model directed beams of nonthermal electrons are a primary means by which energy is transported to the chromosphere. These are thermalised by Coulomb collisions in the relatively dense chromosphere/lower transition region, producing hard X-rays. The energy spectrum of those X-rays can be inverted to obtain properties of the parent electron distribution \citep[e.g.][]{1971SoPh...18..489B,2011SSRv..159..107H,2011SSRv..159..301K}. In addition to a distribution of nonthermal electrons, it is possible that nonthermal protons \citep[e.g.][]{2000AIPC..522..401R,2009ApJ...698L.152S,2012ApJ...759...71E,2023ApJ...945..118K}, thermal conduction \citep[e.g.][]{Antiochos1978,1983ApJ...265.1090C,1986SoPh..103...47M}, and downward propagating high-frequency Alfv\'enic waves \citep[][]{1982SoPh...80...99E,2008ApJ...675.1645F,2013A&A...558A..76R,2016ApJ...818L..20R,2018ApJ...853..101R,2016ApJ...827..101K} also play a role.  As new field lines reconnect, the ribbons and footpoints in the chromosphere undergo apparent motions, as energy is transported to the footpoints of the magnetic loops. The intense plasma heating drives mass flows through pressure imbalance, filling the coronal loops with chromospheric material, known as `chromospheric evaporation.' Conservation of momentum also results in downflows of dense material, known as `chromospheric condensations'. There is a rich observational and theoretical literature of chromospheric evaporation and condensations, and a few select but by no means exhaustive examples are \cite{2006ApJ...638L.117M,2009ApJ...699..968M,2022ApJ...936...85S,2015ApJ...807L..22G,2020ApJ...895....6G,1985ApJ...289..414F,1985ApJ...289..425F,1985ApJ...289..434F}. For reviews that touch of these topics see also: \cite{2011SSRv..159...19F}, \cite{2015SoPh..290.3399M}, and \cite{2022FrASS...960856K}. The dense and  hot flare loops subsequently emit in extreme-UV and soft X-rays, producing the flare arcade

The launch of the slit-scanning Interface Region Imaging Spectrograph \citep[IRIS;][]{2014SoPh..289.2733D} opened a new spectroscopic window on the flaring chromosphere, offering routine observations in the near-UV and far-UV at high spectral, spatial and temporal resolution. For a recent review of the IRIS mission consult \cite{2021SoPh..296...84D}, and for reviews that focus in particular on the marriage of IRIS observations and flare loop models, relevant to the latter half of this study, see \cite{2022FrASS...960856K} and \cite{2023FrASS...960862K}. Though observing several optically thin lines from the transition region or hot ($\sim11$~MK) flare plasma IRIS primarily observes the chromosphere via optically thick spectral lines, like most ground-based observations of the chromosphere.

Consequently, the majority of chromospheric flare studies using IRIS observations have focussed on trying to interpret the optically thick \ion{Mg}{ii} h \& k resonance lines, and the subordinate triplet that form nearby. These lines are seen to brighten, broaden significantly, and exhibit both redshifts and red-wing asymmetries. Additionally, the central reversal often fills in such that the lines are single peaked. The subordinate lines transition from absorption to emission. See \cite{2015A&A...582A..50K}, \cite{2015SoPh..290.3525L}, \cite{2018ApJ...861...62P}, and \cite{2016ApJ...827...38R} for discussions regarding their general flare responses. Using machine learning techniques to cluster \ion{Mg}{ii} NUV profiles from more than 30 flares into representative flare spectra, \cite{2018ApJ...861...62P} also identified that at the leading edge of flare ribbons, also known as ribbon fronts, the h \& k lines appear quite different. They have deep central reversals, a slightly blueshifted line core, and are even broader than `typical' flare profiles that themselves broaden to have a full width at half maximum up to the order of $\sim1$~\AA. Numerical studies of the \ion{Mg}{ii} flare profiles have been successful in identifying that a large electron density ($n_{e} > 5\times10^{14}$~cm$^{-3}$ at the core formation height) can result in single peaked profiles \citep[][]{2017ApJ...842...82R,2019ApJ...879...19Z}, that redshifts and red-wing asymmetries are related to chromospheric condensations \citep[e.g.][]{2020ApJ...895....6G}, and that ribbon front behaviours can be explained by a relatively weak flux of energetic nonthermal electrons \citep[e.g.][who found that an energy flux $F< \sim 1-5\times10^9$~erg~s$^{-1}$~cm$^{-2}$ was more consistent with ribbon front profiles, compared to more commonly used values on the order $F = 10^{10-11}$~erg~s$^{-1}$~cm$^{-2}$]{2023ApJ...944..104P}. 

However, there are some outstanding questions regarding the \ion{Mg}{ii} spectra in flares. Primary amongst these is the source of such large broadening. Radiative hydrodynamic modelling of the \ion{Mg}{ii} spectra in flares is as-yet unable to satisfactorily produce very broad profiles. Several attempts have been made to rectify this. \cite{2019ApJ...879...19Z} improved the treatment of Stark damping, but still found that a factor 30x the Stark damping was necessary to be consistent with the observations. Might this suggest the need for a stronger density enhancement deep in the atmosphere? Other attempts have largely focussed on either unresolved large bi-directional macroscopic flows \citep[e.g.][]{2017ApJ...842...82R}, with downflows in excess of $>150-200$~km~s$^{-1}$ required in the chromosphere, of which is there is scant evidence\footnote{That is, we do not observe such large Doppler shifts of other more narrow chromospheric lines, though these flows could in theory be confined to a small range of altitudes. Typical flows in chromospheric condensations are on the order a few 10-100~km~s$^{-1}$}, or on enhanced microturbulent broadening \citep[e.g.][]{2017ApJ...842...82R,2019ApJ...878L..15H,2019ApJ...879...19Z}. A large magnitude of microturbulent broadening through the line core formation region, and into the deeper chromosphere, of $30-50$~km~s$^{-1}$ was required to approach the width of the observed spectra in those studies. It is not clear if such large values, which approach the chromospheric sound speed, are realistic. 

Although optically thick lines are usually very non-Gaussian in shape, any microturbulence present in the atmosphere could still act to the broaden the core and near wings of the lines, in the same way that both optically thick and thin lines are broadened thermally. The difficulty is that unlike optically thin lines, subsequent scattering and opacity effects redistribute those those photons, making it non-trivial to separate the various mechanisms that result in broadening above the thermal width, and thus difficult to measure turbulence in the lower solar atmosphere.

One means to estimate the actual values of microturbulence that would act to broaden the \ion{Mg}{ii} lines is to measure the nonthermal width of an optically thin spectral line that forms in the same volume or at least very close in altitude to the \ion{Mg}{ii} lines, with the assumption that this nonthermal width is due to microturbulence. 
For the time being, the origin or nature of this turbulence does not matter (e.g. fluid/plasma instabilities, sheared flows), we are simply placing a limit on its role and do not speculate further. Nonthermal widths of optically thin lines in the transition region and corona have been studied extensively, both in flares and other solar phenomenon \citep[see these reviews:][]{2015SoPh..290.3399M,2018LRSP...15....5D,2021SoPh..296...84D}, but this cannot be said of the chromosphere. 

Fortunately, IRIS does observe a small number of optically thin lines in the chromosphere, though these have been studied very little in comparison to the much stronger optically thick lines. Further, it is usually the case that only portions of the detector are read out, and that the optically thin \ion{O}{i} 1355.598~\AA\ line is not fully observed since it is at the edge of most observation window linelists. Here we perform an observational analysis and numerical modelling study of the \ion{O}{i} 1355.598~\AA\ line, along with two other chromospheric lines that may be optically thin. We use the results of that study to infer an observationally constrained value of microturbulence present in the flaring chromosphere and, by including the microturbulence as a free parameter in our modelling, make an assessment of whether this is sufficient to explain the magnitude of \ion{Mg}{ii} line broadening in flares.

We address in this manuscript: (1) what are the observed spectral line properties, in particular the nonthermal widths, within the optically thin flare chromosphere?; (2) is our forward modelling of \ion{O}{i} 1355.598~\AA\ in a flare simulation with a chromospheric condensation consistent with observations?; (3) does \ion{O}{i} 1355.598~\AA\ form near the Mg II NUV lines, and does it remain optically thin?; and (4) is the observed microturbulence added within the condensation in our model sufficient to broaden the \ion{Mg}{ii} lines or do we still need additional physics in our models to explain \ion{Mg}{ii} line widths?

\section{Line selection}
The \ion{O}{i} 1355.598~\AA\ formation properties were studied most recently by \cite{2015ApJ...813...34L}, who used an MHD model to understand quiet Sun properties of this line. They found that it was optically thin and therefore suitable for assessing nonthermal widths in the chromosphere around its formation temperature that in their radiative magnetohydrodynamics (RMHD) model ranged around $T\approx6-10$~kK. The \ion{O}{i}/\ion{O}{ii} ionisation stratification was dependent on charge exchange via neutral hydrogen, meaning that in flares the strongly varying H ionisation stratification will impact the formation of \ion{O}{i}. The intercombination 1355.598~\AA\ line is populated largely by cascades from higher lying levels, and the intensity dependent on the square of the electron density. This means that the non-equilibrium flaring H ionisation and electron density stratification are crucial when predicting \ion{O}{i} 1355.598~\AA\ emission, requiring radiation transfer modelling. To our knowledge the only flare observations at the time of writing include \cite{1980A&A....86..377C} and \cite{2016ApJ...829...35W}. The former found that during flares the ratio of \ion{O}{i} to the nearby \ion{C}{i} 1355.844~\AA\ lines decreases such that the \ion{C}{i} line is much stronger than \ion{O}{i} (possibly due to electron density effects), and the latter noted that in a B class flare the line showed no notable Doppler motions or strong broadening. 

A crucial assumption in our methodology is that the \ion{O}{i} 1355.598~\AA\ line forms near \ion{Mg}{ii} in flares, and that the \ion{O}{i} 1355.598~\AA\ remains optically thin throughout the flare. In their numerical study of \ion{Si}{iv} in flares, \cite{2019ApJ...871...23K} showed the relative formation height of the line core of several lines, including \ion{O}{i} which was indeed nearby \ion{Mg}{ii}. Further, they note that the \ion{O}{i} 1355.598~\AA\ line seems to remain optically thin. However, they did not perform a detailed analysis of the \ion{O}{i} formation in their model. We show a more detailed analysis of \ion{O}{i} in a data-guided flare simulation here. 

Alongside \ion{O}{i} we analyse \ion{Cl}{i} 1351.66~\AA, and \ion{Fe}{ii} 2814.445~\AA. The \ion{Cl}{i} 1351.66~\AA\ line was studied by \cite{1983ApJ...266..882S}, who found that it was stronger than expected due to optical pumping by \ion{C}{ii} lines. They found that the optical depth was around unity at some height in the chromosphere. Opacity likely plays a role in its formation, but \cite{1983ApJ...266..882S} note that the line is `effectively thin', meaning that photons can escape without being absorbed but may be scattered in angle, frequency or between other \ion{Cl}{i} levels. So, although stronger than \ion{O}{i} this line is less ideal in that any nonthermal width measurement must come with the caveat that it is an upper limit due to potential opacity effects. \cite{1983ApJ...266..882S} also note that modelling with a more realistic \ion{C}{ii} resonance line radiation field could result in significant changes to their modelling, and no flare modelling has been performed. Since the flare we observed includes a rare observation of this line and that we noticed a similar red-wing asymmetry in both \ion{O}{i} and \ion{Cl}{i}, we elected to include \ion{Cl}{i} 1351.66~\AA\ in our analysis in case the shifted component in particular was optically thin.

We include the \ion{Fe}{ii} for similar reasons. This line was analysed observationally by \cite{2017ApJ...836...12K}, \cite{2019ApJ...878..135K} and \cite{2020ApJ...895....6G}, who noted that its relative narrowness and smaller optical depth compared to the \ion{Mg}{ii} resonance lines made it better suited to studying chromospheric condensations. The lines included either a red wing asymmetry, or a wholly separate red wing component that subsequently merged with the stationary component (that is, the component near the rest-wavelength). \cite{2020ApJ...895....6G} found that the \ion{Fe}{ii} line wing formed along the leading edge of the chromospheric condensation in their flare model. Further comments on the opacity of the line during flares is included later in this manuscript.

\section{Observations}\label{sec:obs}

\begin{figure*}
	\centering 
	\hbox{
	\hspace{0.5in}
	\subfloat{\includegraphics[width = .275\textwidth, clip = true, trim = 0.cm 0.cm 0.cm 0.cm]{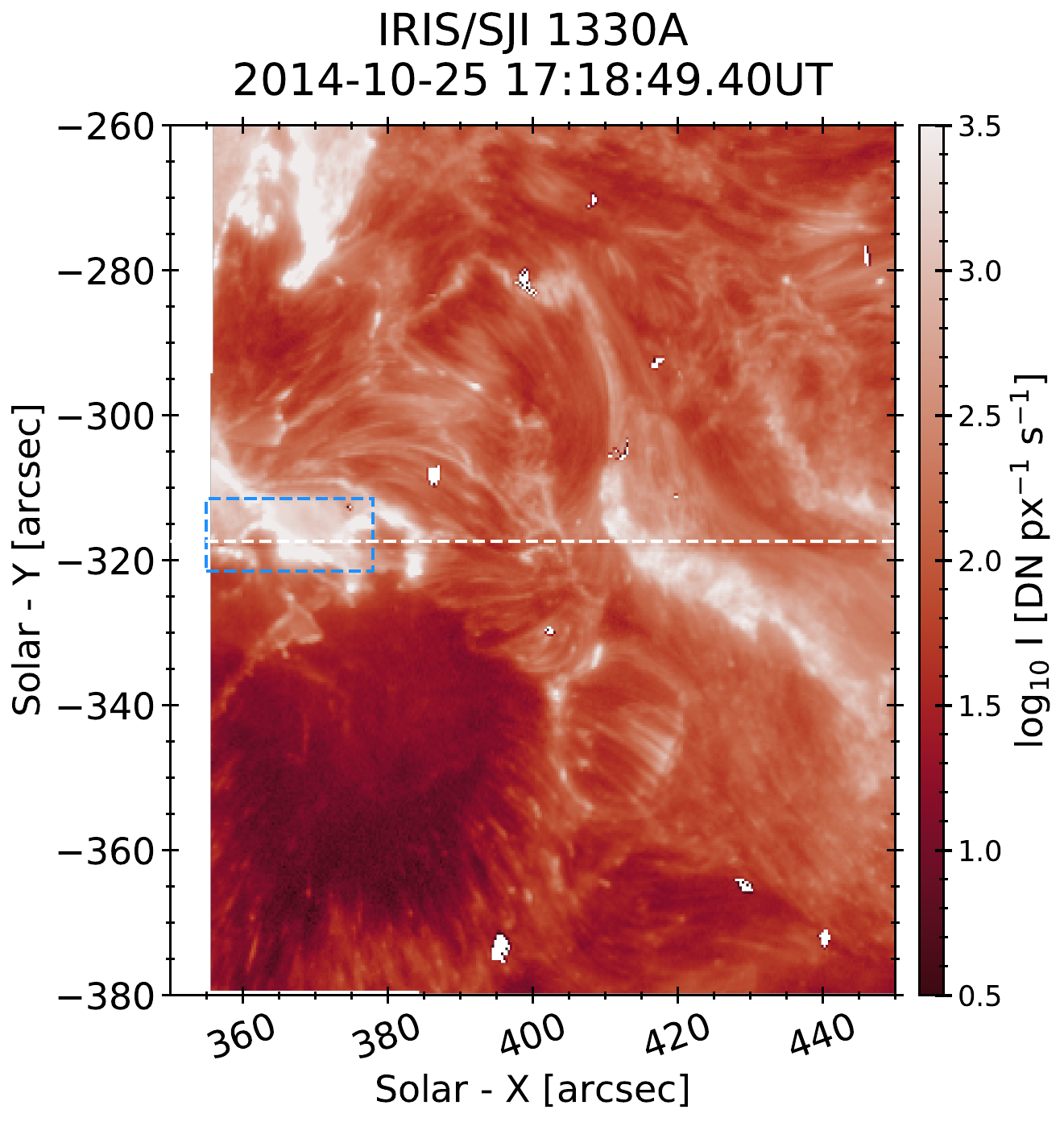}}	
	\subfloat{\includegraphics[width = .275\textwidth, clip = true, trim = 0.cm 0.cm 0.cm 0.cm]{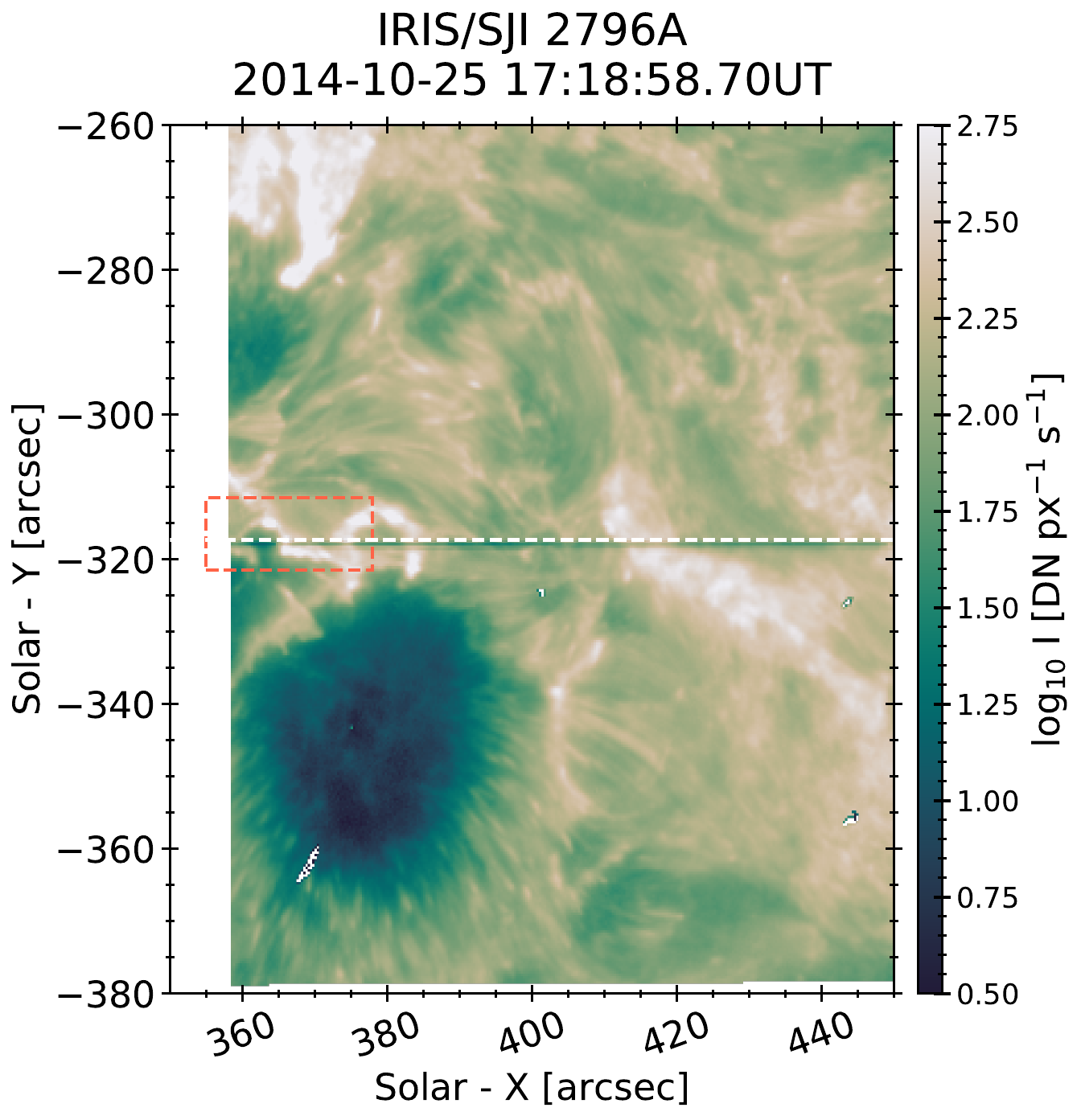}}	
	\subfloat{\includegraphics[width = .275\textwidth, clip = true, trim = 0.cm 0.cm 0.cm 0.cm]{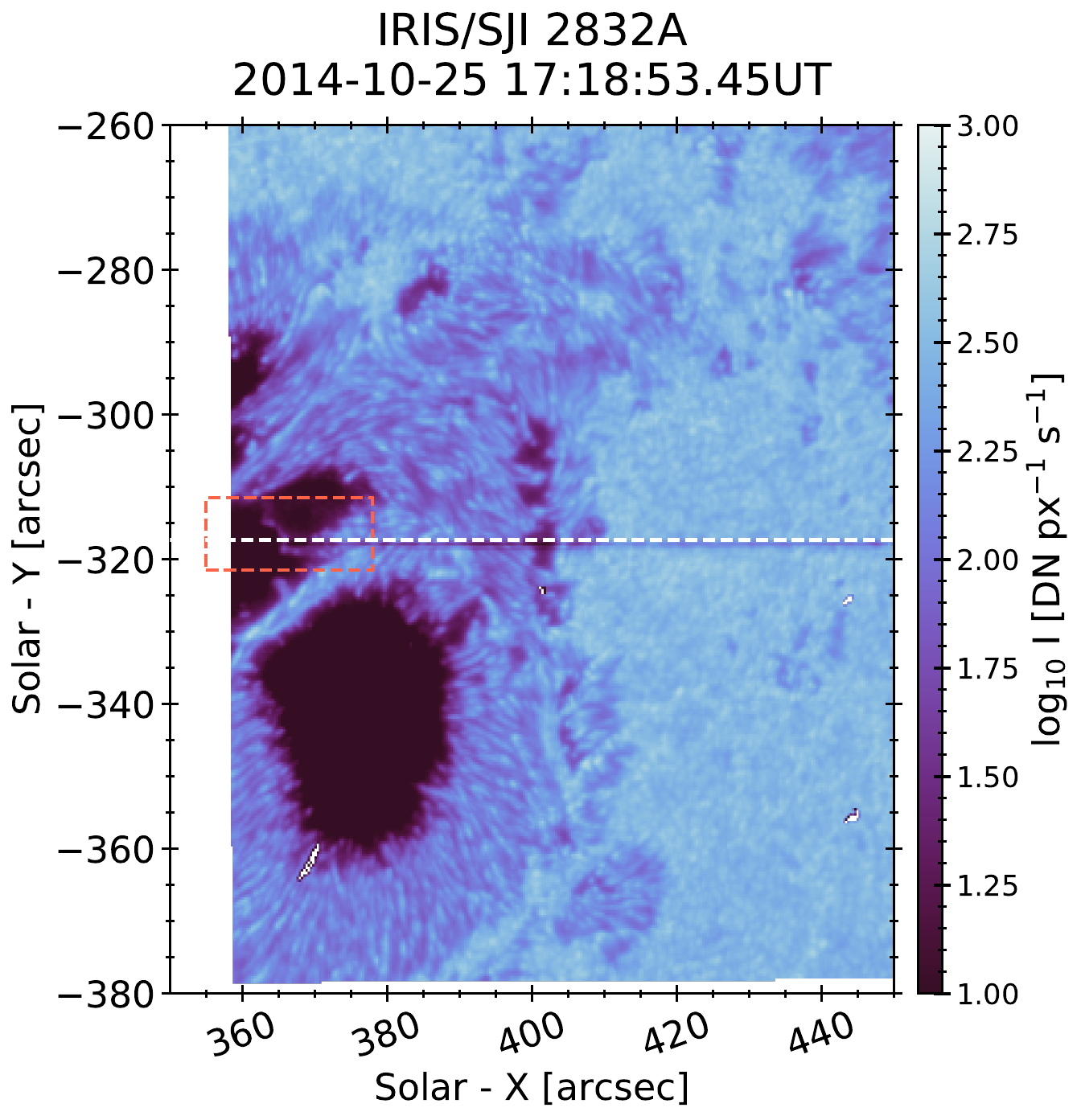}}	
	}
	\caption{\textsl{Context for the X-class 2014-Oct-25 solar flare, focussing on a time near the fourth, and final, hard X-ray burst \citep[see hard X-ray analysis in ][]{2019ApJ...878..135K}. Shown are three IRIS SJI images, $1330$~\AA, $2796$~\AA\ \& $2832$~\AA\ from left to right. The flare ribbons are clear in the first two panels, where emission is dominated by the \ion{C}{ii} resonance lines and \ion{Mg}{ii} k line, respectively. In the final image the flare ribbons are not very discernible, since this passband largely samples the continuum, but there a feature in the sunspot umbra/penumbra is present within the highlighted area. The highlighted area, bounded by a red or blue box, is the main region of interest for the remainder of the study, which is a flare ribbon referred to as XR3 by \citet{2017ApJ...838..134B} and \citet{2019ApJ...878..135K}. The white dashed line is the location of the IRIS SG slit.}}
	\label{fig:iris_sji_overview}
\end{figure*}

The GOES X1-class flare from 2014 October 25 at $\approx$ 17UT (SOL2014-10-25T17:08:00) was observed by IRIS, the Reuven Ramaty High-Energy Solar Spectroscopic Imager \citep[RHESSI; ][]{2002SoPh..210....3L}, and the \textsl{Fermi} Gamma-Ray Burst Monitor \citep[GBM; ][]{2009ApJ...702..791M}, the latter two of which provide important context regarding the presence of high-energy particles. This event has been studied by numerous authors \citep{2017ApJ...838..134B,2017ApJ...837..160K,2022ApJ...926..164A,2019PASJ...71...14L,2019ApJ...878..135K}, in part because IRIS was operating with full spectral readout, a rather uncommon mode. 

The flare had a complicated three-ribbon structure, of which IRIS caught two. \cite{2017ApJ...838..134B} christened the ribbons from solar west to east XR1, XR2, and XR3. \cite{2019ApJ...878..135K} studied continuum emission from XR3, which brightened during the fourth hard X-ray (HXR) peak as observed by \textsl{Fermi}/GBM. This ribbon swept into the penumbra and umbra where it crossed the IRIS slit. Figure~\ref{fig:iris_sji_overview} shows a snapshot of this ribbon near its peak, where XR3 is highlighted by the box near $x = 370$\arcsec, $y=-315$\arcsec. In that figure each image is a different filter from IRIS' Slit-Jaw Imager. The other ribbon present in those images is XR2. Four bursts of HXRs were observed by \textsl{Fermi}/GBM, at 16:59:17UT, 17:00:23UT, 17:03:31UT, and 17:17:27UT. We study UV emission associated in time with the last of these bursts. Though imaging is not possible using \textsl{Fermi}/GBM data both \cite{2017ApJ...837..160K} and \cite{2022ApJ...926..164A} performed imaging of RHESSI HXRs, finding a coronal source with extensions down to footpoints co-temporal with the 1st \textsl{Fermi}/GBM burst, and with a purely coronal source at a time in between the 3rd and 4th \textsl{Fermi}/GBM bursts. We do not present a further analysis of the HXR data here, but comment further in Section~\ref{sec:numerics} where it is relevant to data-guided modelling of this flare. 

For these observations IRIS was operating in sit-and-stare mode, with a full readout of the spectral detectors with a relatively high cadence (averaging $\sim5.36$~s) with an exposure time of $\tau_{\mathrm{exp}}\sim4$~s. The spectrograph observed in the following channels: the far-UV short channel (FUVS) at $\lambda = [1332-1358]$~\AA, far-UV long channel (FUVL) at  $\lambda = [1389-1407]$~\AA, and the near-UV channel (NUV) at $\lambda = [2783-2835]$~\AA. On-board two-pixel summing was performed both in the spectral and spatial dimensions, such that the spectra were obtained along the slit with $0.33$\arcsec pix$^{-1}$, and with spectral spacing of $50.92$~m\AA~pix$^{-1}$ in the NUV,  $25.96$~m\AA~pix$^{-1}$ in the FUVS, and $25.44$~m\AA~pix$^{-1}$ in the FUVL. IRIS' Slit-Jaw Imager provided images in the 1330~\AA\ (\ion{C}{ii}, $T\sim10-30$~kK), 2796~\AA\ (\ion{Mg}{ii} k, $T\sim5-15$~kK), and 2832~\AA\ (\ion{Mg}{ii} wing, $T\sim5-6.5$~kK) filters with a typical cadence of $\sim16$~s. Quoted formation temperatures are the pre-flare values.

\subsection{General Comments on the Flaring Spectra}\label{sec:overview}
\begin{figure*}
	\centering 
	\vbox{
	\hbox{
	\subfloat{\includegraphics[width = 0.475\textwidth, clip = true, trim = 0.cm 0.cm 0.cm 0.cm]{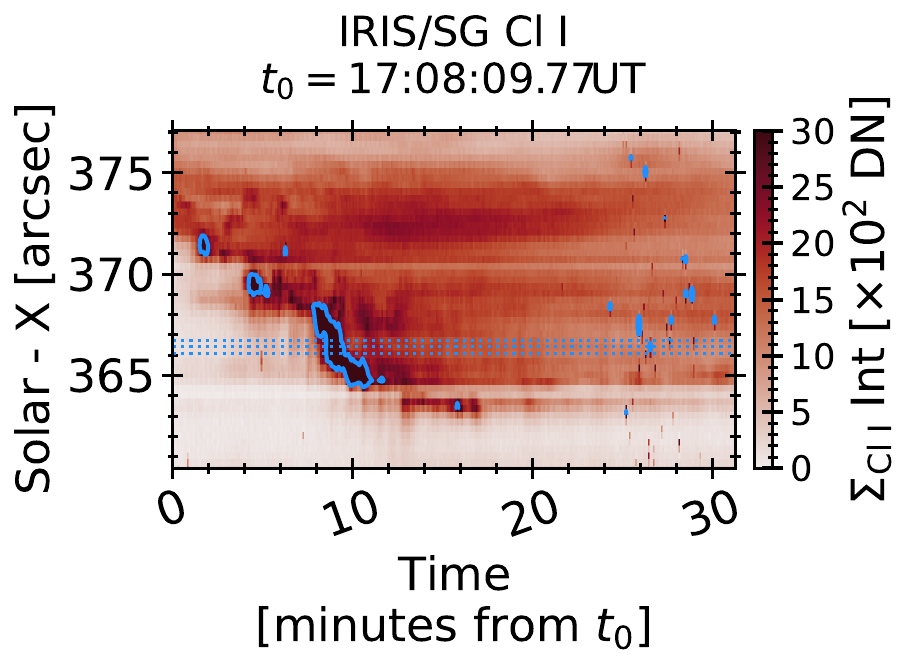}}	
	\subfloat{\includegraphics[width = 0.475\textwidth, clip = true, trim = 0.cm 0.cm 0.cm 0.cm]{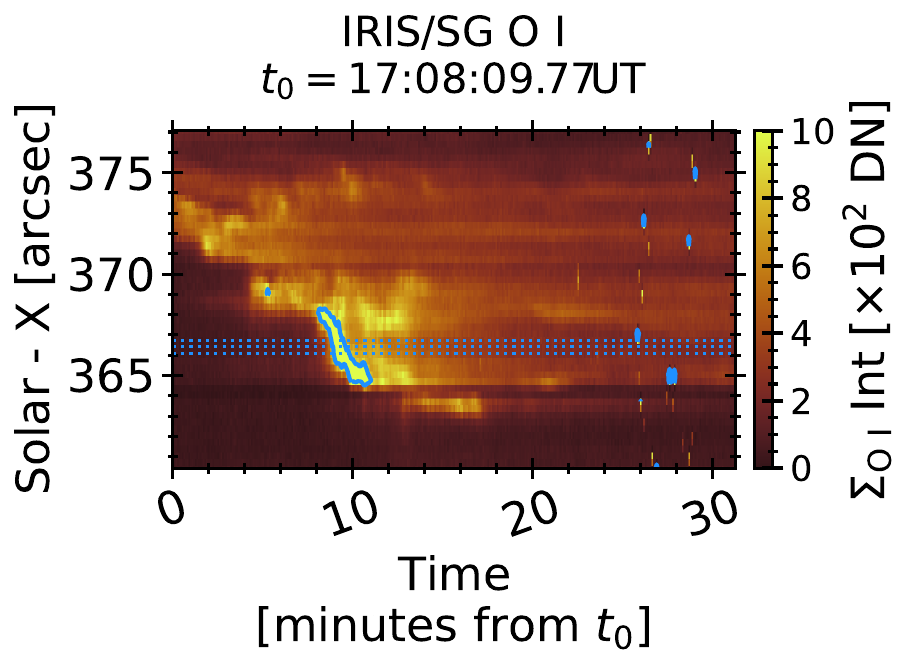}}
	}
	}
	\vbox{
	\hbox{
	\subfloat{\includegraphics[width = 0.475\textwidth, clip = true, trim = 0.cm 0.cm 0.cm 0.cm]{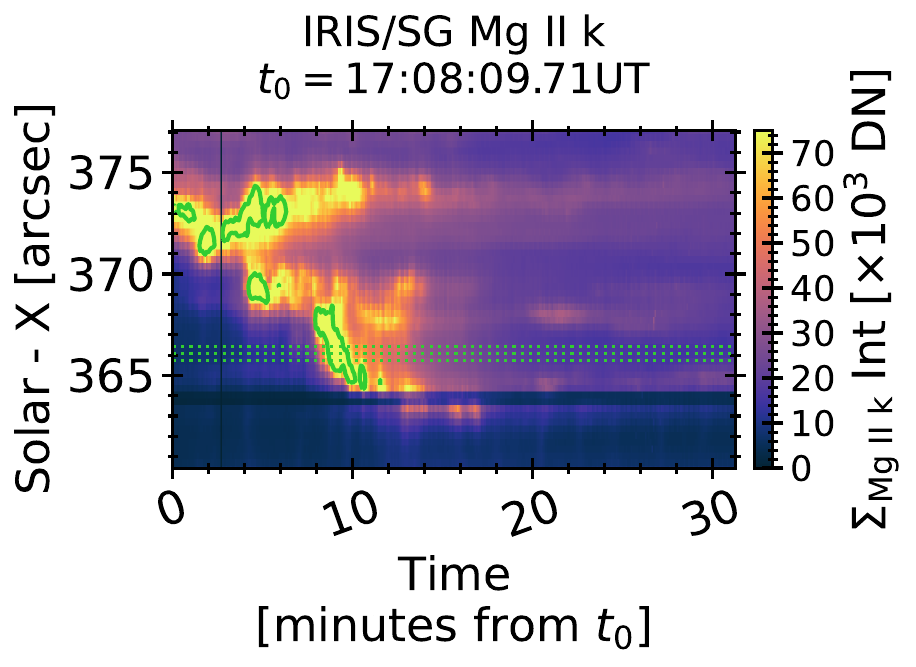}}	
	\subfloat{\includegraphics[width = 0.475\textwidth, clip = true, trim = 0.cm 0.cm 0.cm 0.cm]{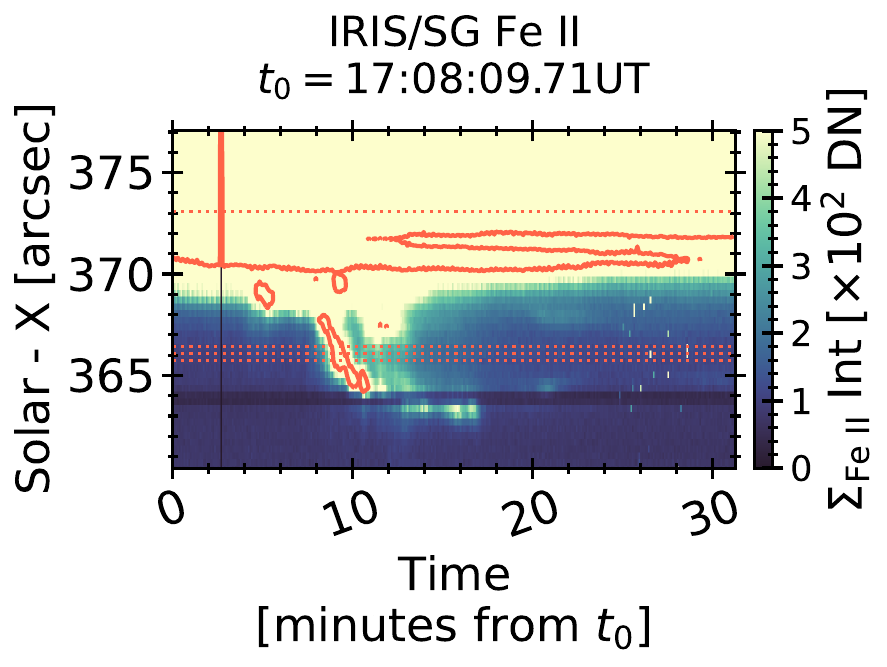}}	
	}
	}
	\caption{\textsl{Spacetime intensity diagrams for the four chromospheric spectral lines studied in detail. Top left: \ion{Cl}{i} $1351.657^{+ 0.343}_{-0.257}$~\AA\ (the blue contours show 3000~DN). Top right: \ion{O}{i} integrated $1355.598^{+ 0.135}_{-0.075}$~\AA\ (the blue contours show 1000~DN). Bottom left: \ion{Mg}{ii} k integrated $2796.34^{+0.5}_{-0.5}$~\AA\ (green contours show $10^{5}$~DN). Bottom right:  \ion{Fe}{ii} integrated $2814.445^{+ 0.500}_{-0.125}$~\AA\ (the red contours show 1000~DN). Note that in this panel the region above the ribbon of interest is much brighter in the NUV continuum, such that it is orders of magnitude larger than the intensity of the ribbon appearing in the penumbra making the \ion{Fe}{ii} line there indistinguishable with this scaling. In each panel the dashed horizontal lines indicate the pixels along the slit studied in detail. The y-axis on this figure is solar-X direction (the x-axis along the rectangle shown on Figure~\ref{fig:iris_sji_overview}).}}
	\label{fig:iris_spacetime_overview}
\end{figure*}

Spacetime diagrams of XR3, around the time of the fourth large scale HXR burst and heating episode in the transition region/chromosphere (inferred from brightenings of the IRIS SJI observations), are shown in Figure~\ref{fig:iris_spacetime_overview}. Each panel is a different spectral line, where the intensity is integrated over the extent of the line. Top left is \ion{Cl}{i} 1355.66~\AA, top right is \ion{O}{i} 1355.598~\AA, bottom left is the \ion{Mg}{ii} k line, and the bottom right is the \ion{Fe}{ii} 2814.445~\AA\ line. In each panel the strongest sources are highlighted by contours, the level of which is stated in the figure caption. Flaring emission in each line follows a similar evolution and morphology, with the ribbon propagating east along the slit with time (which is downward in the maps in Figure~\ref{fig:iris_spacetime_overview}). The flare sources generally persist at a very strong level for less than two minutes, after which their intensity falls below the contour threshold. They do, however, have a more gradual decay following this initial decrease, staying brigher than the pre-flare for many minutes. Since outside of the penumbra/umbra the background continuum level is very much stronger, the \ion{Fe}{ii} line is in absorption or only weakly in emission and difficult to discern, only showing as a prominent emission line in the lower part of the maps. From these maps of the source evolution three pixels were selected for detailed study later in the manuscript, indicated by the dotted horizontal lines.

\begin{figure*}
	\centering 
	\hbox{
	\hspace{0.85in}
	\subfloat{\includegraphics[width = 0.65\textwidth, clip = true, trim = 0.cm 0.cm 0.cm 0.cm]{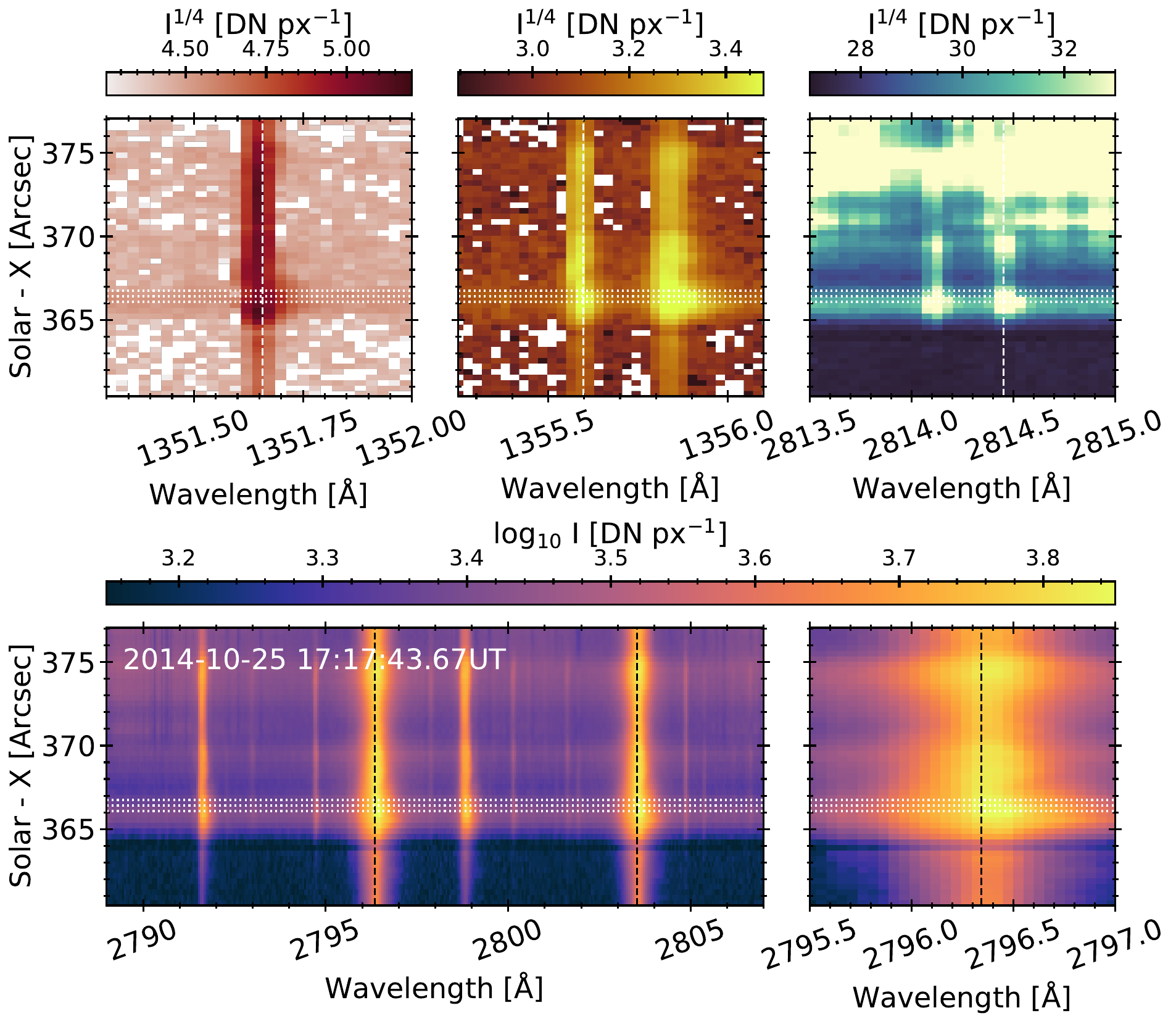}}	
	}
	\caption{\textsl{An overview of chromospheric spectra in the X-class 2014-Oct-25 solar flare, focussing on the region of interest indicated in Figure~\ref{fig:iris_sji_overview}. Each panel is a different spectral transition, with the rest wavelength of the lines of interest indicated by vertical dashed lines. In each panel the white dotted lines indicated three pixels that are studied in detail. Shown are: \ion{Cl}{i} 1351.657~\AA\  (top left; intensity scaled to 1/4 power), \ion{O}{i} 1355.598~\AA\ (top middle; intensity scaled to 1/4 power), \ion{Fe}{ii} 2814.445~\AA\ (top right; intensity scaled to 1/4 power), \ion{Mg}{ii} near-UV resonance and subordinate lines (bottom left; intensity on a logarithmic scale), and a more detailed view on the \ion{Mg}{ii} k line (bottom right; intensity on a logarithmic scale). It is clear during the flare the lines brighten, broaden, and are asymmetric with red wing enhancements. The dark band most clearly seen in the NUV data is the fiducial mark on the IRIS detector, used to align the FUV and NUV observations. Note that since IRIS was rotated by 90$\degree$ the slit position on the y-axis reflects position E-W on the Sun (solar-X).}}
	\label{fig:iris_sg_overview}
\end{figure*}

Inspecting the spectra near the peak of the this heating episode reveals intensity enhancements, some line broadening, and modest Doppler shifts that mostly manifest as red wing asymmetries. Figure~\ref{fig:iris_sg_overview} shows spectrograms of the \ion{Cl}{i} 1351.66~\AA\ line, the \ion{O}{i} 1355.598~\AA\ line and the adjacent \ion{C}{i} 1355.844~\AA\ line, the \ion{Fe}{ii} 2814.445~\AA\ line, and the \ion{Mg}{ii} NUV spectra including the h \& k resonance lines (2796.34~\AA, 2803.53~\AA) and subordinate triplet (2791.6~\AA, 2798.75~\AA, 2798.82~\AA). A zoomed-in view of the \ion{Mg}{ii} k line is included also. The flare portion near $x = 365-369$\arcsec\ exhibits clear asymmetries, though we note that the overall width of the \ion{Mg}{ii} lines are not as broad as some other flares \citep[e.g.][]{2015A&A...582A..50K,2015SoPh..290.3525L,2018ApJ...861...62P,2019ApJ...878L..15H}.

\begin{figure*}
	\centering 
	\vbox{
	\hbox{
	\hspace{0.5in}
	\subfloat{\includegraphics[width = 0.4\textwidth, clip = true, trim = 0.cm 0.cm 0.cm 0.cm]{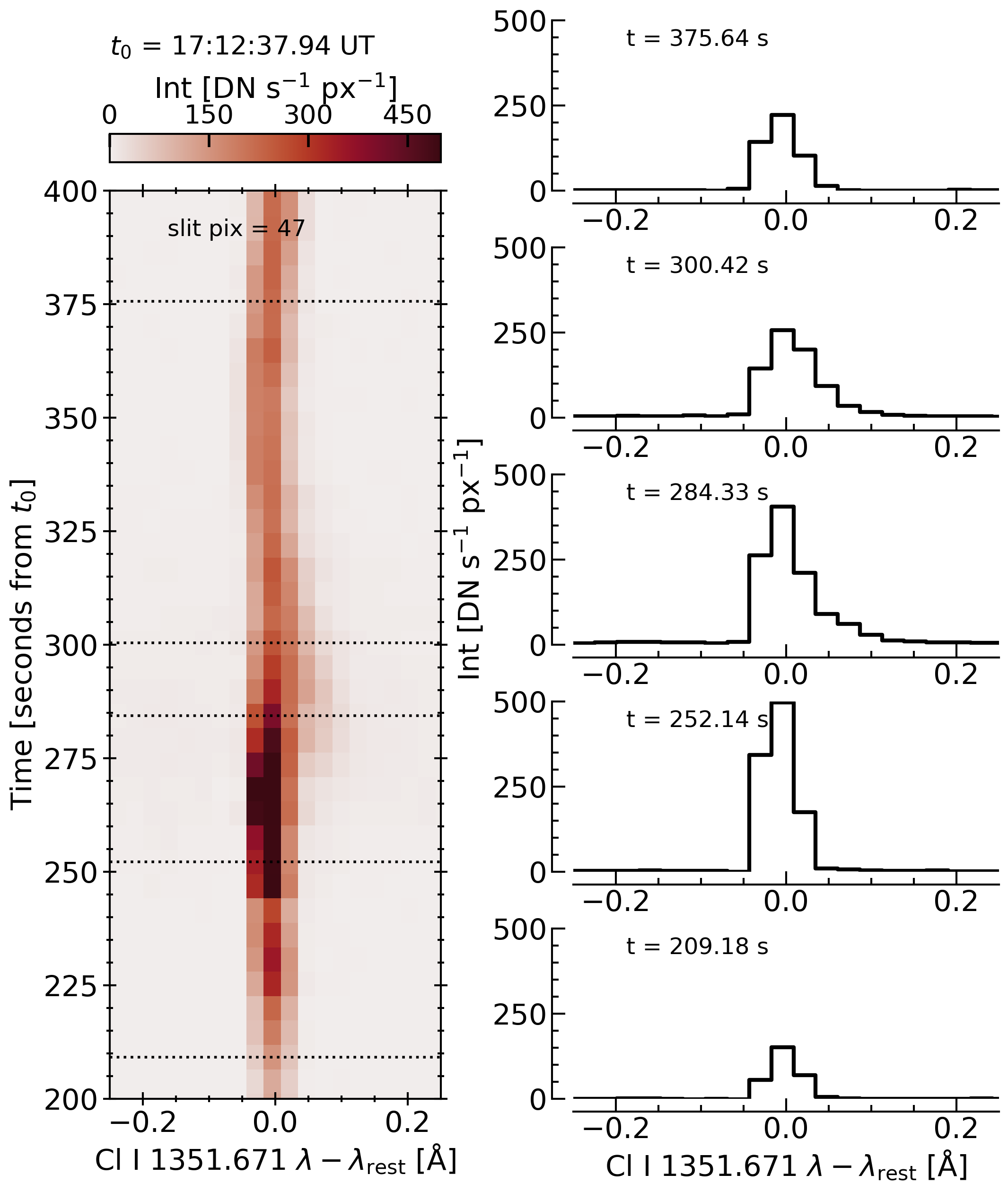}}
	\subfloat{\includegraphics[width = 0.4\textwidth, clip = true, trim = 0.cm 0.cm 0.cm 0.cm]{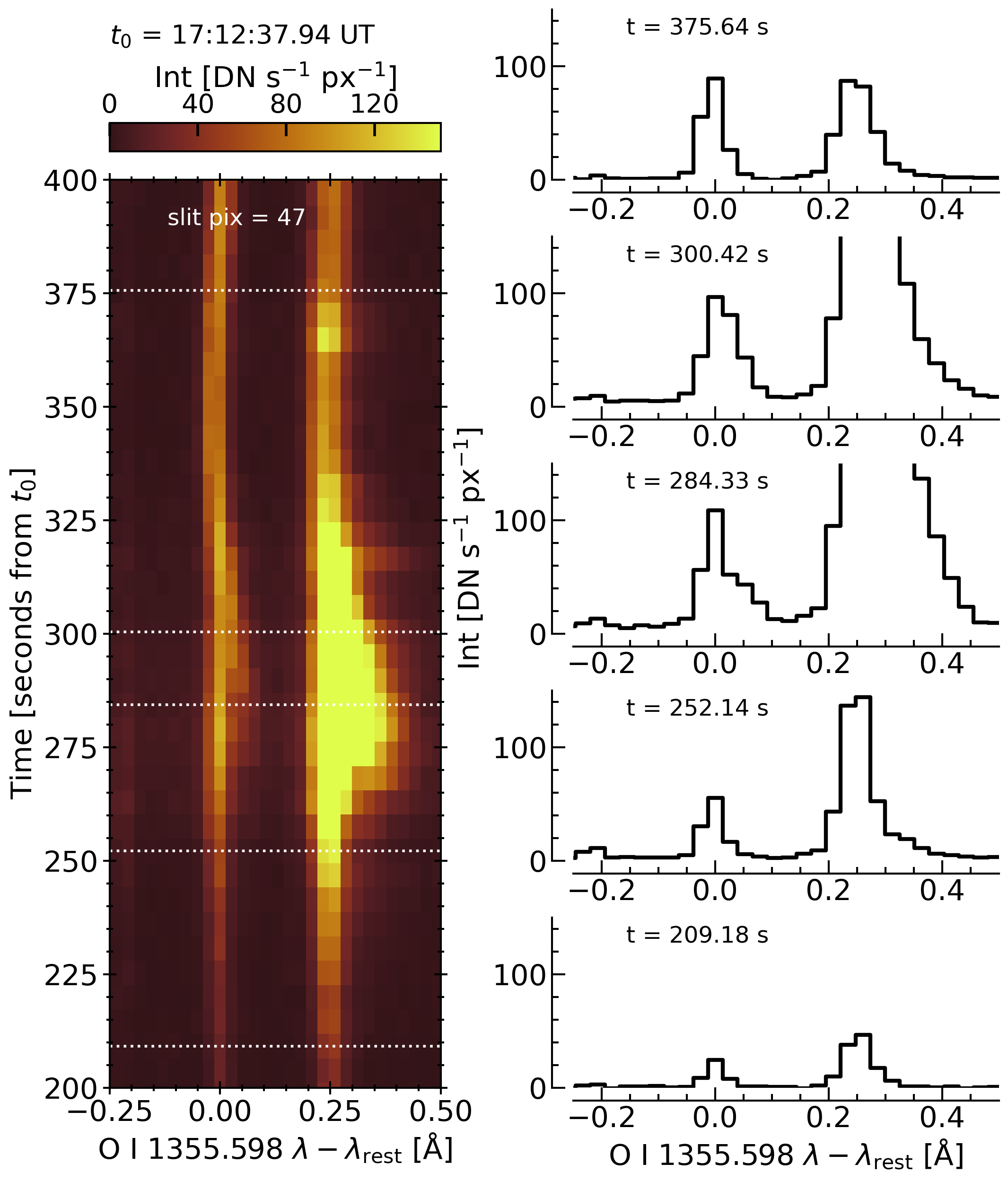}}		
	}
	}
	\vbox{
	\hbox{
	\hspace{0.5in}
	\subfloat{\includegraphics[width = 0.4\textwidth, clip = true, trim = 0.cm 0.cm 0.cm 0.cm]{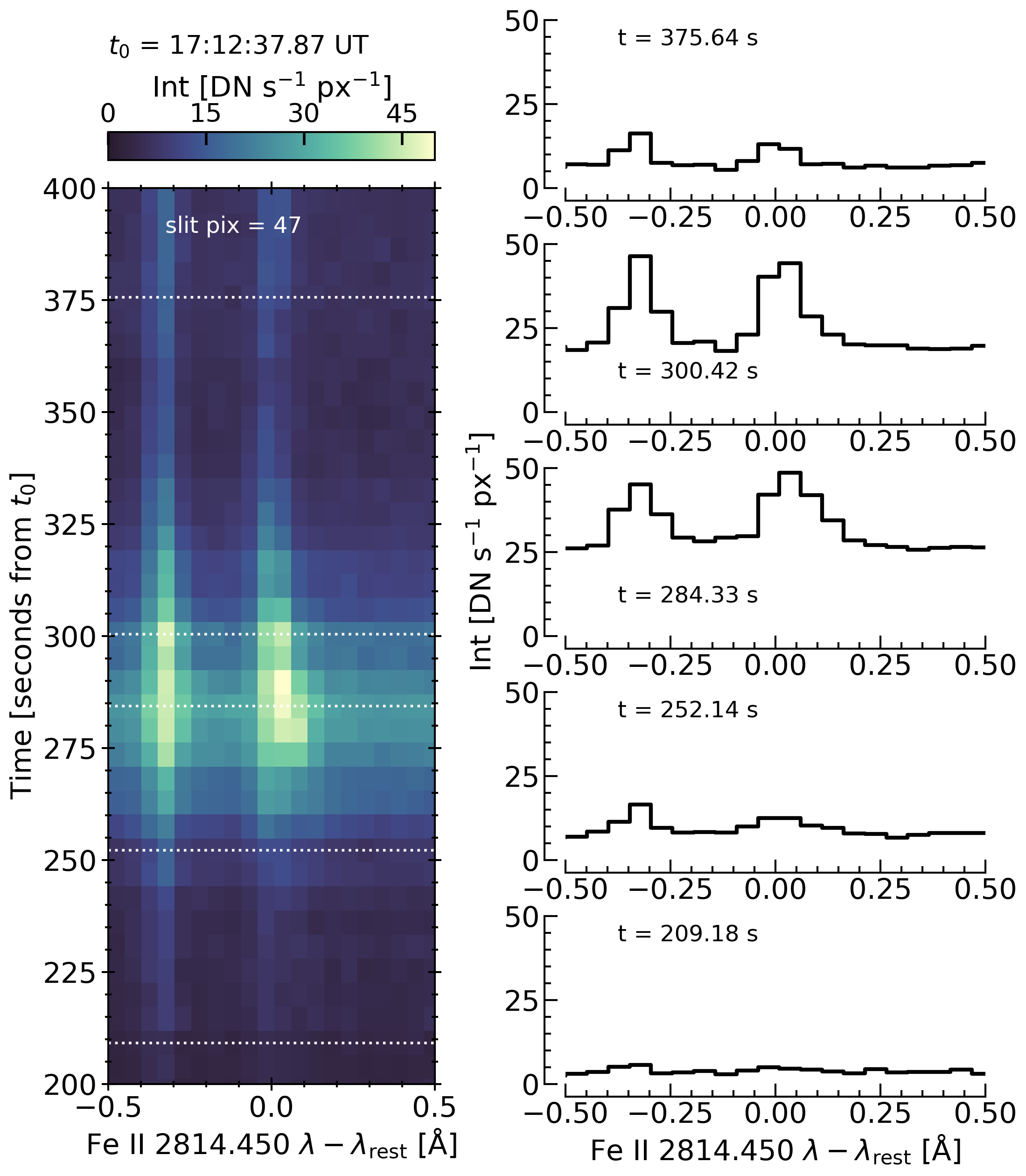}}	
	\subfloat{\includegraphics[width = 0.4\textwidth, clip = true, trim = 0.cm 0.cm 0.cm 0.cm]{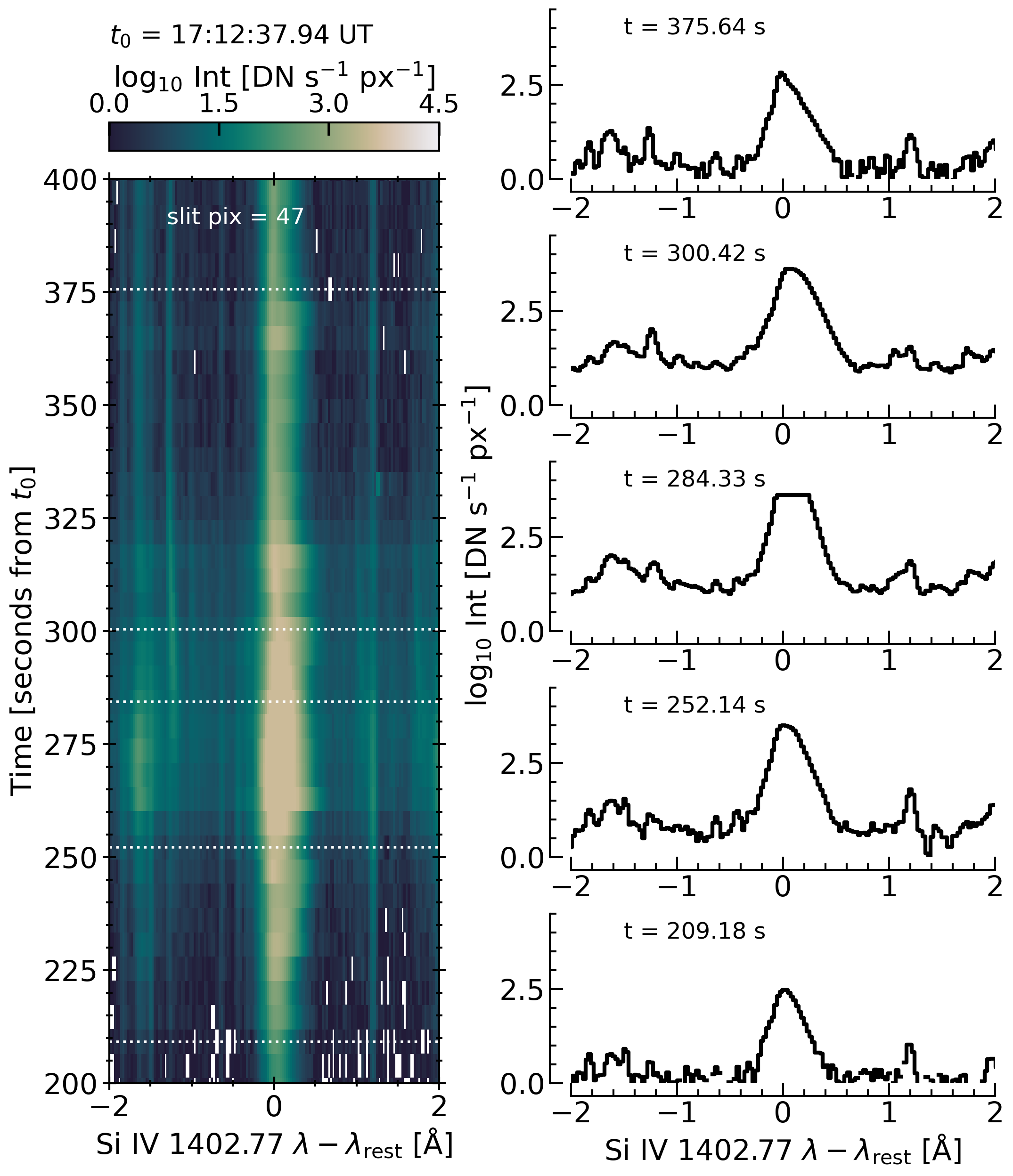}}	
	}
	}
	\caption{\textsl{The temporal evolution of emission from the source of interest (slit pixel $\#47$, which is the bottom horizontal line shown on Figure~\ref{fig:iris_spacetime_overview}.), with cut-outs showing individual spectral in more detail. Shown are the \ion{Cl}{i} 1351.657~\AA\  (top left), \ion{O}{i} 1355.598~\AA\ (top right),  \ion{Fe}{ii} 2814.445~\AA\ (bottom left), and \ion{Si}{iv} 1402.77~\AA\ (bottom right). In each wavelength-time diagram the horizontal lines indicate the times of the cut-outs, and the times are shown as seconds from $t_{0}$. A weak red-wing component is present in each spectra, e.g. around $t=284$~s, which shows up more clearly in the narrower lines.}}
	\label{fig:iris_spectra_overview1}
\end{figure*}

\begin{figure}
	\centering 
	\vbox{
	\hbox{
	\hspace{0.0in}
	\subfloat{\includegraphics[width = 0.4\textwidth, clip = true, trim = 0.cm 0.cm 0.cm 0.cm]{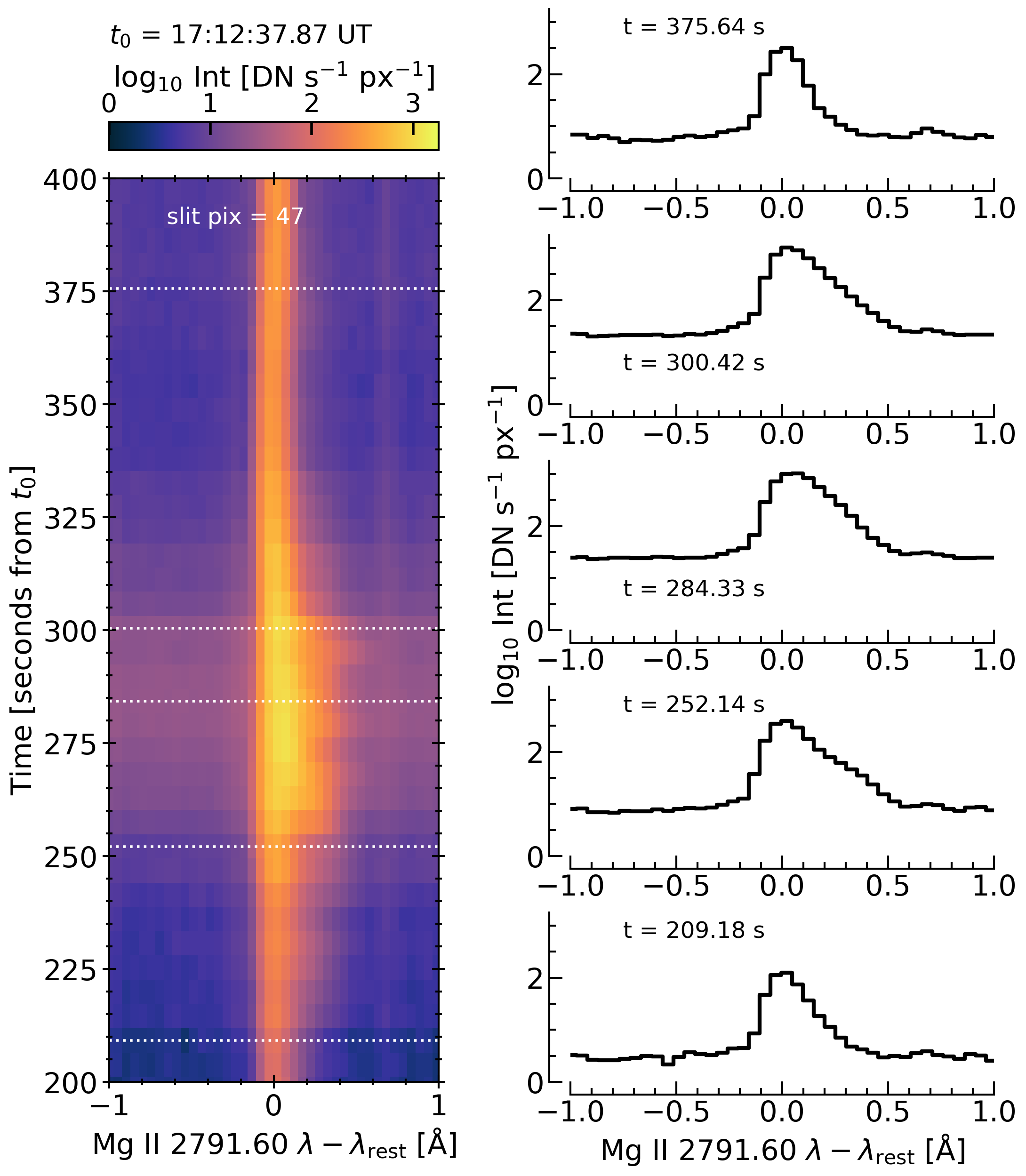}}
	}
	}
	\vbox{
	\hbox{
	\hspace{0.0in}
	\subfloat{\includegraphics[width = 0.4\textwidth, clip = true, trim = 0.cm 0.cm 0.cm 0.cm]{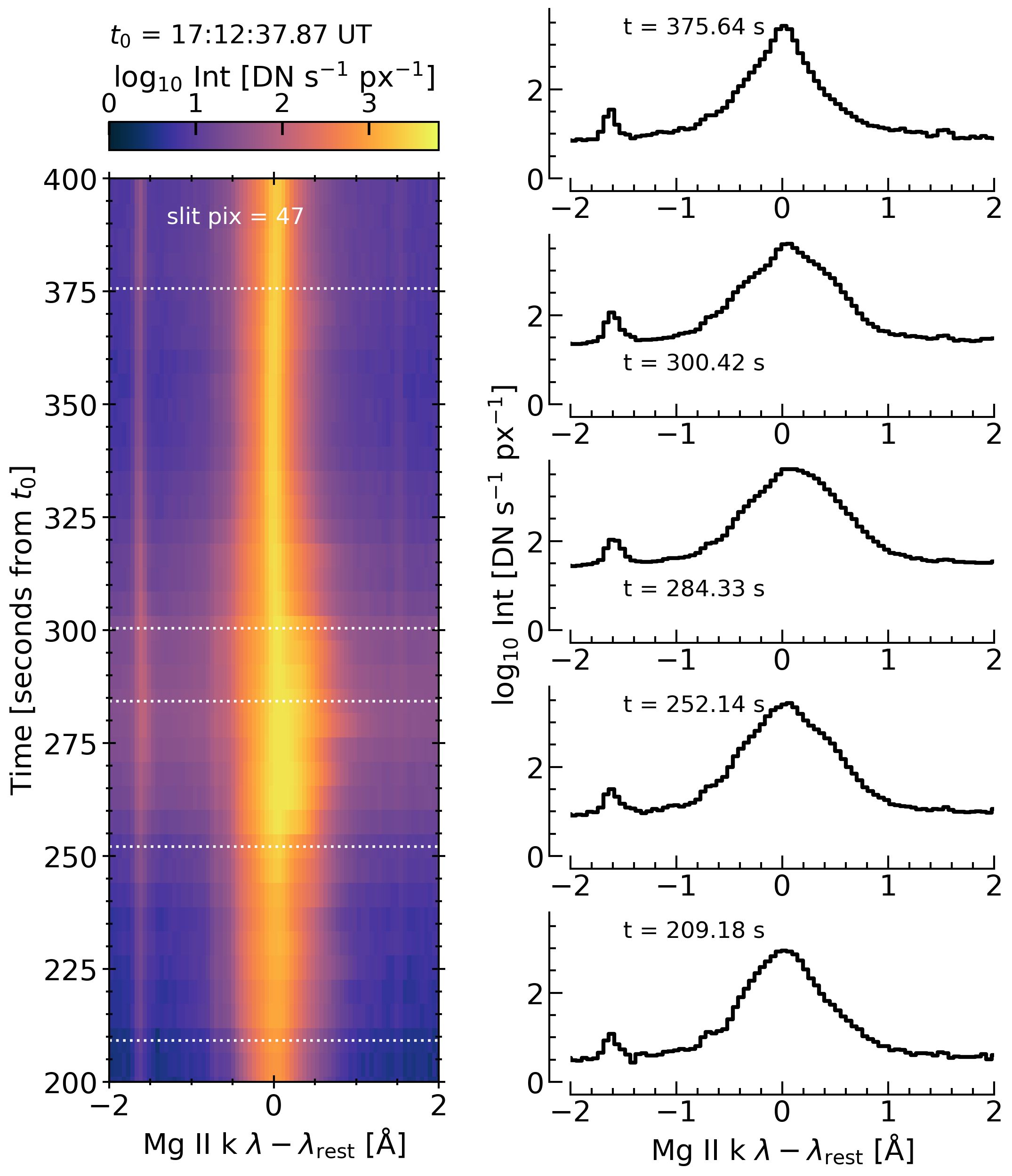}}	
	}
	}
	\caption{\textsl{Same as Figure~\ref{fig:iris_spectra_overview1} but showing the \ion{Mg}{ii} 2791.6~\AA\ and \ion{Mg}{ii} k lines}}
	\label{fig:iris_spectra_overview2}
\end{figure}

Selecting one of our sources within the region of interest, we show the time evolution of the response of each line to the flare in Figure~\ref{fig:iris_spectra_overview1} and Figure~\ref{fig:iris_spectra_overview2}, which includes the \ion{Si}{iv} 1402.77~\AA\ line for added context. For each line an image shows the intensity as a function of wavelength (x-axes) and time (y-axes), where time is shown in seconds from a zero point that is stated above the colour bar of each panel. Cut-outs show examples of the spectral lines before, during, and after the flare, with dotted horizontal dashed lines in the images indicating the time of each cut-out. A description of each line is included in Appendix~\ref{sec:figure4disc}.

\section{Nonthermal Widths in the Flare Chromosphere}
\begin{figure*}
	\centering 
	\vbox{
	\hbox{
	\hspace{0.75in}
	\subfloat{\includegraphics[width = 0.375\textwidth, clip = true, trim = 0.cm 0.cm 0.cm 0.cm]{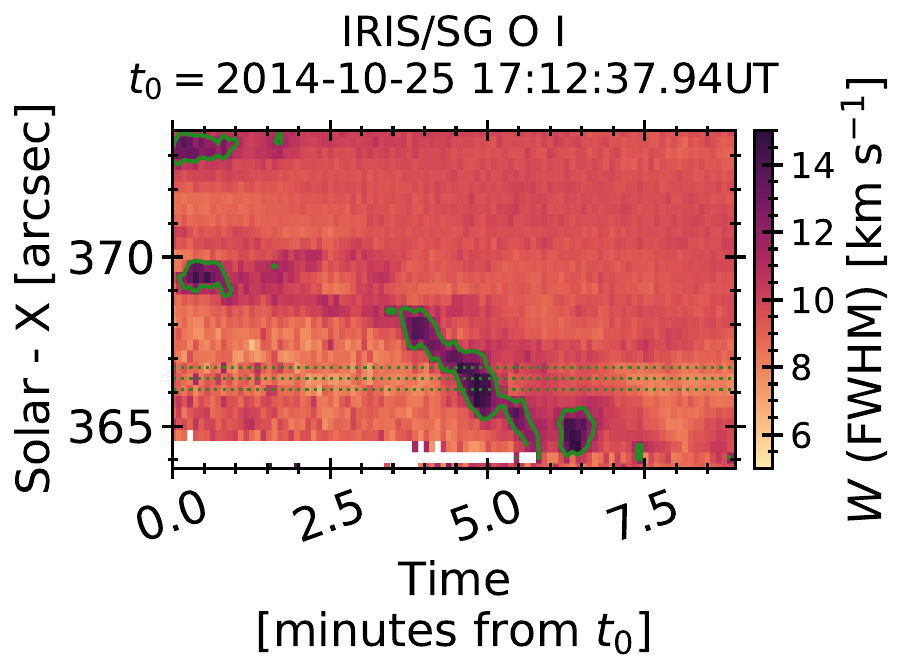}}
	\subfloat{\includegraphics[width = 0.375\textwidth, clip = true, trim = 0.cm 0.cm 0.cm 0.cm]{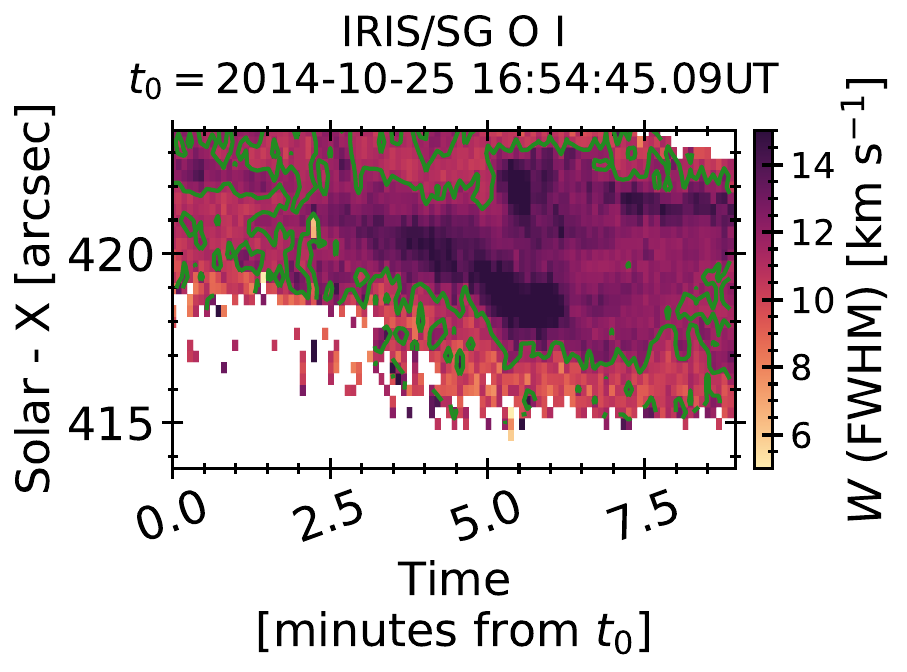}}	
	}
	}
	\vbox{
	\hbox{
	\hspace{0.75in}
	\subfloat{\includegraphics[width = 0.375\textwidth, clip = true, trim = 0.cm 0.cm 0.cm 0.cm]{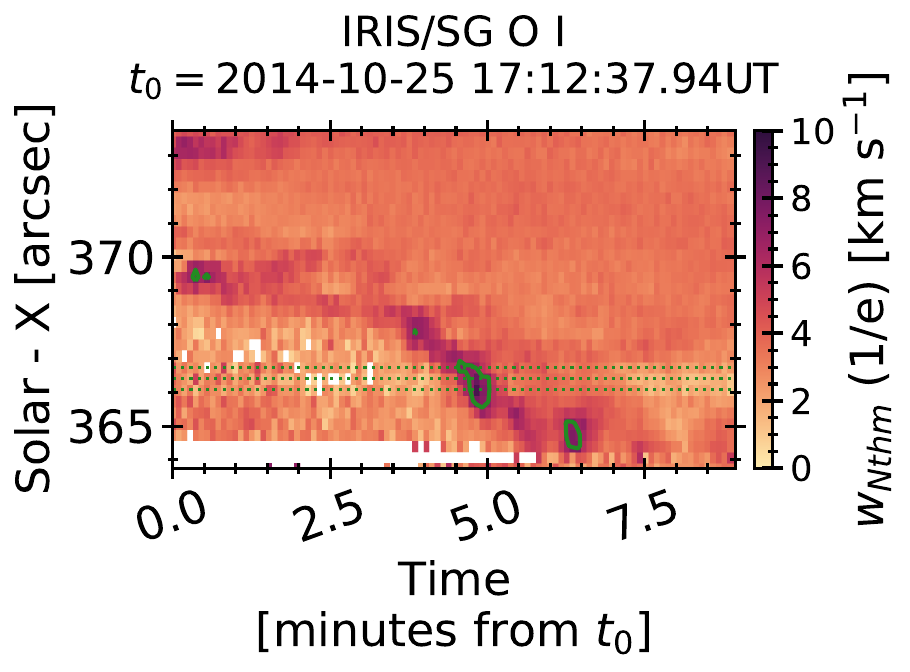}}
	\subfloat{\includegraphics[width = 0.375\textwidth, clip = true, trim = 0.cm 0.cm 0.cm 0.cm]{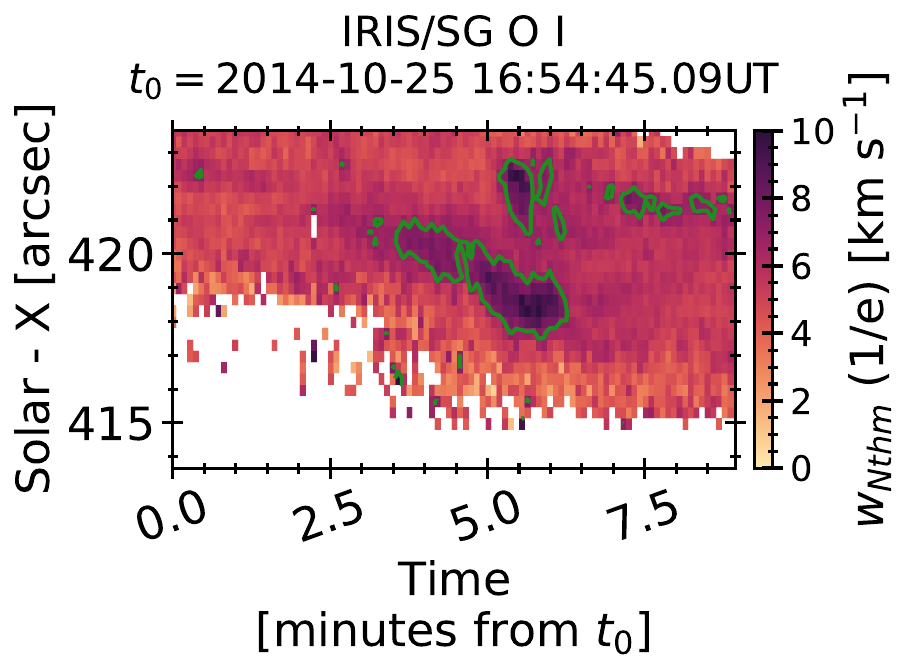}}	
	}
	}
	\caption{\textsl{Spacetime diagrams of the \ion{O}{i} 1355.598~\AA\ line's FWHM, $W$ (top row; green contours are at 11.5~km~s$^{-1}$) and nonthermal line width, $w_{\mathrm{Nthm}}$ (bottom row; green contours are at 7~km~s$^{-1}$), both obtained from fitting a single Gaussian component.  The nonthermal width is expressed as the $1/e$ half width. 
 The first column for each shows the main region of interest, and the second column shows another segment of flare ribbons further west in the field of view. Here we are primarily illustrating that regions of flare-induced broadening are rather localised and transient, and showing typical magnitudes of $w_{\mathrm{Nthm}}$ outside of the flare, so assume a fairly low formation temperature of $T = 6$~kK when calculating $w_{\mathrm{Nthm}}$.}}
	\label{fig:iris_widths_sg_overview}
\end{figure*}

\subsection{Single Gaussian Fitting}\label{sec:singlegauss}
Though the spectra exhibit red-wing asymmetries during the flare, we initially fit a single Gaussian component to each line. This enables us to measure the background values of the nonthermal widths, as well as to get a sense of the duration of any broadening. Even if the increase of the width of the Gaussian function fit to the lines was largely due to the presence of red-wing components (since the standard deviation, $\sigma$, would increase to accommodate the red wing), the lifetimes of the wider single Gaussian models are reflective of the lifetimes of the red wing asymmetry and are a useful metric to obtain.

A single Gaussian component was fit to the \ion{O}{i} 1355.598~\AA\ line and the nearby \ion{C}{i} 1355.884~\AA\ line, which is optically thick but we include that line solely to better constrain the continuum level. The \ion{O}{i} line is narrow, and since IRIS performed on-board spectral summing, there are times at which the entire line is only a few pixels across. During the flare, however, the line spans a larger number of pixels. 

Importantly, although it is typical to calculate the value of an analytic function only at the center wavelength of each pixel when fitting an analytic function such as a Gaussian profile to spectral data, in fact, the signal is the integrated flux across the wavelength range of each pixel. The typical approach thus assumes a linear approximation to the function over an individual wavelength bin. If there are only a few spectral pixels across the full width at half maximum (FWHM) of the profile, using only the center wavelength of each pixel can introduce significant errors, particularly with respect to line width and asymmetry parameters. For a detailed discussion of such errors, see \cite{2016SoPh..291...55K}, who provide an iterative algorithm to adjust the data to compensate for these errors when fitting analytic functions to the data. Here, we take a more direct approach: when fitting an analytic function to the observed spectral intensity, we numerically calculate the average of the analytic function across each bin, specifically:
\begin{equation}
\bar F(x_j)=\frac{1}{N} \sum_{n=-\frac{N-1}{2}}^{\frac{N-1}{2}} F(x_j + n \frac{\Delta x}{N}) ,  
\end{equation}
where $F(x)$ is the analytic function, $x_j$ is the center wavelength of pixel $j$, $\Delta x$ is the pixel width, $N$ is the number of values of $x$ to use for numerically calculating the average, and the average over each pixel, $\bar F(x)$, is fit to the data (i.e., is used to calculate $\chi^2$) rather than $F(x)$. We use $N=5$, that is, we perform the calculation of the profile at five times higher sampling than the data in order to rapidly calculate the average across each pixel for the fitting procedure, therby obtaining accurate width information.

\begin{figure*}
	\centering 
	\vbox{
	\hbox{
	\hspace{0.75in}
	\subfloat{\includegraphics[width = 0.375\textwidth, clip = true, trim = 0.cm 0.cm 0.cm 0.cm]{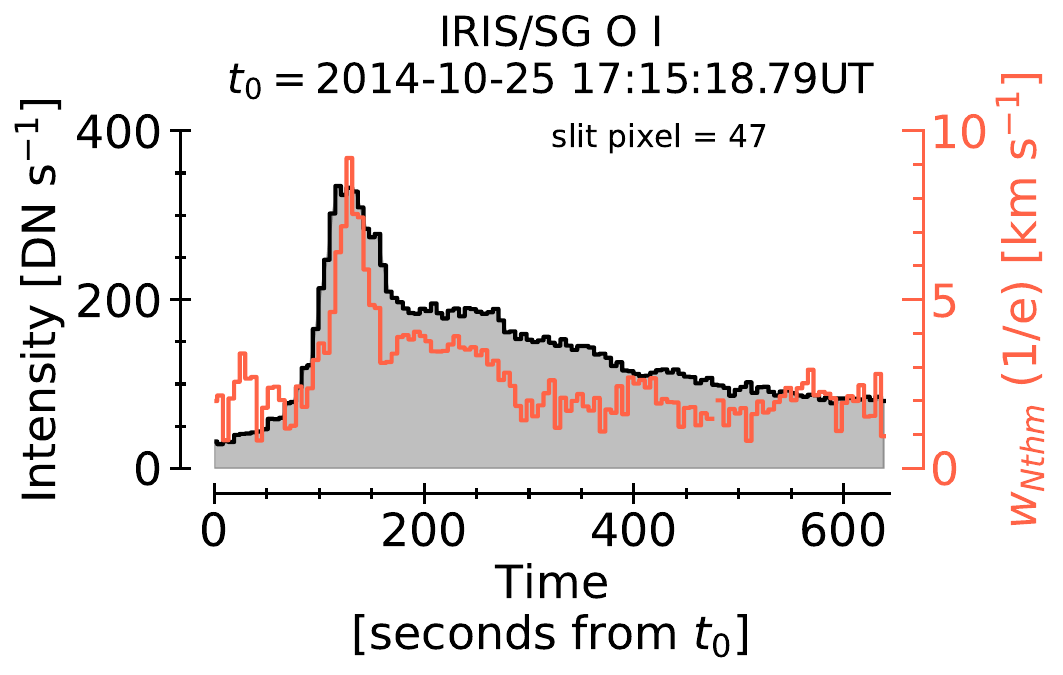}}
	\subfloat{\includegraphics[width = 0.4015\textwidth, clip = true, trim = 0.cm 0.cm 0.cm 0.cm]{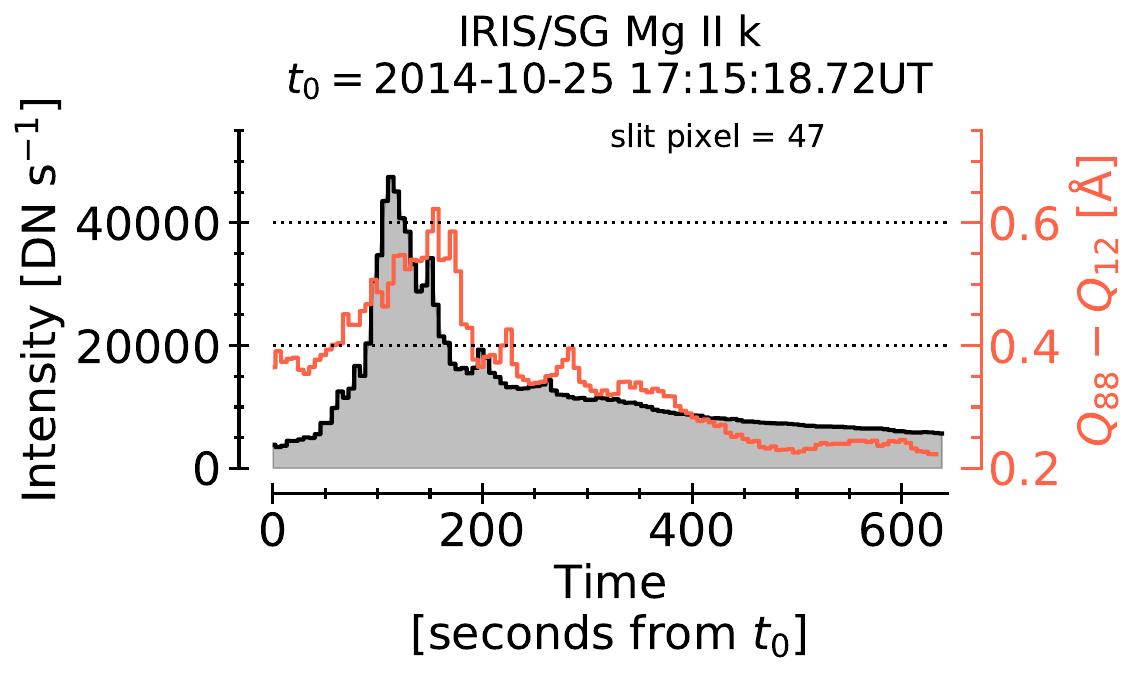}}
	}
	}
	\vbox{
	\hbox{
	\hspace{0.75in}
	\subfloat{\includegraphics[width = 0.375\textwidth, clip = true, trim = 0.cm 0.cm 0.cm 0.cm]{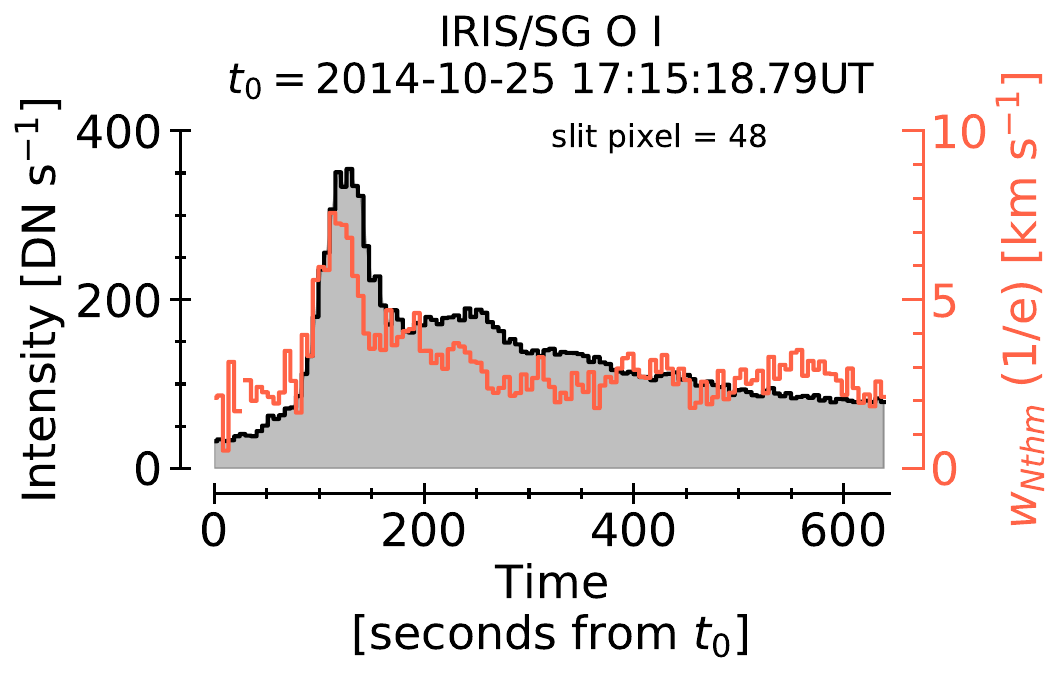}}
	\subfloat{\includegraphics[width = 0.4015\textwidth, clip = true, trim = 0.cm 0.cm 0.cm 0.cm]{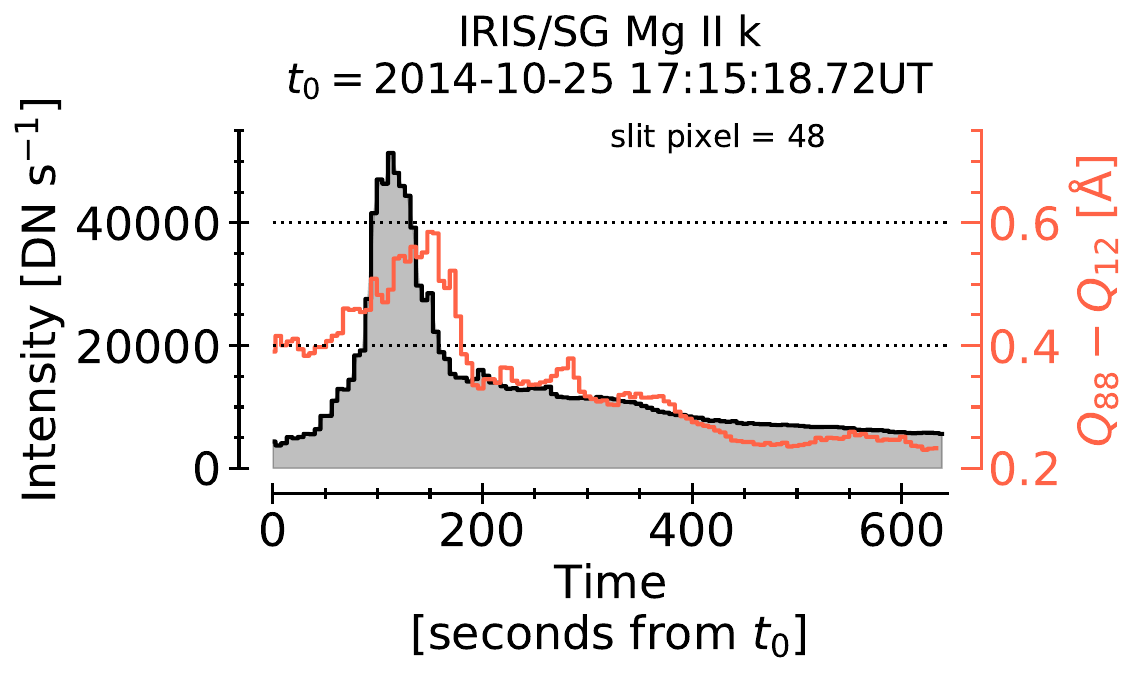}}	
	}
	}
	\vbox{
	\hbox{
	\hspace{0.75in}
	\subfloat{\includegraphics[width = 0.375\textwidth, clip = true, trim = 0.cm 0.cm 0.cm 0.cm]{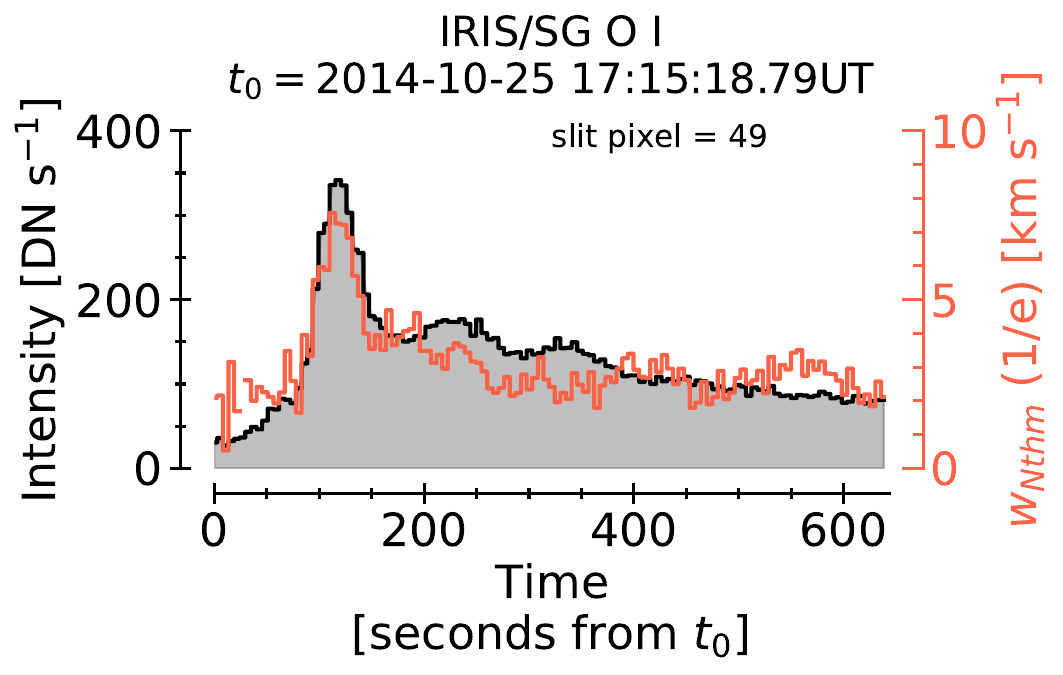}}
	\subfloat{\includegraphics[width = 0.4015\textwidth, clip = true, trim = 0.cm 0.cm 0.cm 0.cm]{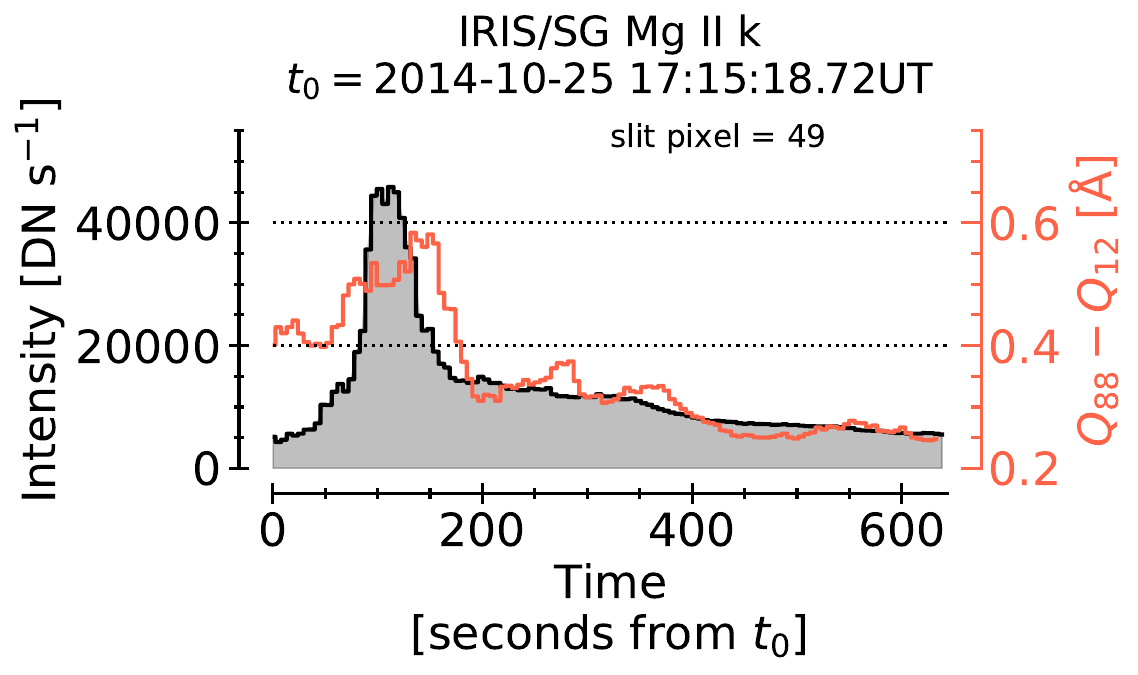}}	
	}
	}
	\caption{\textsl{Evolution of the widths (red lines) and wavelength integrated intensities (black lines and grey shaded region) of the \ion{O}{i} 1355.598~\AA\ line (left column) and \ion{Mg}{ii} k lines (right column), for the three pixels within the region of interest (indicated by dashed lines in previous figures). The nonthermal width of the \ion{O}{i} was obtained from single Gaussian component fits. Here we are primarily illustrating the duration of flare-induced intensity enhancement and related line width increases, and the magnitudes of $w_{\mathrm{Nthm}}$ before and after the flare, so assume a fairly low formation temperature of $T = 6$~kK. For \ion{Mg}{ii} the line width is measured as the difference between the 88-th and 12-th quantiles ($Q_{88} - Q_{12}$). On those panels the dotted horizontal lines are the typical values during the flare.}}
	\label{fig:iris_widths_sg_overview2}
\end{figure*}

For a nice overview of line width descriptions see \cite{2015ApJ...811...80R}. In brief, line widths are generally expressed by several conventions: the standard deviation of the Gaussian ($\sigma$), the FWHM ($W$), or by the (1/e) half-width, ($w$). These are related to each other in the following way:

\begin{equation}
w = \sqrt{2}\sigma = \frac{W}{2\sqrt{\ln{2}}}
\end{equation}

\noindent The total width of a spectral line is comprised of various components. Hereafter we refer to any components of the widths expressed as (1/e) half-widths as $w_{\mathrm{comp}}$, and those expressed as FWHM as $W_{\mathrm{comp}}$. Assuming a Maxwellian plasma, and no opacity effects, the (1/e) half-width of the line profile can be described by 

\begin{equation}\label{eq:width}
w = \sqrt{\frac{2k_{b}T}{m_{i}} + \xi^{2}},
\end{equation}

\noindent  where $k_{b}$ is Boltzmann's constant, $T$ is the plasma temperature, $m_{i}$ is the ion/atom mass, and $\xi$ is some additional broadening mechanism that is referred to as the nonthermal width.  We make the assumption in this study that the nonthermal width is wholly due to microturbulence velocity $V_{\mathrm{turb}}$, such that $\xi = V_{\mathrm{turb}}$. The first part of the sum is the thermal Doppler width of the line. The masses for the atoms/ions studied here were $m_{i,\mathrm{O}} = 15.999~u$, $m_{i,\mathrm{Cl}} = 35.453~u$, and $m_{i,\mathrm{Fe}} = 55.845~u$, where $u = 1.6605\times10^{-27}$~kg is the atomic mass unit.

An observed spectral line width also includes broadening due to the instrument itself, $W_\mathrm{I}$. The total observed FWHM line width is $W_\mathrm{obs} =   2\sqrt{2\ln{2}}~\sigma_{\mathrm{obs}}$, and the nonthermal width of an optically thin line is

\begin{equation}
w_\mathrm{{Nthm}} = \frac{1}{2\sqrt{\ln{2}}}\sqrt{W_\mathrm{obs}^2 - \left(2\sqrt{\ln{2}} \sqrt{\frac{2k_bT}{m_i}}\right)^2 - W_\mathrm{I}^2}.
\end{equation}

\noindent Though sometimes reported as a FWHM in the literature, we will refer to nonthermal widths as a (1/e) half-width\footnote{Though in figure labels we drop the `half-width' for brevity}, $w_\mathrm{Nthm}$, since this is appropriate when comparing to values to be included in numerical modelling. We typically discuss widths in velocity units, but when we refer to widths in wavelength units this was done by multiplying by $\frac{\lambda_{0}}{c}$, where $c$ is the speed of light, $\lambda_{0}$ is the central wavelength of the transition.

 Note that for an optically thick line opacity effects result in widths and shapes that are not solely defined by the thermal and nonthermal widths as in Eq~\ref{eq:width}, but those values are still important in the line's formation and characteristics.

Figure~\ref{fig:iris_widths_sg_overview} shows spacetime maps of the observed FWHM (top row)  and of the $w_\mathrm{Nthm}$ (bottom row), both in velocity units. The first column is our main region of interest, near the easternmost ribbon, and the second column shows a patch of the western ribbon. Patches of increased width are present, with morphology consistent with the enhanced intensity of flare ribbon sources seen in Figure~\ref{fig:iris_spacetime_overview}. When measuring the nonthermal widths we assumed a value of $T = 6$~kK \citep[consistent with the range of values in the modelling work of][]{2015ApJ...813...34L}, which is likely too small for the flare, but at the moment we are primarily interested in obtained estimates of the non-flaring $w_\mathrm{Nthm}$ and the duration of the wider lines. For the instrumental width we assumed that the spectral point spread function was a Gaussian with a FWHM of 2.2 unbinned pixels in the FUVS channel, from the notes in the IRIS SSWIDL routine \texttt{iris\_nonthermalwidth.pro}\footnote{\url{https://hesperia.gsfc.nasa.gov/ssw/iris/idl/sao/util/tian/iris_nonthermalwidth.pro}}, giving $W_\mathrm{I} = 28.6$~m\AA. In  Appendix~\ref{sec:varyformt} we demonstrate that for a reasonable range of values of formation temperature or of $W_\mathrm{I}$ the derived  $w_\mathrm{Nthm}$ does not change drastically. 

 In the region of interest, which is in the penumbra/umbra, the background values are $w_\mathrm{Nthm}\sim2-4$~km~s$^{-1}$, whereas farther west in a quiet-Sun region the background values are somewhat higher, $w_\mathrm{Nthm}\sim4.5-6$~km~s$^{-1}$. These are both somewhat lower than the values measured by \cite{2015ApJ...809L..30C}, who reported nonthermal widths of $7.3$~km~s$^{-1}$, but that study looked at plage. The lifetime of each enhancement was around $\Delta t \sim 30-90$~s. Selecting three pixels in the region of interest for closer study we show lightcurves in Figure~\ref{fig:iris_widths_sg_overview2}, in which the red curves are $w_\mathrm{Nthm}$, and the black curves/grey shaded areas the integrated intensities across the line (from the data, not from the Gaussian model). Recall that here the increased widths of the Guassian functions fit to the \ion{O}{i} line in the flare ribbon are largely due to the presence of the red-wing component.  Despite initially decaying with a similar timescale as the width of the Gaussian model, the intensity does not return to background levels for many minutes. The lifetime of the intensity peak is around $60-90$~s, and exhibits a slow rise some 1.5-2.5 minutes before reaching its peak. 
 
The \ion{Mg}{ii} transitions also broaden, which are shown on the second column of Figure~\ref{fig:iris_widths_sg_overview2}. Instead of fitting a Gaussian to the \ion{Mg}{ii} spectra, which are optically thick and very non-Gaussian in shape, we measure the line's quantiles \citep[e.g.][where for example the wavelength corresponding to the 88-th quantile is written $Q_{88}$]{2015A&A...582A..50K,2018ApJ...865..123R,2021A&A...653A...5P}, and define the width as the distance between 88-th and 12-th quantiles, $W_{Q} = Q_{88} - Q_{12}$. This is analogous to the FWHM if the line were Gaussian in shape. The width begins to increase during the flare, reaching typical values of $W_{Q} = [0.4-0.6]$~\AA\ around the time of peak intensity. Interestingly, $W_{Q}$ peaks shortly \textsl{after} the intensity peak.

\subsection{Spectral Line Metrics From the Chromospheric Condensation}\label{sec:detailfit}
\begin{figure}
	\centering 
	\vbox{
	\subfloat{\includegraphics[width = 0.42\textwidth, clip = true, trim = 0.cm 0.cm 0.cm 0.cm]{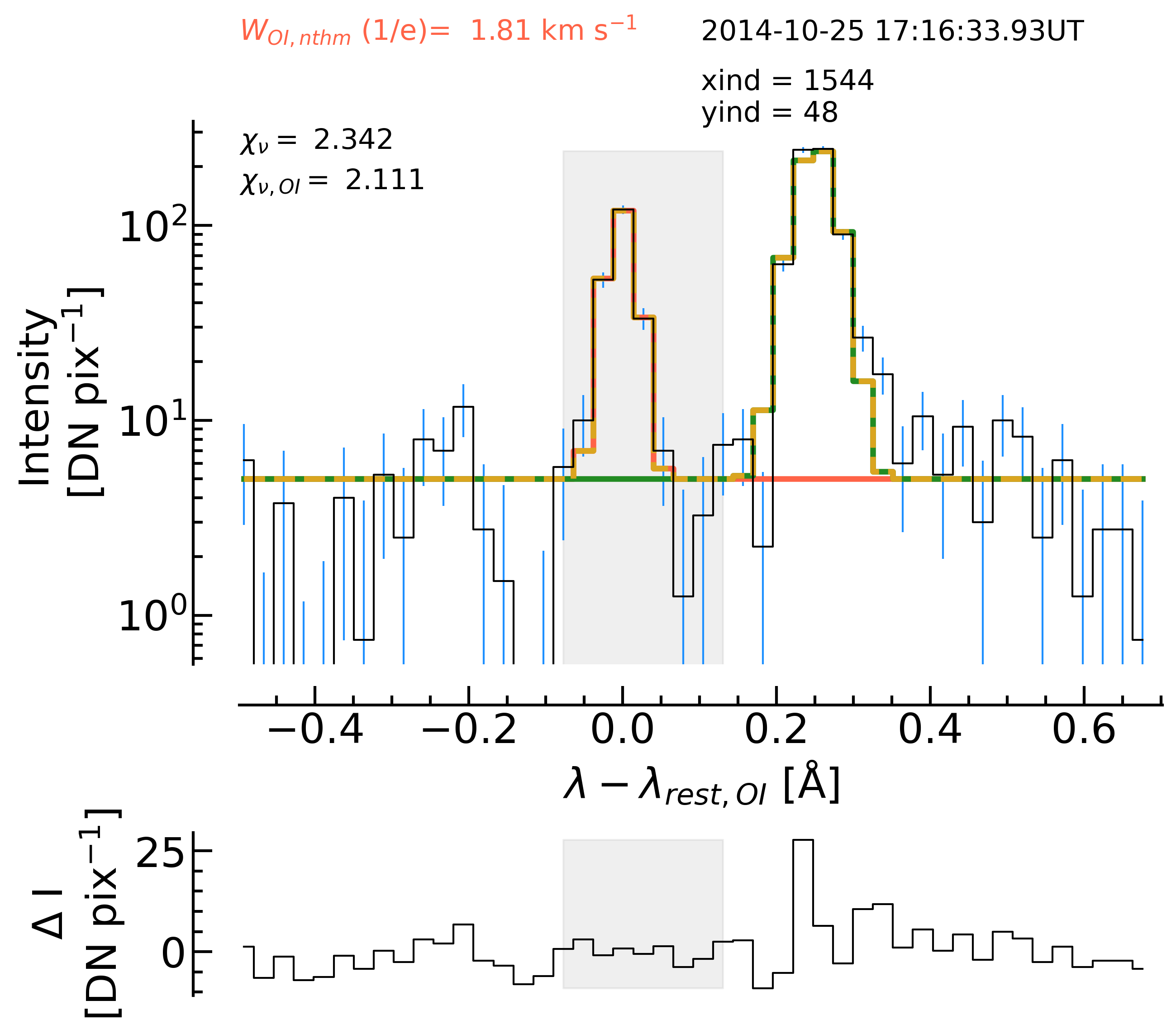}}
	}
	\vbox{
	\subfloat{\includegraphics[width = 0.42\textwidth, clip = true, trim = 0.cm 0.cm 0.cm 0.cm]{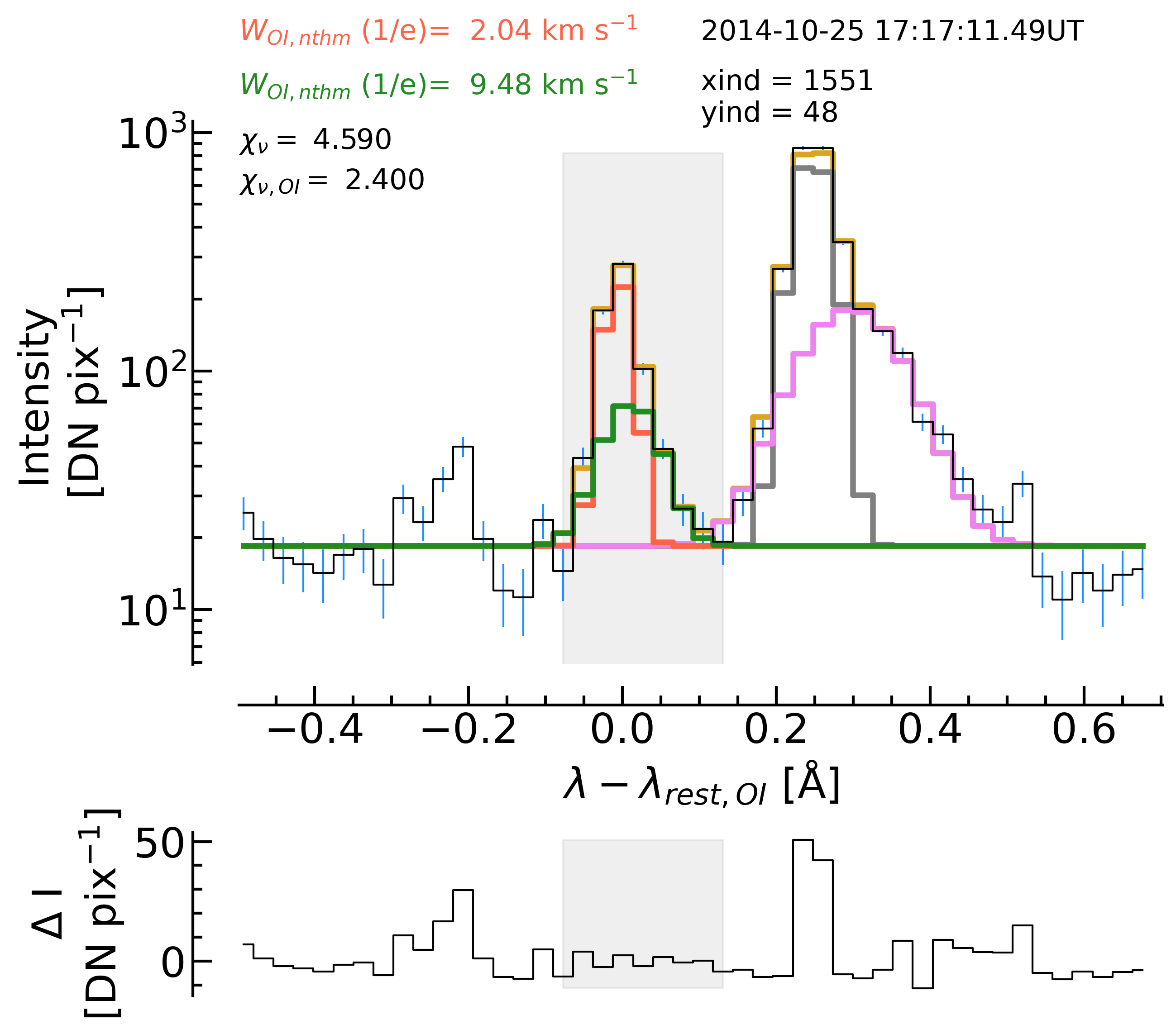}}	
	}
	\vbox{
	\subfloat{\includegraphics[width = 0.42\textwidth, clip = true, trim = 0.cm 0.cm 0.cm 0.cm]{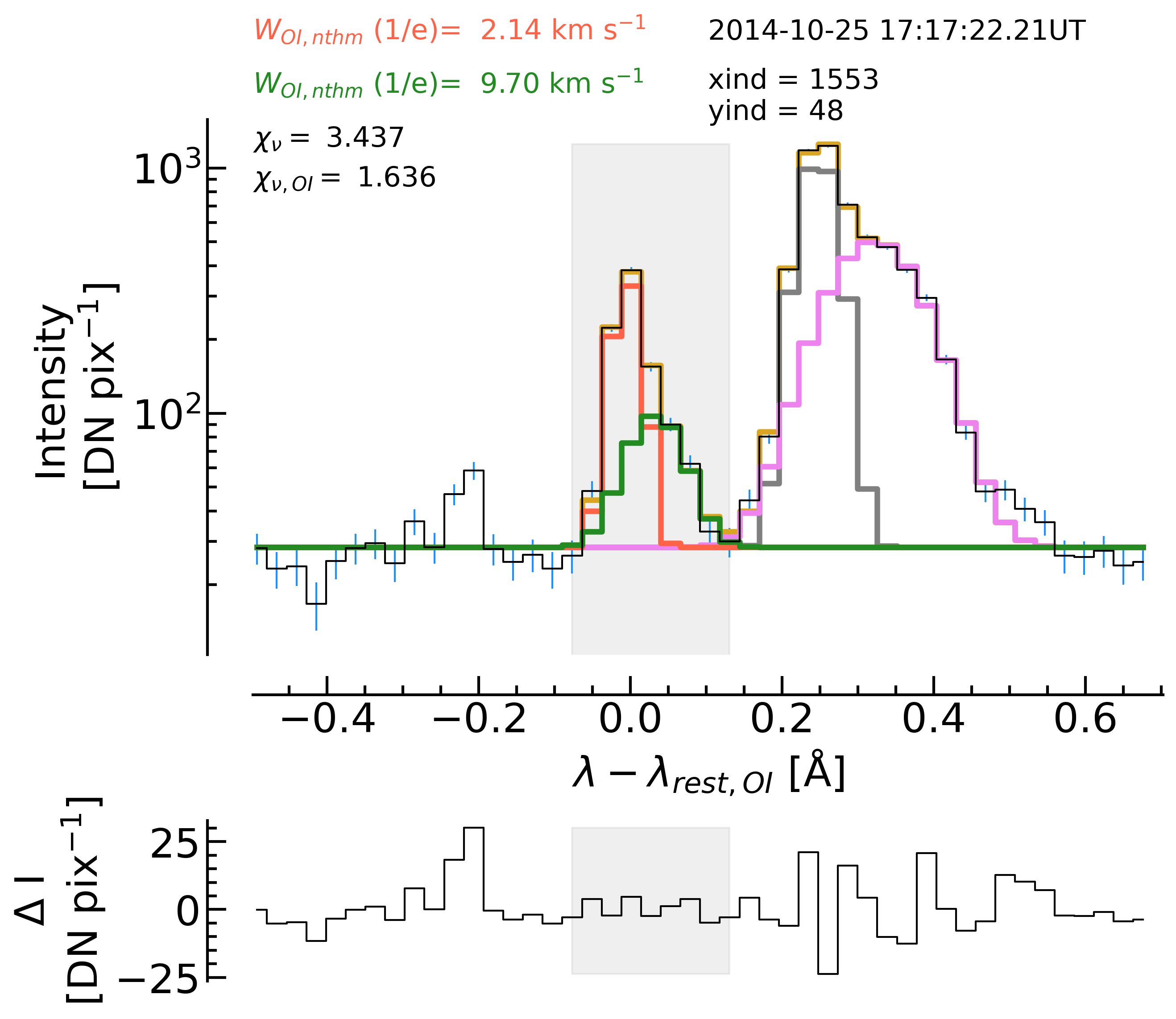}}
	}
	\caption{\textsl{Examples of fitting Gaussian models to \ion{O}{i} 1355.598~\AA. Each panel shows a different time for a single IRIS pixel, with residuals below and the shaded grey portion shows the region used to calculate the \ion{O}{i} $\chi^{2}_{\nu,\mathrm{OI}}$. In the top panel a single component is fit to each line (red for \ion{O}{i} and green for \ion{C}{i}, with the full model in gold). In the other panels two components are fit to each line, one stationary (red for \ion{O}{i} and grey for \ion{C}{i}) and one redshifted (green for \ion{O}{i} and violet for \ion{C}{i}), with the total model in gold. Error bars on the data are in blue.}}
	\label{fig:iris_oi_fits}
\end{figure}

As indicated earlier, many of the spectra exhibited a red-wing asymmetry (resulting in very large $\chi_{\nu}^2$ values), so for three pixels within the region of interest (slit pixel \# $[47, 48, 49]$) we fit two Gaussian components each to \ion{O}{i} 1355.598~\AA, \ion{C}{i} 1355.844~\AA, \ion{Cl}{i} 1351.66~\AA, and \ion{Fe}{ii} 2814.445~\AA, around the times of the flaring emission. A single Gaussian component was fit for every exposure, but if the asymmetry of the line was above a certain threshold\footnote{A measure of asymmetry for each profile was calculated using the metric described in \cite{2019ApJ...879L..17P}, where the intensity difference in the red and blue wings is compared to the peak intensity (in our case we measured the intensity between $\lambda = [0.04 - 0.14]$~\AA\ in each wing).  If the asymmetry was above the typical background value for each line then the profile was flagged for multiple-component fitting} an attempt to fit two Gaussian components was performed. One component was labelled the `stationary' component, and was restricted to have a small centroid shift typically $\pm5$~km~s$^{-1}$ from the rest wavelength, but varied sometimes after a visual inspection of the results. The shifted component that produced the red-wing asymmetry was labelled the `condensation' component, and was restricted to have a centroid larger than the rest wavelength of the line. A visual inspection confirmed both the presence of red wing component as well as the quality of the fit. Reduced $\chi^{2}$ metrics, $\chi_{\nu}^{2}$, were calculated also, and for the \ion{O}{i} window $\chi_{\nu}^{2}$ was calculated for both the full window (i.e. including \ion{C}{i}), and separately for \ion{O}{i}. If the quality of the fit for a particular exposure was not good (based on a larger $\chi^{2}_{\nu}$, or a fit that did not capture an obvious red wing component upon visual inspection), the bounds and initial guess were varied to obtain a better result with smaller $\chi^{2}_{\nu}$.

Examples of fitting the \ion{O}{i} 1355.598~\AA\ line are shown in Figure~\ref{fig:iris_oi_fits}, where each panel shows a single exposure of \ion{O}{i} 1355.598 and \ion{C}{i} 1355.844~\AA\ from IRIS (black lines), with the pixel along the slit (`yind') and time index (`xind') noted, both of which refer to the pixel numbers in the level-2 IRIS data. Error bars (photon counting plus dark current noise) are included as blue vertical lines at each wavelength. For each exposure the fit residuals (model - data) are shown below (note that the data and model fits are shown on a logarithmic scale, but residuals are on a linear scale). 

In some instances the \ion{C}{i} 1355.844~\AA\ appeared to exhibit an asymmetry for quite some time prior to the onset of a measurable asymmetry in \ion{O}{i} 1355.598~\AA\ or the other lines. This may be because \ion{C}{i} 1355.844~\AA\ is a stronger line than \ion{O}{i} 1355.598~\AA, but \ion{Cl}{i} is fairly strong also and tended to be co-temporal with \ion{O}{i}. Perhaps optical depth effects are playing a role here. Once present, the condensation component grew in strength, but never had a peak intensity dominant over the stationary component. It then decreased in peak intensity until it was no longer discernible as a component. Typical lifetimes were $60-90$~s. After its disappearance, there were occasional re-emergences after a short gap ($\sim5-30$~s), that were weak and persisted only a few exposures ($\sim5-15$~s). As mentioned earlier the \ion{Cl}{i} line may have a non-negligible opacity in the stationary component \citep[e.g.][]{1983ApJ...266..882S}, but we operate under the assumption that the condensation component was optically thin in order to fit that part of the line and extract metrics from it. Similarly, \ion{Fe}{ii} 2814.445~\AA\ may be optically thick in the pre-flare \citep[e.g.][]{2017ApJ...836...12K,2020ApJ...895....6G}, though we investigate this in the modelling sections of this paper.

Using the guess and bounds for each initial successful two-component Gaussian fit to the profiles, we then employed a Monte Carlo approach: 10,000 realisations were made for each spectral line from each exposure, drawing randomly from a Poisson distribution with mean equal to the observed intensity within each wavelength bin plus a random uniform sampling of the dark current plus readout noise. The total line intensity, nonthermal widths, and Doppler shift of both components were measured for each realisation. Going forward, quoted values of each of those line metrics are the 50\% quantile of an exposure's distribution, with the upper (lower) limits being the 84\% (16\%) quantiles.  

\begin{figure*}
	\centering 
	\subfloat{\includegraphics[width = 0.75\textwidth, clip = true, trim = 0.cm 0.cm 0.cm 0.cm]{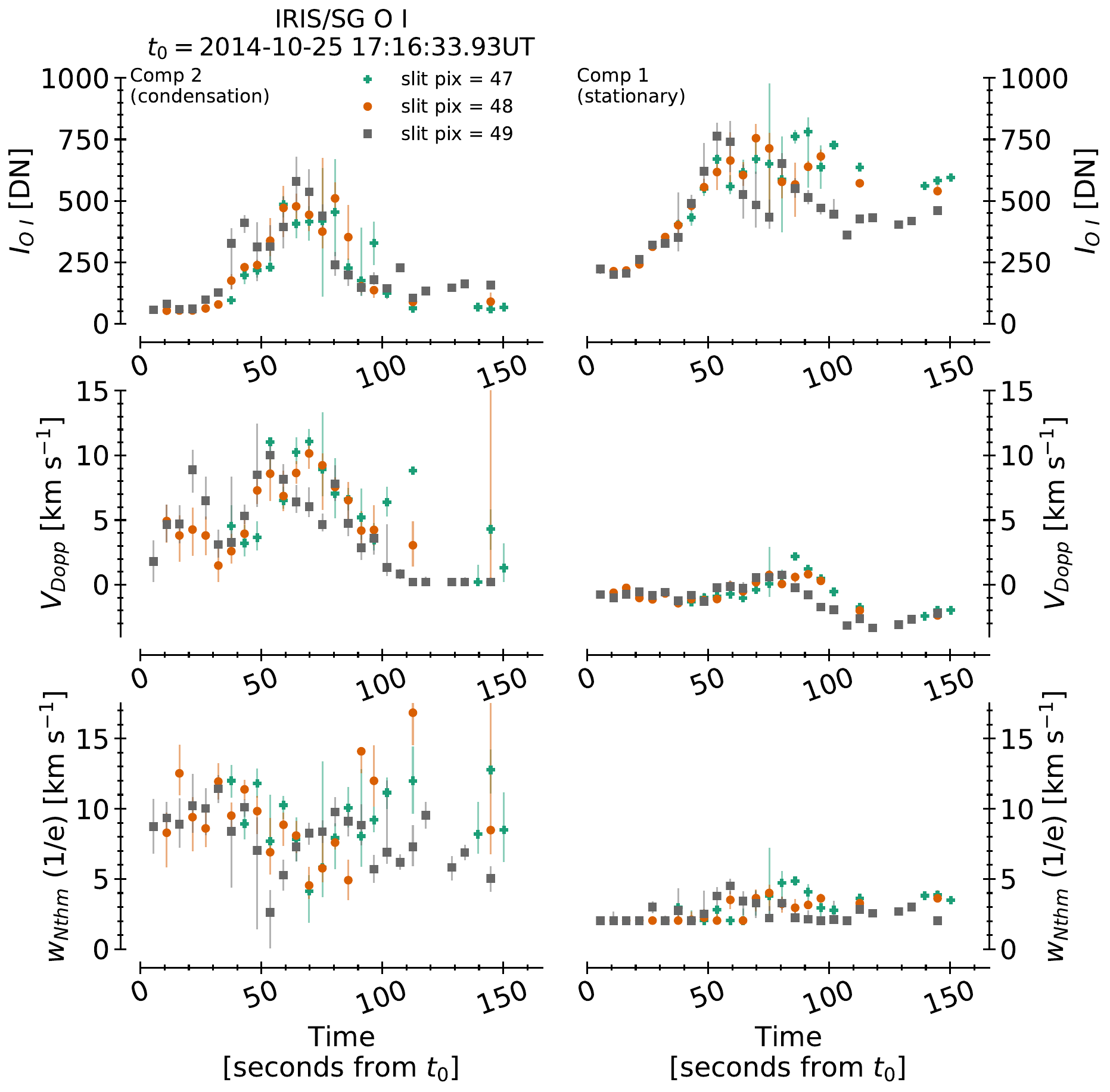}}	
	\caption{\textsl{The results of fitting two components to the \ion{O}{i} 1355.598~\AA\ line during the flare. The left column is the shifted component (the condensation) and the right column is the `stationary' component. For each the total intensity, Doppler shift, and nonthermal line widths ((1/e) half-widths) are shown. Three pixels were selected from the region of interest (indicated by the horizontal lines in several prior figures), separated by different colours and symbols. Error bars are from a Monte Carlo analysis of each profile.}}
	\label{fig:iris_oi_dgfitmetrics}
\end{figure*}

Comparing the \ion{O}{i} 1355.598~\AA\ fit results for each of the three pixels within the region of interest in Figure~\ref{fig:iris_oi_dgfitmetrics} we can see fairly consistent behaviour within each source. In that figure the total line intensity is the top row, with the total line intensity defined as $I_{OI} = \sqrt{2\pi} I_{p}\sigma/\Delta\lambda$, where $I_{p}$ is the amplitude and $\sigma$ the standard deviation of the Gaussian component (in wavelength units), and $\Delta\lambda$ is the wavelength grid spacing  (required to normalise $\sigma$ to be a number of pixels rather than in units of \AA). 
The second row is the Doppler shift of the line, $V_{\mathrm{Dopp}}$, calculated from the shift of the centroid of the Gaussian component from the rest wavelength, and the third row is the nonthermal width $w_{\mathrm{Nthm}}$. The first column is the condensation component, and the second column is the stationary component. 

The total intensity of the condensation component grows over $\sim40$~s to a peak, and decays over approximately a further $40$~s (plus or minus a few exposures for each pixel). While the stationary component grows in intensity over a similar timescale, it does not decrease following the peak, remaining bright over the time frame analysed here. Doppler shifts are between $2-12$~km~s$^{-1}$ for the condensation component (with the range based on a histogram of all three pixels). That is actually rather slow in comparison to general flare-induced chromospheric downflows which can be up to several 10s of km~s$^{-1}$. The stationary component shows only marginal Doppler shifts around the rest wavelength, including slight blueshifts, which is not unexpected if this is indeed representing a mostly stationary layer. \cite{2016ApJ...829...35W} report that in the B-class flare they studied that \ion{O}{i} 1355.598~\AA\ showed almost no change in Doppler motions, and generally exhibited a small blueshift of $V_\mathrm{{Dopp}} \sim -2$~km~s$^{-1}$, which the authors attributed to uncertainty with the wavelength calibration. Similarly, \cite{2016ApJ...829...35W} note almost no change in line width, which is consistent with the stationary component that we observe, where $w_{\mathrm{Nthm}}$ is comparable to the background values discussed earlier. However, the condensation component does show an increase above the background, reaching $w_{\mathrm{Nthm}}\sim5-12$~km~s$^{-1}$ (with the range based on a histogram of all three pixels), though with some slightly larger or smaller values at times. There is a fair amount of scatter within that range, which is likely due to the assumption of a constant $T=10$~kK and due to the narrowness of the line which makes fitting difficult. In a future study, the ideal would be to not include spectral summing.

\begin{figure*}
	\centering 
	\subfloat{\includegraphics[width = 0.75\textwidth, clip = true, trim = 0.cm 0.cm 0.cm 0.cm]{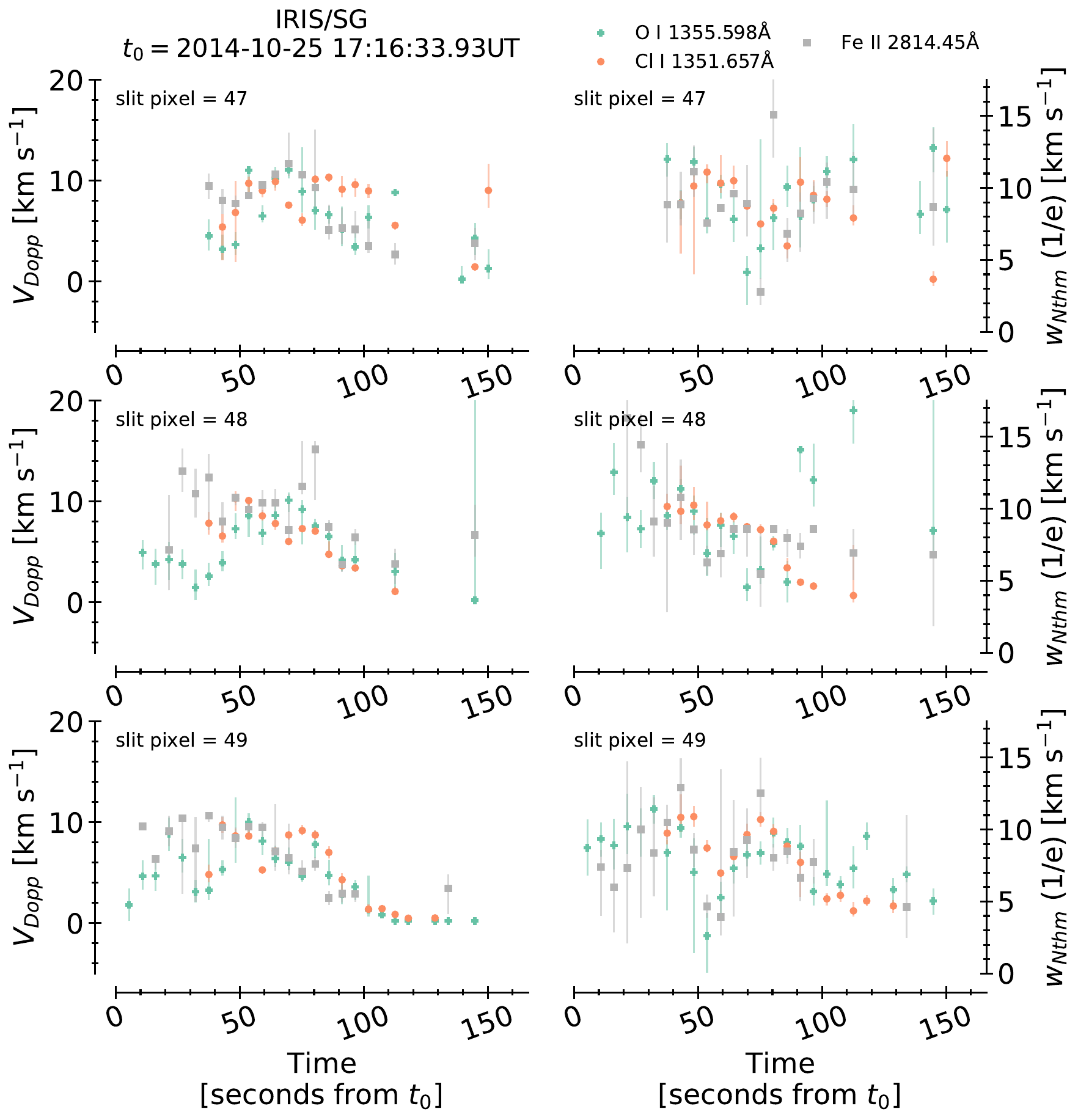}}	
	\caption{\textsl{Comparing the Doppler shift (first column) and nonthermal line width (second column) of the shifted component (the condensation) obtained for the  \ion{O}{i} 1355.598~\AA\ (green `$+$' symbols), \ion{Cl}{i} 1351.657~\AA\ (orange `$\bullet$' symbols), \& \ion{Fe}{ii} 2814.445~\AA\ (grey `$\blacksquare$' symbols). Each row shows the metrics for a different slit position.  Error bars are from a Monte Carlo analysis of each profile.}}
	\label{fig:iris_alllines_dgfitmetrics}
\end{figure*}

We made the assumption that the shifted components of \ion{Cl}{i} 1351.66~\AA\ and \ion{Fe}{ii} 2814.445~\AA\ were optically thin (allowing us to extract similar metrics as we did for \ion{O}{i} 1355.598~\AA), and that they also form at the same temperature as \ion{O}{i} since they originate within the chromospheric condensation. These assumptions seem to be largely borne out, with comparable values of $V_{\mathrm{Dopp}}$ and $w_{\mathrm{Nthm}}$ for each line. Considering all three lines there was a typical nonthermal velocity around $w_{\mathrm{Nthm}} = 10$~km~s$^{-1}$. Where there are differences, such as \ion{Cl}{i} and \ion{Fe}{ii} having a larger initial Doppler shift, these could be due to those species forming earlier within the condensation's evolution, or being generally brighter and thus detectable earlier. Also, there could be a small temperature gradient between the lines such that a fixed $T=10$~kK is not fully appropriate. Figure~\ref{fig:iris_alllines_dgfitmetrics} shows the comparisons of $V_{\mathrm{Dopp}}$ (first column) and $w_{\mathrm{Nthm}}$ (second column) for the three lines, which are represented by different colours and symbols. Each row is a different slit pixel. 


\section{Modelling IRIS Spectra in Solar Flares}
As outlined in the Introduction, \ion{Mg}{ii} resonance line wings are anomalously broad compared to model predictions, and  one potential resolution or mitigation is to increase the turbulent velocity in flare simulations. We performed data-guided flare simulations to model these lines, and when synthesising the spectral line emission included an enhanced turbulent velocity of 10~km~s$^{-1}$ inside the condensation, as suggested by the preceding observational analysis. Given the widths of the \ion{O}{i}, \ion{Fe}{i}, and \ion{Cl}{i} stationary components it is unlikely that a turbulent velocity $>>10$~km~s$^{-1}$ is present at depths greater than the condensation. The simulations and the model-data comparisons are presented below.

\subsection{Solar Flare Modelling}\label{sec:numerics}
Two flare simulations were produced using the \radynfp\ codes, driven by injected nonthermal electron distributions inferred from observations, but with the caveat that the pre-flare atmosphere is not wholly representative of the pre-flare state in the observations. These are intended as a general guide, and not to represent an apples-to-apples model-data comparison of these flare sources. 

\radynfp\ is the combination of the \radyn\ radiation hydrodynamics code \citep{1995ApJ...440L..29C,1997ApJ...481..500C,1999ApJ...521..906A,2005ApJ...630..573A,2015ApJ...809..104A} and the nonthermal particle transport code \fpc\ \citep{2020ApJ...902...16A}, which together model the response of the solar atmosphere to energy injection via a distribution of nonthermal particles. \radyn\ solves the coupled equations of non-local thermodynamic equilibrium radiation transfer, non-equilibrium atomic level populations, and hydrodynamics, including the feedback among those equations. These equations are solved for one half of a symmetric loop, on an adaptive grid capable of resolving the strong shocks and gradients that typically appear during flares. Energy injection via the injected nonthermal particles then heats the plasma, which evolves and subsequently has a feedback on the energy losses of the particles as the simulation progresses. Nonthermal collisional excitation and ionisation of hydrogen and helium by the particle beam is included in the solution, which are important sources of ionisation during the flare impulsive phase. Backwarming by the optically thin flare-heated corona and transition region are all included in the solution, which can heat and photoionise the chromosphere. \radyn\ was recently upgraded with an improved treatment of electric-pressure (Stark) damping of the Hydrogen lines, with the exception of the Ly$\alpha$ \& Ly$\beta$ transitions \citep{2022ApJ...928..190K}. Though \radyn\ now has the ability to suppress conduction via non-local effects and turbulence \citep{2022ApJ...931...60A} we do not model those effects in our experiments, to avoid adding additional unconstrained parameters. Instead, thermal conduction is Spitzer, saturated at the free-streaming limit. \fpc\ can model the propagation and thermalisation of a distribution of particles of any mass and charge (i.e. electron, proton, or heavy ion beams, or combinations thereof), including processes such the beam-neutralising return current. The injected distribution can be of any form, but here we use the standard assumption of a power-law with slope $\delta$, low-energy cutoff $E_{c}$ and total energy flux, $F$. 

\radyn, and now \radynfp, has become a workhorse of the flare modelling community, where it has been used to understand atmospheric evolution, energy transport mechanisms, and radiative processes during solar flares. For an in-depth discussion of how \radyn\ and other flare loop models have been employed over the last decade to help understand IRIS observations see \cite{2022FrASS...960856K} and \cite{2023FrASS...960862K}. 
\begin{figure*}
	\centering 
	\vbox{
	\hbox{
	\hspace{0.35in}
	\subfloat{\includegraphics[width = 0.45\textwidth, clip = true, trim = 0.cm 0.cm 0.cm 0.cm]{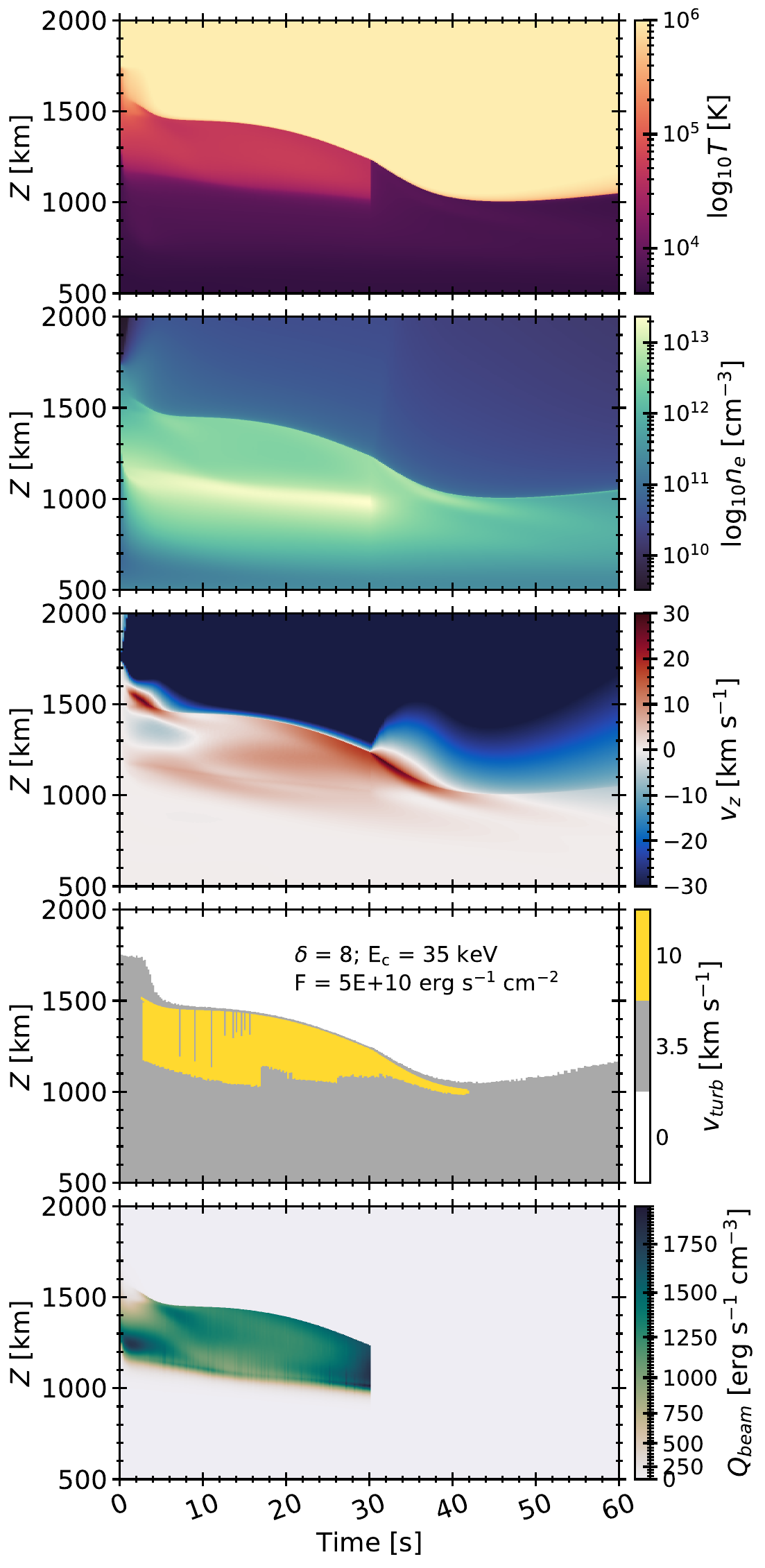}}
	\subfloat{\includegraphics[width = 0.446\textwidth, clip = true, trim = 0.cm 0.cm 0.cm 0.cm]{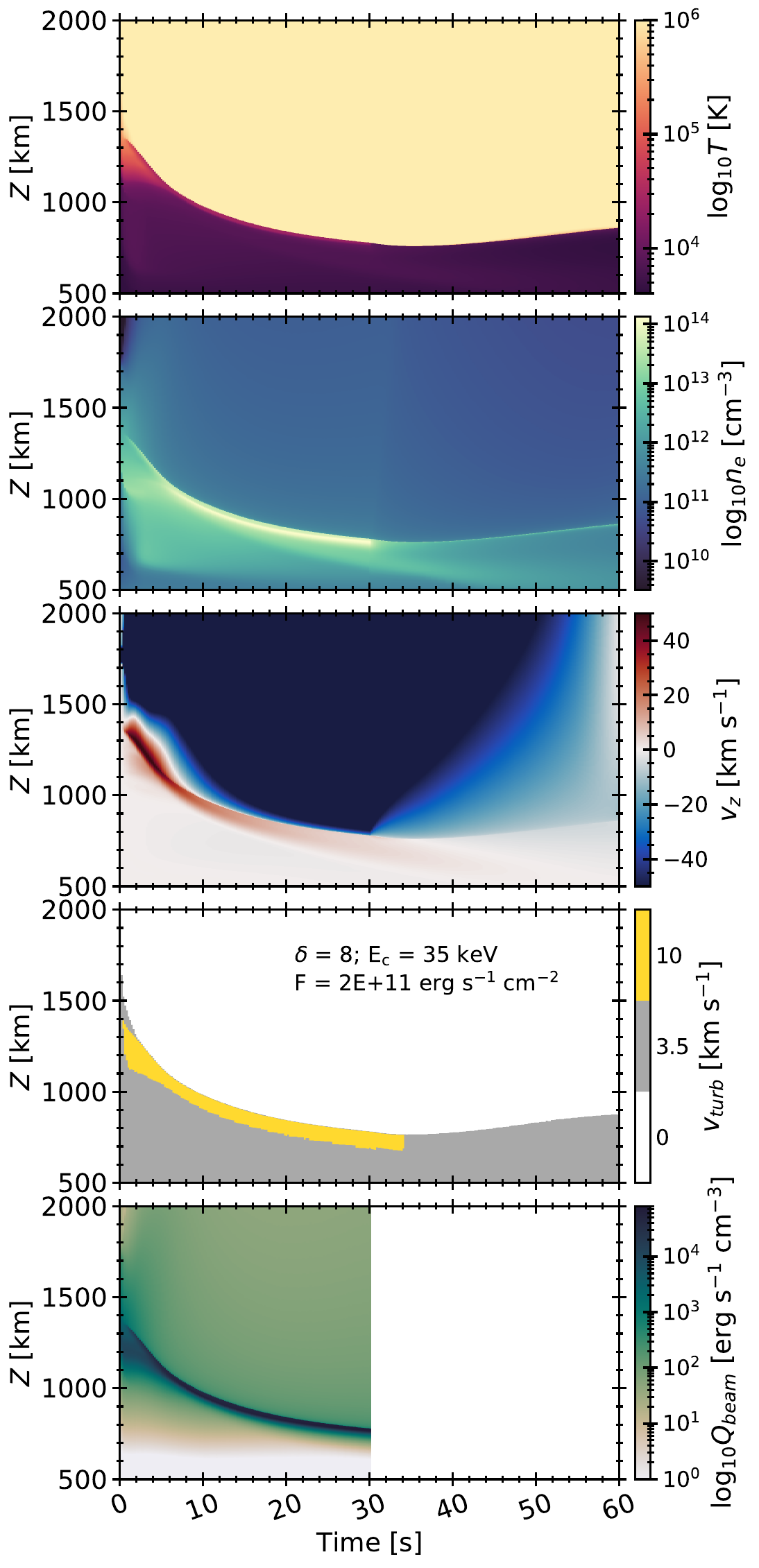}}	
	}
	}
	\caption{\textsl{The evolution of the model flare atmospheres from the \radynfp\ code: each simulation was driven by a nonthermal electron distribution, with $\delta = 8$ and $E_{c} = 35$~keV, and injected energy fluxes of $F = 5\times10^{10}$~erg~s$^{-1}$~cm$^{-2}$ (left column) or $F = 2\times10^{11}$~erg~s$^{-1}$~cm$^{-2}$ (right column), both heated for $t = 30$~s. For each the temperature (top row), electron density (second row), bulk velocity (third row) and turbulent velocity (fourth row), and the heating rate of the nonthermal electron distribution, $Q_\mathrm{beam}$ (bottom row; note the logarithmic scale in the 2F11 simulation) are shown. The stratification of turbulent velocity was manually defined only for use in the \rhpar\ simulations, and was set such that $V_\mathrm{turb} = 3.5$~km~s$^{-1}$ everywhere except $T>500$~kK where $V_\mathrm{turb} = 0$~km~s$^{-1}$, or within the chromospheric condensation where $V_\mathrm{turb} = 10$~km~s$^{-1}$.}}
	\label{fig:radyn_flare_atmos}
\end{figure*}

\cite{2022ApJ...926..164A} studied this same flare, from which they determined a loop of length $L = 85$~Mm of a source nearby the ribbon that we analyse here. We adopt this same length in our experiments such that the \radyn\ pre-flare atmosphere was a $L_{1/2} = 42.5$~Mm half-loop. A reflecting boundary condition was used at the top of the loop, to mimic incoming disturbances from the other leg of the loop. The pre-flare structure was a VAL3C-like atmosphere \citep{1981ApJS...45..635V}, with 300 grid cells and apex values of $T = 1$~MK and $n_{e} = 6.2\times10^{8}$~cm$^{-3}$, constructed using the atmosphere described in \cite{2023arXiv230402618C}, stretched to $L_{1/2} = 42.5$~Mm from an initial 10~Mm. Bound-bound transitions of \ion{H}{i}, \ion{He}{i}, \ion{He}{ii}, and \ion{Ca}{ii} were solved assuming complete frequency redistribution (CRD), though the Lyman series were treated as pure Doppler \citep[though with Stark broadening from][]{2015ApJ...809..104A} profiles to mimic partial frequency redistribution (PRD) of the Lyman lines for the purposes of energy balance \citep[see][]{2012ApJ...749..136L}. We acknowledge here that the observations were of umbral/penumbral flare sources, which would have a cooler lower atmosphere than we model here. To treat such an atmosphere properly, particularly as concerns continuum emission, would require careful consideration of molecular opacity, which we have not yet included in \radyn. Therefore our models represent more general flare simulations, with other properties guided by the observations. 
 
Analysis of the hard X-ray observations from this event, using RHESSI or \textsl{Fermi}/Gamma Ray Burst Monitor \citep[GBM;][]{2009ApJ...702..791M} data, imply a very soft spectrum, with a slope of $\delta = 7-8$ \citep[e.g.][]{2017ApJ...837..160K,2019ApJ...878..135K}. Imaging the RHESSI 25-50~keV X-rays suggests a strong coronal source present near $\sim$17:06UT \citep[][]{2022ApJ...926..164A}, though there are hints that this extends down to footpoints \citep{2017ApJ...837..160K}. There are additional bursts of hard X-rays observed by \textsl{Fermi} around the time of our source of interest $\sim$17:16-17:20~UT. Due to the strong coronal source and apparent lack of strong footpoints observed at $25-50$~keV, \cite{2022ApJ...926..164A} suggested that this flare was mostly thermal in nature, and proceeded to model the event by injecting energy directly at their loop apex (with timescales guided by the lifetime and intensity of AIA 1600~\AA\ pixels co-spatial with the IRIS ribbon). However, for a few reasons we decided to model this event by injecting a distribution of nonthermal particles\footnote{Also, since we aim test the impact of including enhanced microturbulent broadening inside a condensation, so all we really need are flares that produce downflows, without being overly concerned as to how the energy ultimately reached the chromosphere}. See \cite{2019ApJ...878..135K} for details, but in summary: (1) the \textsl{Fermi} data suggested a somewhat high value of the low-energy cutoff, $E_{c} = 35$~keV, so that, indeed, images of 25-50~keV X-rays may have been largely thermal in nature, with the dominant source in the corona (the dynamic range of RHESSI means that in the presence of a strong source, weaker sources are not detectable); (2) features in the \textsl{Fermi} $35-41$~keV and $58-72$~keV lightcurves are co-temporal with the appearance of newly brightened source areas in the chromosphere, suggestive of a direct link between energetic electrons and deposition of energy into the lower atmosphere \citep[see also][]{2020ApJ...895....6G}; and (3) there was a rapid variability in $>35$~keV HXR lightcurves, so they are more likely to to be nonthermal in nature. The total power carried by nonthermal electrons at the times of interest were derived from the \textsl{Fermi} observations, which can be divided over the flaring area to obtain the energy flux. \cite{2019ApJ...878..135K} determined a range of possible energy fluxes, $F = 2\times10^{10} - 2\times10^{11}$~erg~s$^{-1}$~cm$^{-2}$. We chose two values of energy flux within this range, such that the two flare simulations were driven by injection of a nonthermal electron distribution with $\delta = 8$, $E_{c} = 35$~keV, and $F = [5\times10^{10}, 2\times10^{11}]$~erg~s$^{-1}$~cm$^{-2}$ (hereafter we refer to these as 5F10 and 2F11 experiments, respectively), with a constant injection rate lasting 30~s.

The evolution in space and time of the two flares is shown in Figure~\ref{fig:radyn_flare_atmos}, where the lefthand column is the 5F10 simulation, and the righthand is the 2F11 simulation. Though the temperature panels are saturated at $T=1$~MK, this is just to allow more details to be seen in the cooler lower atmosphere, both simulations resulted in coronal temperatures exceeding $T=10$~MK (including in the lower corona near the flare transition region), and the 2F11 simulation produced $T>45$~MK. As would be expected, the 2F11 simulation drove more rapid and dramatic changes, with a much more compressed flare transition region and chromosphere. 

\subsection{Synthesising the Flare Spectra}
\subsubsection{\ion{O}{i} \& \ion{Mg}{ii} Spectral Line Synthesis Using \rhpar}
\radyn\ does not typically synthesise by default the species observed by IRIS, so we instead used the \rhpar\ radiation transfer code \citep{2015A&A...574A...3P,2001ApJ...557..389U}, with \radyn\ flare atmospheres used as input. Using \rhpar\ also allowed us to include overlapping transitions and the effects of PRD \citep[important for the treatment of the \ion{Mg}{ii} line wings, even in flares][]{2013ApJ...772...90L,2019ApJ...883...57K}. The stratification of temperature, electron density, bulk velocity, hydrogen level populations, and the depth scale on which they are defined, from \radyn\ are used by \rhpar\ to solve the NLTE radiation transport and statistical equilibrium atomic population equations. Any number of atomic or molecular species can be included in the solution. It does, however, not take into account non-equilibrium ionisation so that the temporal history of the atmosphere is lost when solving each snapshot. Since the hydrogen ionisation stratification is important for the formation of \ion{O}{i} due to charge exchange, the lack of non-equilibrium effects can be mitigated by using the non-equilibrium hydrogen populations from \radyn. We have modified \rhpar\ to keep the populations of hydrogen from \radyn\ fixed and to just solve the radiation transfer problem for those transitions. This also allows the inclusion of nonthermal collisional effects (excitation and ionisation of H) on the \radyn\ hydrogen populations to be included.  We have previously demonstrated that, aside from the initial onset and cessation of flare heating, non-equilibrium effects can be safely ignored for \ion{Mg}{ii} in moderate-to-strong flares in favour of modellng PRD effects \citep{2019ApJ...885..119K}\footnote{\cite{2022A&A...668A..96T} compared recipes of radiative losses in flares, and when doing so found that employing statistical equilibrium rather than non-equilibrium ionisation could result in intensity differences throughout the heating phase. However, their heating rate was more gradual than in our simulations here (hence dynamics were also more gradual). Also, \cite{2019ApJ...883...57K} found that the differences between a CRD and PRD solution were significantly greater and, we believe, pose a larger problem than non-equilibrium effects.}. Another relevant modification made to \rhpar\ was to use the updated treatment of \ion{Mg}{ii} Stark damping implemented by \cite{2019ApJ...879...19Z} in the non-parallelised version of \rh. Both \ion{Mg}{ii} and \ion{O}{i} are synthesied in non-LTE.

When processing our flare atmosphere snapshots (at $0.2$~s cadence) through \rhpar\ we truncated the atmosphere above $T = 1$~MK, and solved the following in detail: \ion{H}{i}, \ion{C}{i}+\ion{C}{ii}, \ion{O}{i}\footnote{The 16-level model atom from \cite{2015ApJ...813...34L}.}, \ion{Si}{i}+\ion{Si}{ii}, \& \ion{Mg}{ii}\footnote{The 11-level model atom from \cite{2013ApJ...772...90L}.}. Other species included as sources of background opacity in LTE were: \ion{Al}{i}+\ion{Al}{ii}, \ion{Ca}{ii}, \ion{He}{i}+\ion{He}{ii}, \ion{N}{i}+\ion{N}{ii}, \ion{Na}{i}+\ion{Na}{ii}, \ion{S}{i}+\ion{S}{ii}, \ion{Ba}{ii}, \ion{Ni}{i}+\ion{Ni}{ii}, \ion{K}{i}, \& \ion{Sr}{i}. To account for the UV line haze, and the opacity of transitions not included in the model atoms, we employed the  `Kurucz' linelists\footnote{\url{http://kurucz.harvard.edu/linelists.html}}, which model thousands of lines in the range $\lambda = [20-8000]$~\AA, assuming LTE. Finally, we set an additional value of microturbulence at each depth point (varying in time) to mimic the nonthermal width from the observations, that we assume was wholly due to microturbulence:

\begin{equation}
V_\mathrm{turb}=\begin{cases}
10~\mathrm{km~s^{-1}},& T<500~\text{kK}, v_{z} \ge 3~\mathrm{km}~\mathrm{s}^{-1},\\
3.5\mathrm{km~s^{-1}},& T<500~\text{kK}, v_{z} < 3~\mathrm{km}~\mathrm{s}^{-1}\\
0~\mathrm{km~s^{-1}},              & \text{otherwise},
\end{cases}
\end{equation}

\noindent A value of $3.5$~km~s$^{-1}$ was chosen outside of the condensation as it lies in the range for the pre-flare and flaring stationary component nonthermal widths, and a value of $10$~km~s$^{-1}$ was chosen as it was typical of the nonthermal widths in the red-wing component during the flare. 
. 
\begin{figure*}
	\centering 
	\vbox{
	\hbox{
	\hspace{0.5in}
	\subfloat{\includegraphics[width = 0.375\textwidth, clip = true, trim = 0.cm 0.cm 0.cm 0.cm]{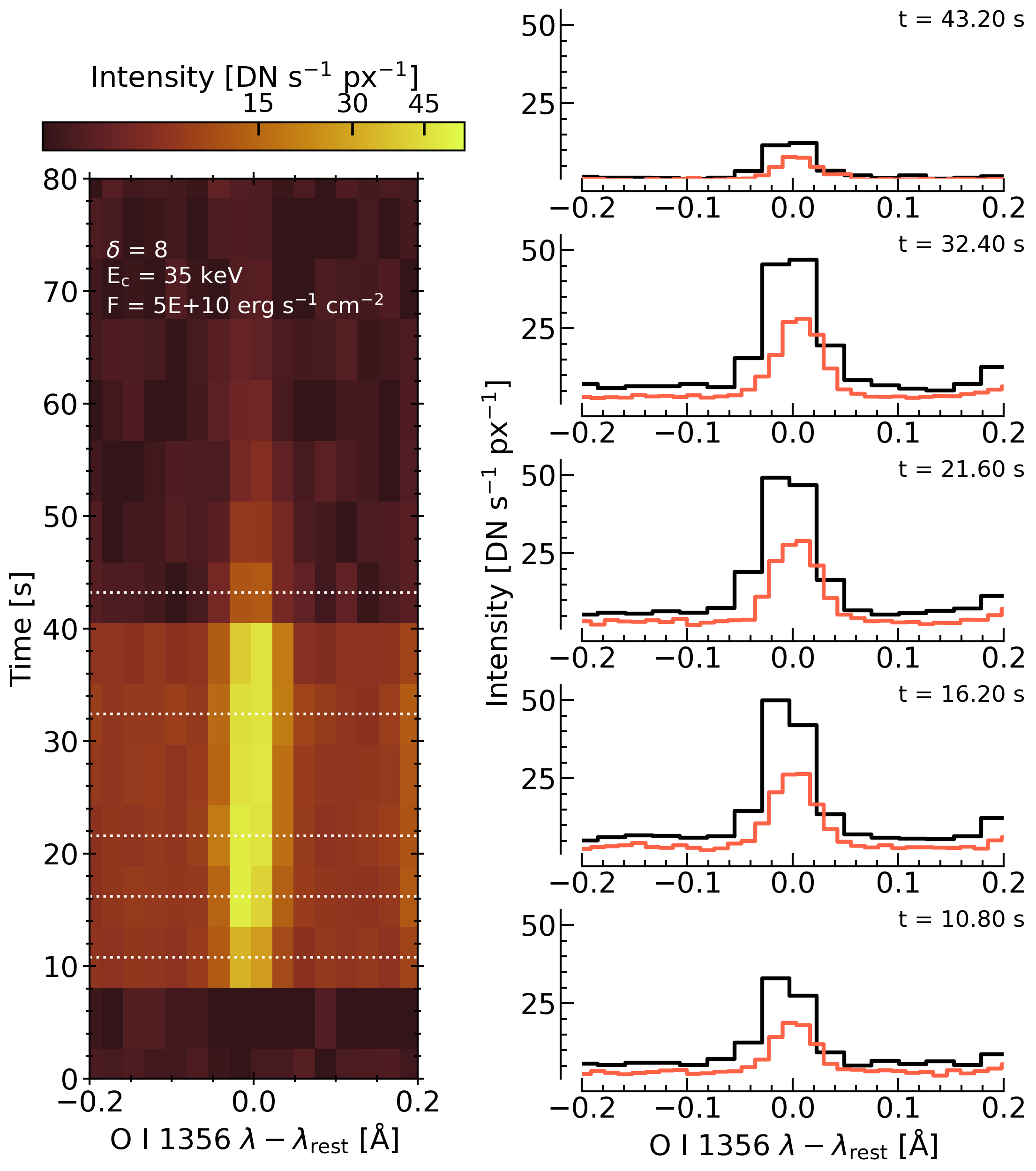}}
	\subfloat{\includegraphics[width = 0.375\textwidth, clip = true, trim = 0.cm 0.cm 0.cm 0.cm]{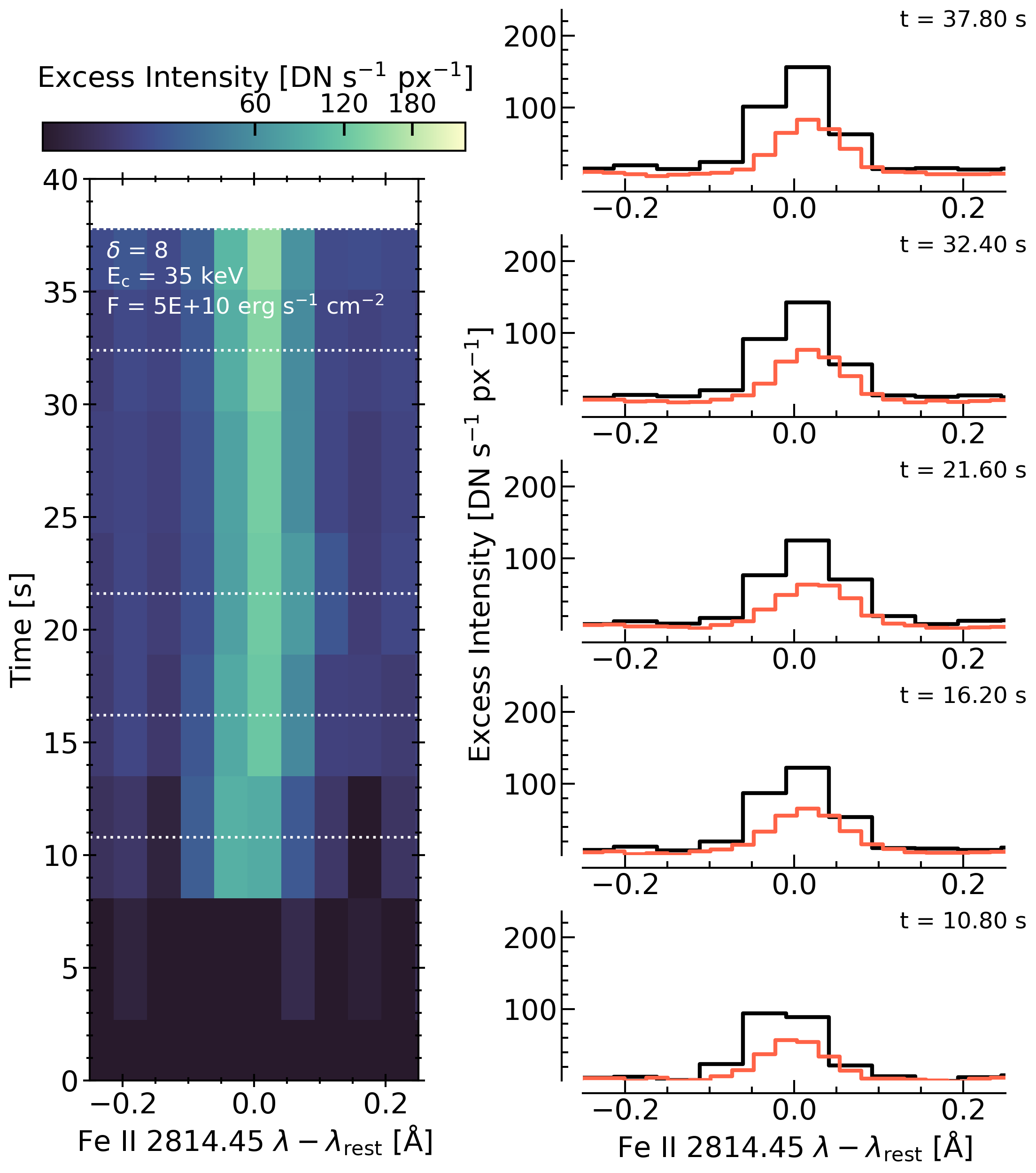}}	
	}
	}
	\vbox{
	\hbox{
	\hspace{0.5in}
	\subfloat{\includegraphics[width = 0.375\textwidth, clip = true, trim = 0.cm 0.cm 0.cm 0.cm]{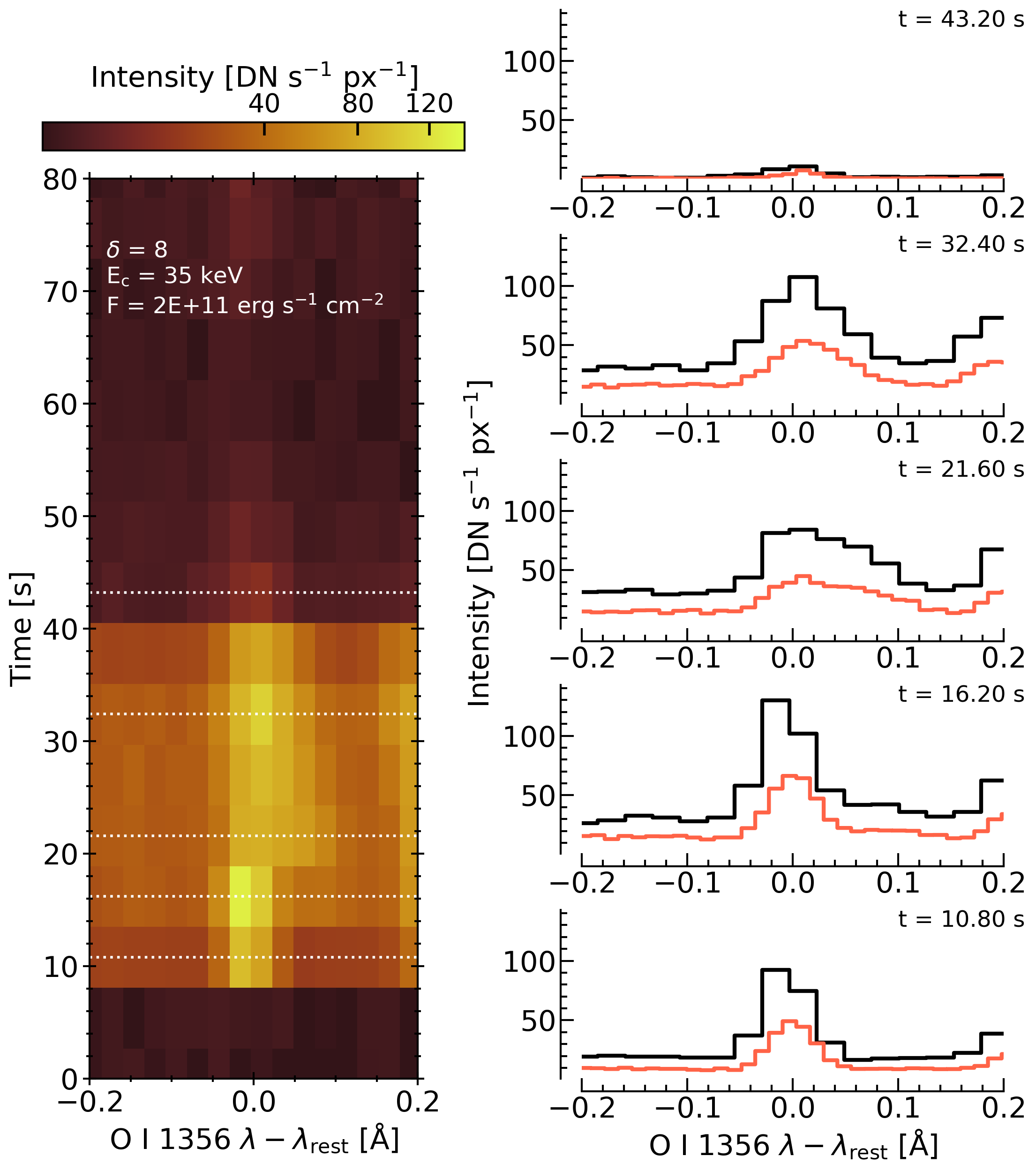}}	
	\subfloat{\includegraphics[width = 0.375\textwidth, clip = true, trim = 0.cm 0.cm 0.cm 0.cm]{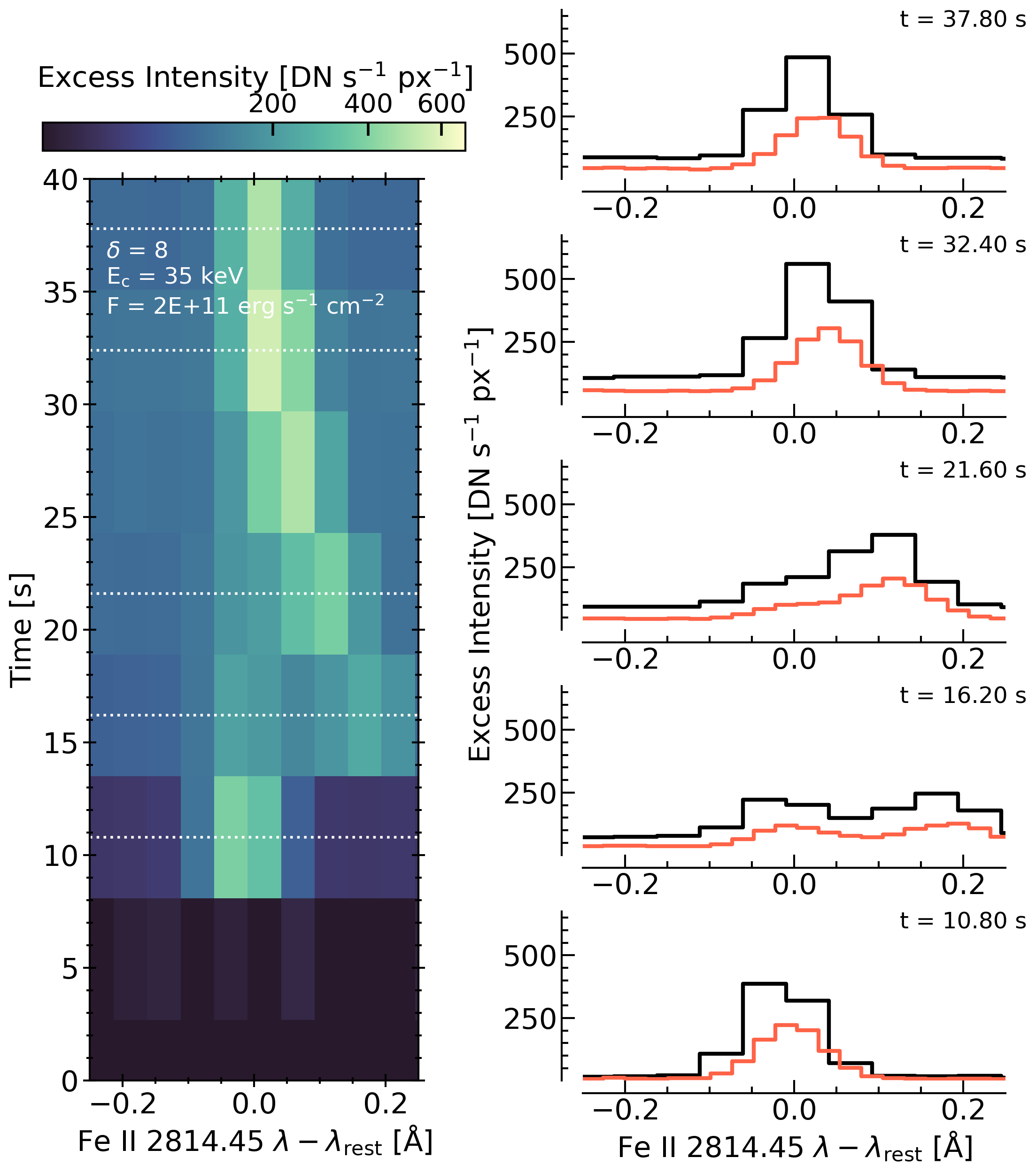}}	
	}
	}
	\caption{\textsl{The temporal evolution of \ion{O}{i} (left) and \ion{Fe}{ii} (right) synthetic emission, with cut-outs showing individual spectra in more detail. In each wavelength-time diagram the horizontal lines indicate the times of the cut-outs. The top row shows the 5F10 simulation, and the bottom row the 2F11 simulation. The red lines assume no spectral binning, and black lines (and the background) assume 2x spectral binning.  For \ion{Fe}{ii} spectra we show the flare excess, which better illustrates the line's response to the flare in comparison to the observations, given the very different background emission from the quiet photosphere (model) and umbra (observation)}}
	\label{fig:rh_spectra_overview}
\end{figure*}

\subsubsection{\ion{Fe}{ii} Spectral Line Synthesis}
\radyn\ includes one wavelength in the IRIS NUV range at $\lambda=2830$ \AA\ in its detailed calculations.  Thus, we follow \cite{2017ApJ...836...12K} and calculate the \ion{Fe}{ii} 2814.445~\AA\ line to compare to the IRIS spectra of this line from the umbral flare brightenings \citep{2019ApJ...878..135K}.   The calculations employ the non-equilibrium electron densities from \radyn\ at every time step in the LTE ionization ratios among \ion{Fe}{i}, \ion{Fe}{ii}, and \ion{Fe}{iii} and LTE population densities of the upper and lower levels of the \ion{Fe}{ii} 2814.445~\AA\ transition.  We also include opacity from all known continuous sources, including the far-wing opacity from the \ion{Mg}{ii} h \& k lines \citep{2020ApJ...895....6G}.  The photospheric wing opacity at the wavelengths of \ion{Fe}{ii} 
 2814.445~\AA\ allows for a more accurate subtraction of the nearby continuum radiation at relatively low intensity values during the flare.  One aspect of the atomic data (statistical weights) for Fe has been updated since these modeling papers, but the details of these changes will be described in a forthcoming publication (Kowalski et al. 2023, in preparation).  The changes result in fainter overall \ion{Fe}{ii} line intensity  but  similar profile shapes.  We also calculate excess continuum intensity values at 2826 \AA\ for comparisons of the continuum-to-line ratio to those calculated from the observations, which reached values as high as $\approx 8$ in this flare \citep{2019ApJ...878..135K}.  We checked that the excess values of the continuum around 2826\AA\ in our calculation with our approximate formulation of the \ion{Mg}{ii} h far-wing opacity are consistent with RADYN's calculation at $\lambda = 2830$ \AA.  

\subsubsection{Comments on Plasma Conditions in line Formation Region}
From an analysis of the line formation properties we can asses if the fundamental assumptions of our methodology are reasonable. That is, that the shifted component of the \ion{O}{i} and \ion{Fe}{ii} lines\footnote{We also make this assumption for \ion{Cl}{i} but cannot test this with our current modelling setup.} are optically thin in the flare, that they have a formation temperature on the order $T\sim10$~kK, and that they form close to \ion{Mg}{ii}; for more details of this analysis see Appendix~\ref{sec:formprops}.

The \ion{O}{i} 1355.598~\AA\ can be considered fully optically thin even during the flare, both the core and shifted component. Though the line core of \ion{Fe}{ii} 2814.445~\AA\ suffered from some opacity effects at certain times, the shifted component was optically thin. Our assumption of optically thin formation conditions for the shifted features of those lines was therefore found to be valid, both forming inside the high-lying condensation several hundred km from the point at which optical depth $\tau_{\lambda} = 1$. Further, during the flare both the core and especially the shifted components of those lines form close to the \ion{Mg}{ii} lines (only a few 10s~km separation), though there is a strong gradient in temperature and density present. Still, it is reasonable to apply the observationally derived values of nonthermal width as a microturbulent broadening within the condensations in our simulations. The formation temperature of \ion{O}{i} 1355.6~\AA\ ($T\sim10-18$~kK) may be slightly larger than assumed in our observational analysis ($T = 10$~kK). This difference of formation temperature in the condensation component should not have a very large impact (recall Appendix~\ref{sec:varyformt}), but would mean that the value of $V_\mathrm{turb} = 10$~km~s$^{-1}$ in this particular observation, and hence any impact on \ion{Mg}{ii} line broadening, is an \textsl{upper} limit. Similarly, \ion{Fe}{ii} and \ion{Cl}{i} are heavier than \ion{O}{i} so relatively small differences in temperature have even less of an impact. Though the stationary components of both \ion{O}{i} and \ion{Fe}{ii} form higher during the flare, near \ion{Mg}{ii}, they have contributions from a width swathe of the chromosphere, from $z > \sim600$~km for \ion{O}{i}, and $z > \sim 400$~km for \ion{Fe}{ii}. Thus, the microturbulence through the mid-lower chromosphere is constrained by the nonthermal widths of the stationary components.

\subsection{Behaviour of the Synthetic Flare Spectra}
\subsubsection{General Characteristics}
To better compare the synthetic-to-observed spectra, the output of the numerical models were processed in the following manner: (1) 10~s of pre-flare were added to the spectra so that when including an IRIS exposure time the synthetic observation did not start immediately at flare onset (i.e. in the synthetic IRIS observations the flare now starts at $t=10$~s, not at $t=0$~s); (2) the intensity, in physical units of erg~s$^{-1}$~cm$^{-2}$~sr$^{-1}$~\AA$^{-1}$, were recast onto a wavelength grid with bin size equal to the IRIS plate scale (NUV: $\Delta\lambda = 25.46$~m\AA, and FUVS: $\Delta\lambda = 12.98$~m\AA), conserving the intensity when doing so; (3) the spectra were then convolved with a point spread function (PSF), assumed to be a Gaussian with FWHM of 2 IRIS spectral pixels; (4) spectra were converted from units of energy to number of photons; (5) the solid angle over which a single IRIS SG pixel subtends was calculated and the spectra multiplied by this value ($\Delta X = 0.33$\arcsec, and $\Delta Y = 0.166$\arcsec, such that we assume a filling factor of the flare source of one); (6) the spectra were then multiplied by the IRIS effective area, obtained from the IDL program \texttt{IRIS\_get\_response.pro}, from the SolarSoftware tree for IRIS\footnote{\url{https://hesperia.gsfc.nasa.gov/ssw/iris/idl/lmsal/util/iris_get_response.pro}} (we used the time-dependent values to get the effective area from the 25-October-2014); (7) the spectra were integrated over an exposure time of $\tau_\mathrm{exp} = 4$~s ,
and a readout time of $1.4$~s was adopted (the average cadence of the observations was $\tau_\mathrm{cad} = 5.36$~s, which when rounded to the cadence of the \rhpar\ output of $0.2$~s gives us the $1.4$~s readout time); (8) Poisson noise was added by sampling from a Poisson distribution with mean value equal to the intensity within each bin; (9) spectral summing was applied to mimic the observations, providing new wavelength grids with $\Delta\lambda = 50.92$~m\AA\ in the NUV, and $25.96$~m\AA\ in the FUV; (10) count rates in photons were converted to Data Numbers (DN), assuming 18 photons DN$^{-1}$ in the NUV and 4 photons DN$^{-1}$ in the FUV. 

\begin{figure*}
	\centering 
	\vbox{
	\hbox{
	\hspace{0.5in}
	\subfloat{\includegraphics[width = 0.375\textwidth, clip = true, trim = 0.cm 0.cm 0.cm 0.cm]{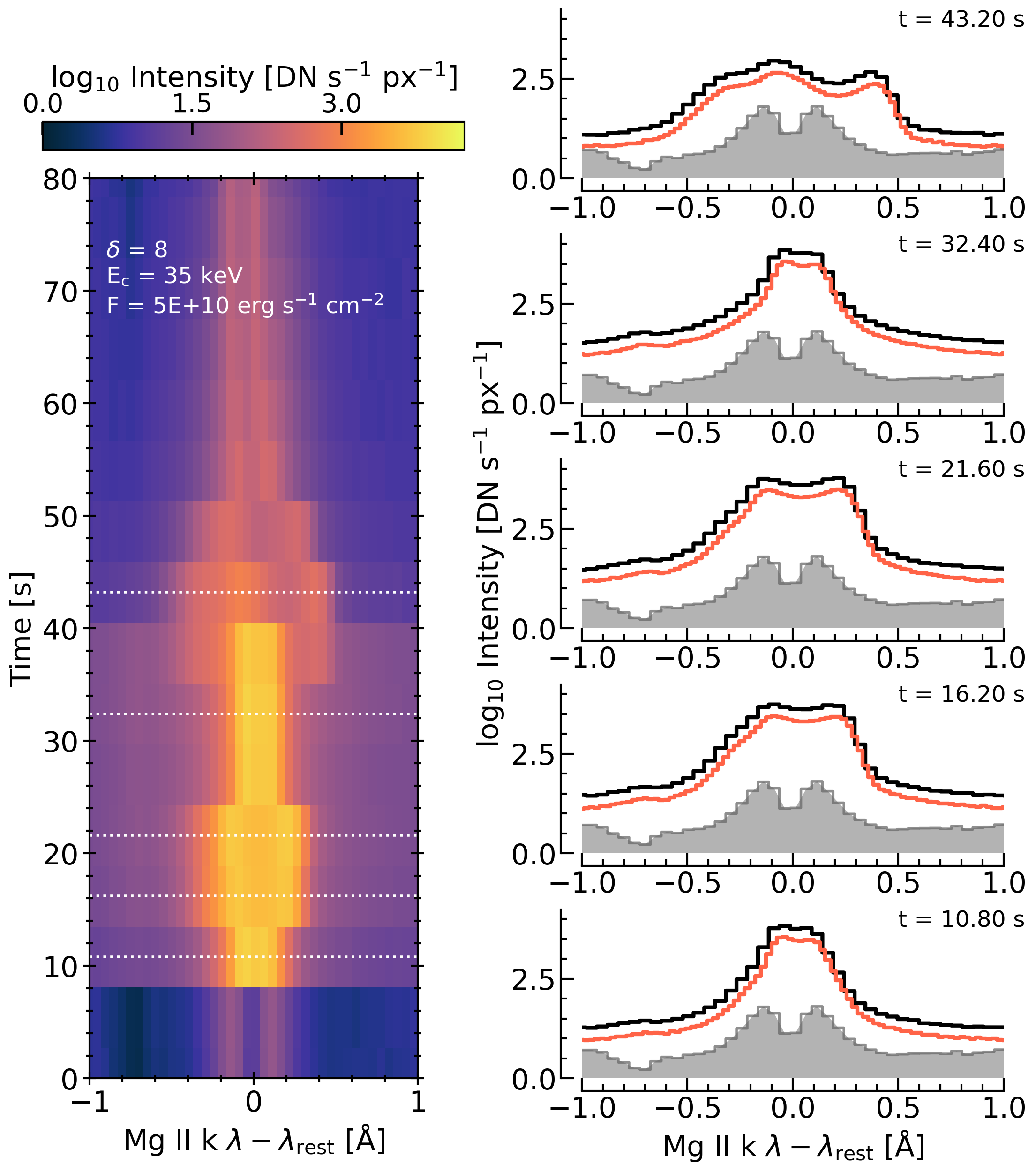}}
	\subfloat{\includegraphics[width = 0.375\textwidth, clip = true, trim = 0.cm 0.cm 0.cm 0.cm]{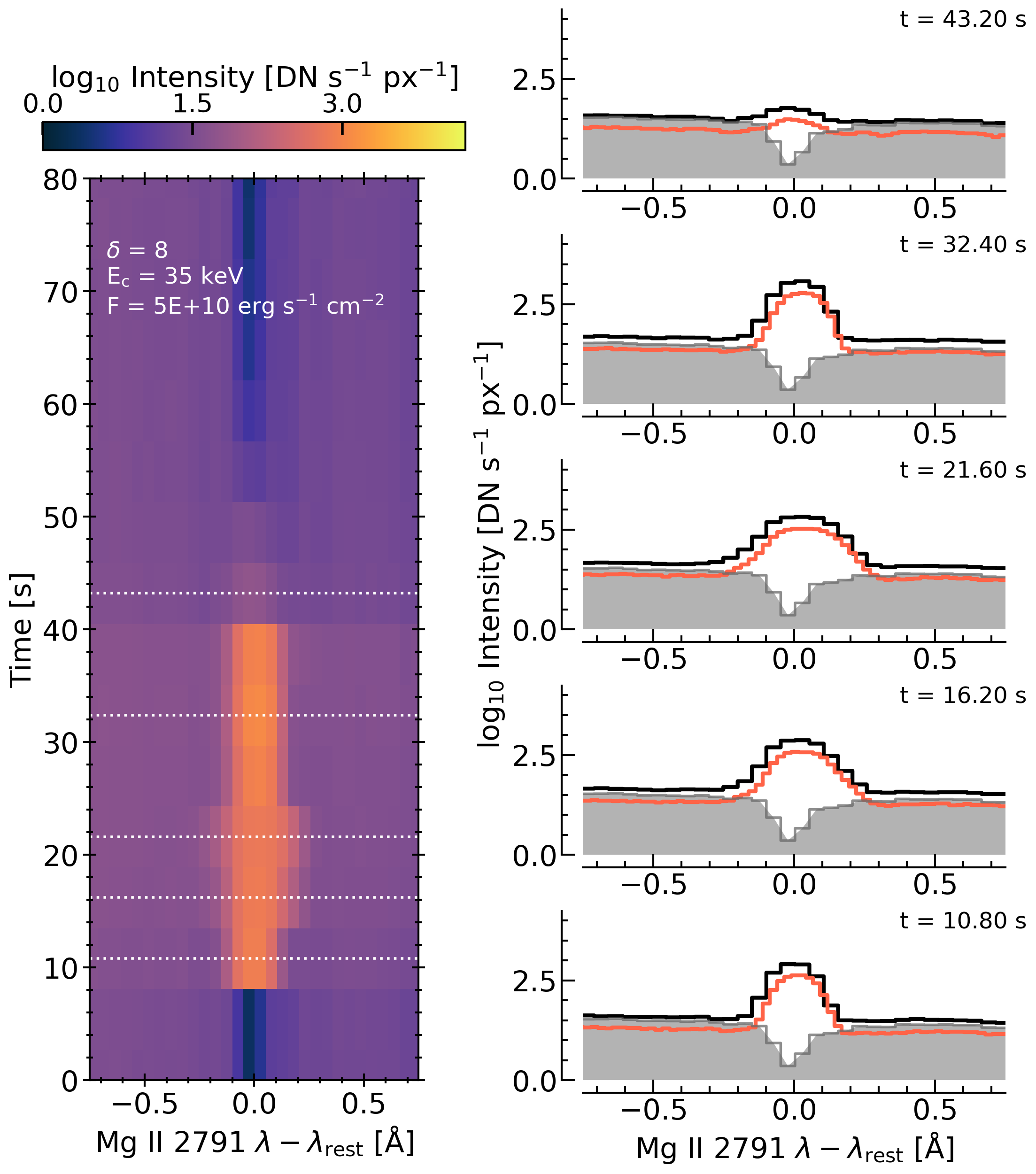}}	
	}
	}
	\vbox{
	\hbox{
	\hspace{0.5in}
	\subfloat{\includegraphics[width = 0.375\textwidth, clip = true, trim = 0.cm 0.cm 0.cm 0.cm]{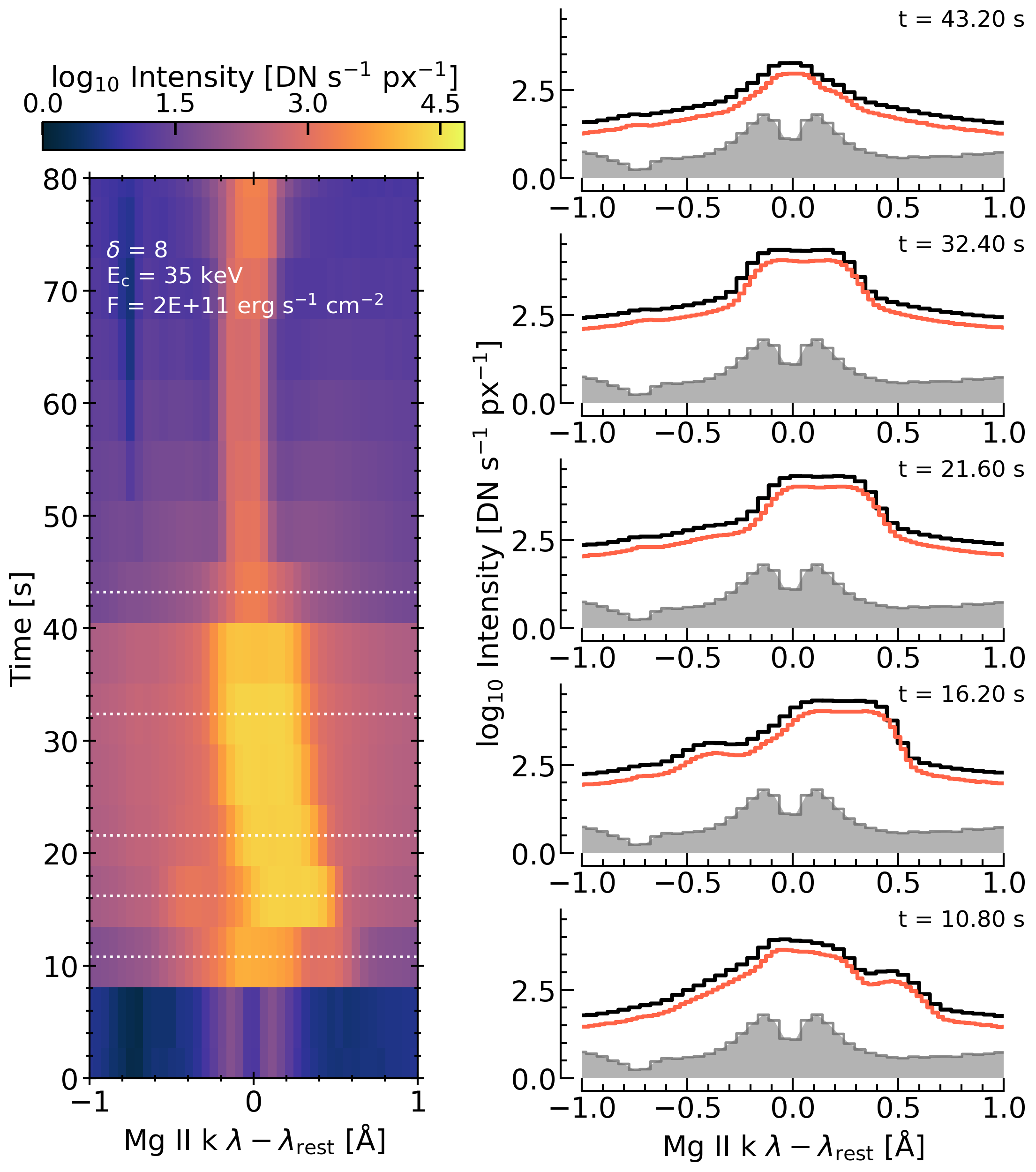}}	
	\subfloat{\includegraphics[width = 0.375\textwidth, clip = true, trim = 0.cm 0.cm 0.cm 0.cm]{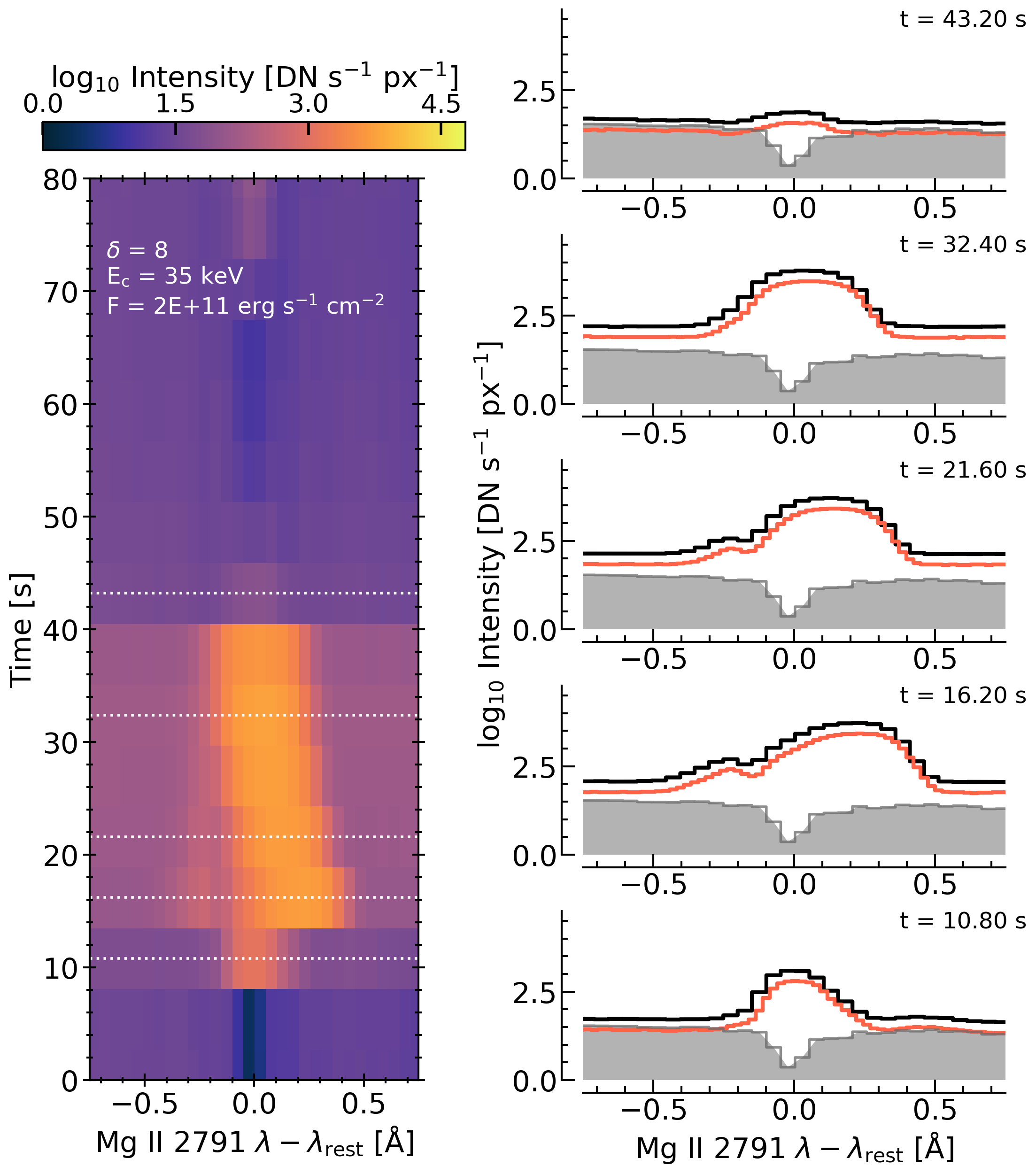}}	
	}
	}
	\caption{\textsl{Same as Figure~\ref{fig:rh_spectra_overview} but showing the \ion{Mg}{ii} k line (left) and \ion{Mg}{ii} 2791~\AA\ line (right). Grey-shaded areas are the pre-flare spectra.}}
	\label{fig:rh_spectra_overview2}
\end{figure*}

Synthetic IRIS spectra are shown in Figures~\ref{fig:rh_spectra_overview} (\ion{O}{i} \& \ion{Fe}{ii}) and \ref{fig:rh_spectra_overview2} (\ion{Mg}{ii}), for both the 5F10 and 2F11 simulations; example \ion{O}{i} and \ion{Mg}{ii} k spectra at the native \rhpar\ resolution, without degradation to IRIS-like observations, are shown in Appendix~\ref{sec:spectranative}. On those figures an image shows the temporal evolution as a function of wavelength, along with several cutouts showing snapshots. For the \ion{Fe}{ii} 2814.445~\AA\ line we show the flare excess spectra, which is a better model-data comparison.  In the observations the flare source appears in the umbra, whereas the modelled pre-flare is a VAL3C-like quiet Sun atmosphere. The latter will have a substantially stronger photospheric contribution to the opacity and emission at the NUV wavelengths near 2814.445~\AA\ such that the background line is in absorption. If we subtract that background then the synthetic spectra appears much closer in shape to the observations, with the line in emission. From those figures we see though the intensities are not markedly different, the behaviour of line widths and asymmetries can be quite different between the two flares.

While the 5F10 simulation does produce a condensation, it is rather weak with relatively small effects on the lines. The \ion{O}{i} line only exhibits a very small red wing asymmetry, which appears after the main heating phase. A red wing asymmetry is more apparent in the \ion{Mg}{ii} lines. From those spectra it is also very clear when the additional turbulent velocity was included or not, with obviously wider \ion{Mg}{ii} profiles between for example 5-15~s. At native resolution the \ion{Mg}{ii} k line contains a central reversal at all times in the simulation, though one that is shallower than in the pre-flare, whereas the \ion{Mg}{ii} 2791~\AA\ line is flat-topped. This is typical of flare simulations unless extremely large electron densities are present \citep[in excess of a few $\times10^{14}$~cm$^{-3}$, see discussions in][]{2017ApJ...842...82R,2019ApJ...879...19Z,2023FrASS...960862K}. However, when degrading to IRIS resolution the central reversal does, at times, seem to disappear or become hard to identify (even more so in the 2F11 simulations).  In the 5F10 simulation, there is a second peak in the condensation speed such that the bulk velocity where \ion{Mg}{ii} forms increases (c.f. $t>30$~s in the left column of Figure~\ref{fig:radyn_flare_atmos}) resulting in the addition of turbulent broadening in the post-heating phase and a wider line profile. At those times the subordinate line is very weak, barely lying above the continuum level.

It is immediately clear from all of the spectral lines, but in particular the narrower lines of \ion{O}{i} and \ion{Fe}{ii}, that the 2F11 simulation drives a stronger condensation, which appears as both Doppler shifts of the line cores as well as prominent red wing asymmetries. At native model resolution these appear in the \ion{O}{i} \& \ion{Fe}{ii} lines initially as a separable components in the red wings, that merge with the stationary component over the duration of the heating phase, occurring more rapidly for the \ion{O}{i} line. This is reminiscent of the red-wing components discussed by \cite{2020ApJ...895....6G}. They noted that the shifted-to-stationary component \ion{Fe}{ii} intensity was too large in their model-data comparison, which may have been due to an overdense condensation in their simulation. Our results are similar when inspecting the full line intensity, but when performing pre-flare subtraction then it is apparent that the core intensity in our simulation increased more than the wing, which does agree well with the general behaviour of our observations. When degrading to IRIS resolution this initial separation is more difficult to identify. There are times when the whole line appeared redshifted, rather than just exhibiting an asymmetry. The observations more typically showed the latter behaviour. Similarly, the \ion{Mg}{ii} lines at times appeared to have fully Doppler shifted line cores, rather than just red-wing asymmetries. This may indicate that the condensation produced in the 2F11 model was too fast or too dense, such that it accrued enough matter to shift the opacity of the NUV lines somewhat higher in altitude to form inside the condensation. 

\subsubsection{Model-data Comparison}
\begin{figure*}
	\vbox{ 
	\hspace{0.2in}
	\hbox{
	\subfloat{\includegraphics[width = 0.55\textwidth, clip = true, trim = 0.cm 0.cm 0.cm 0.cm]{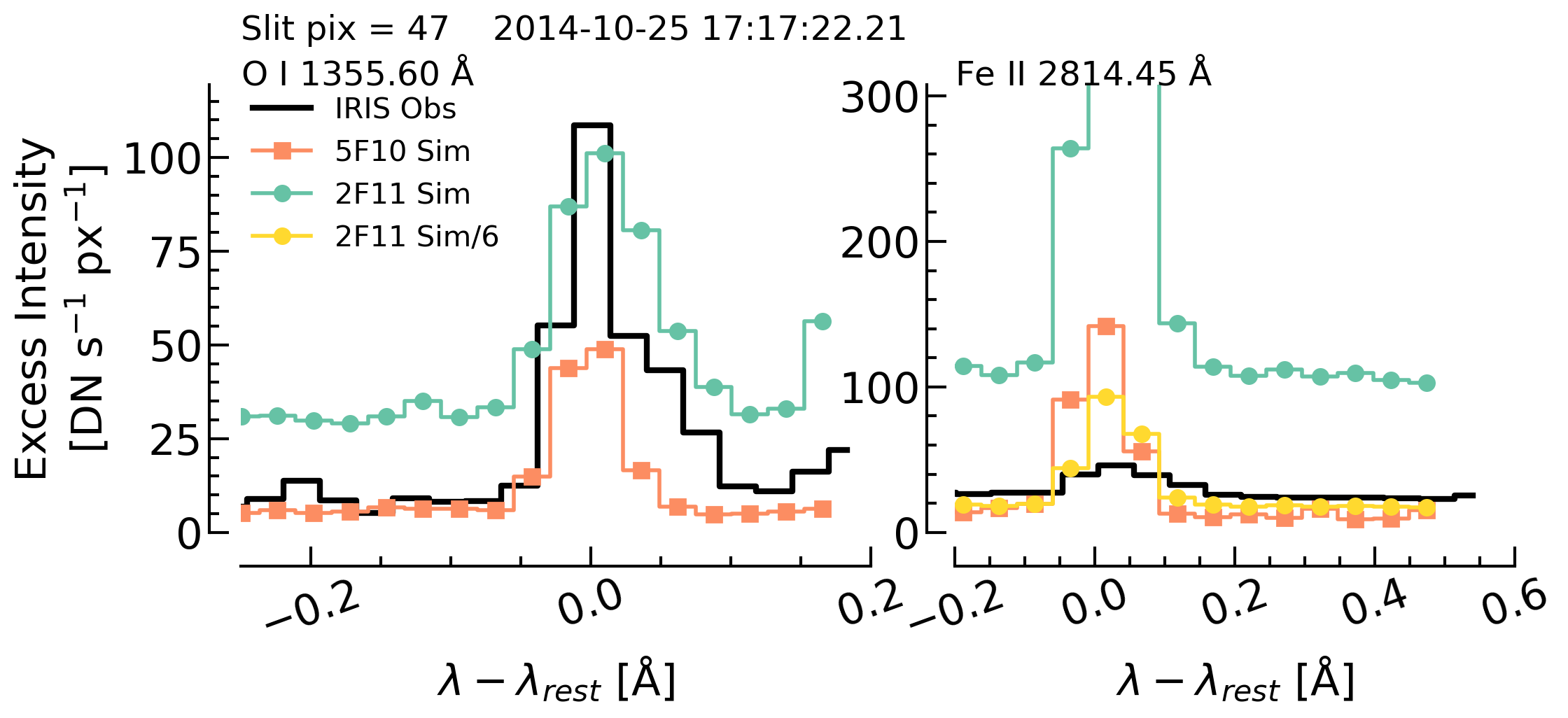}}
	} 
	}
	\vbox{ 
	\hspace{0.2in}
	\hbox{
	\subfloat{\includegraphics[width = 0.55\textwidth, clip = true, trim = 0.cm 0.cm 0.cm 0.cm]{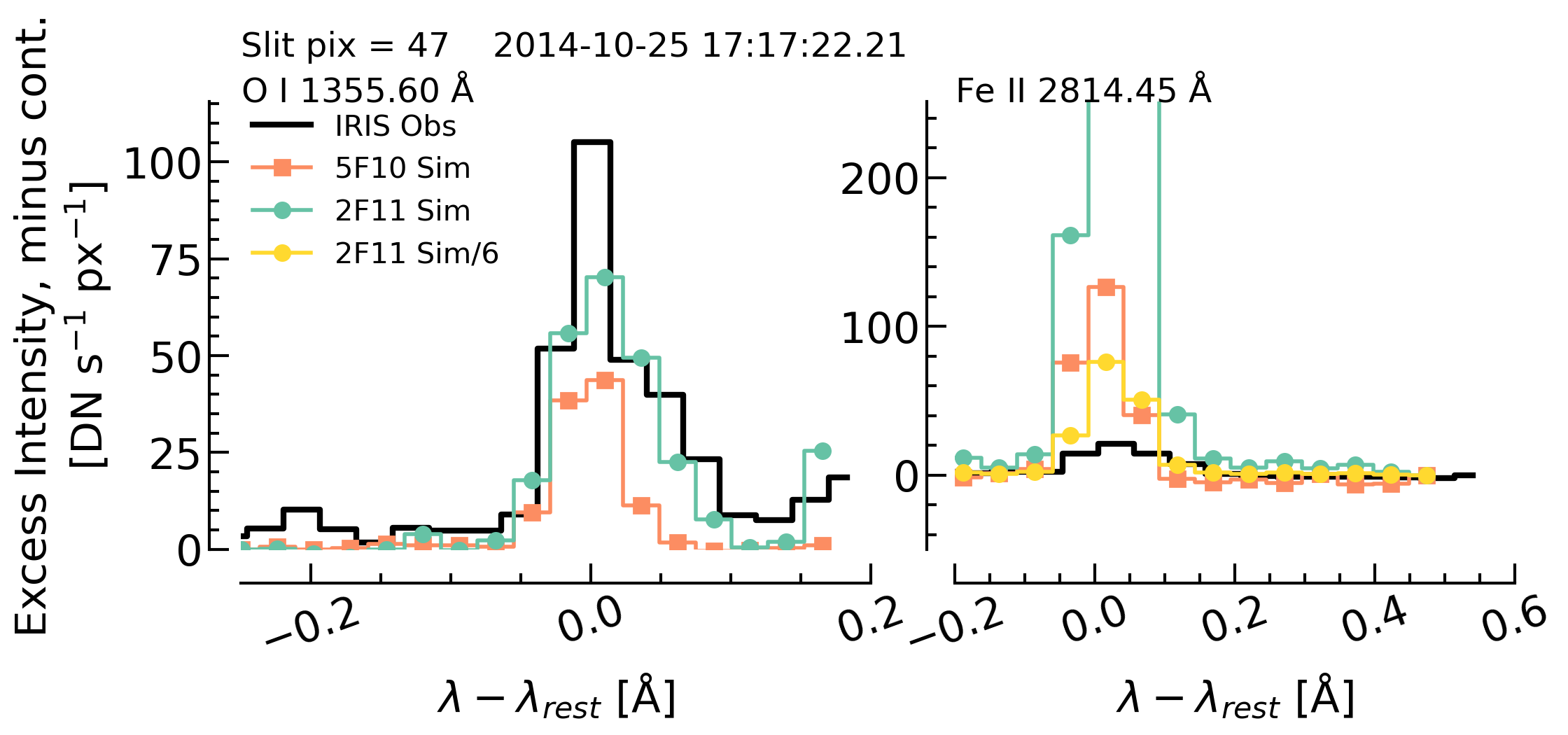}}
	} 
	}
	\vbox{
	\hspace{0.25in}
	\hbox{
	\subfloat{\includegraphics[width = 0.6\textwidth, clip = true, trim = 0.cm 0.cm 0.cm 0.cm]{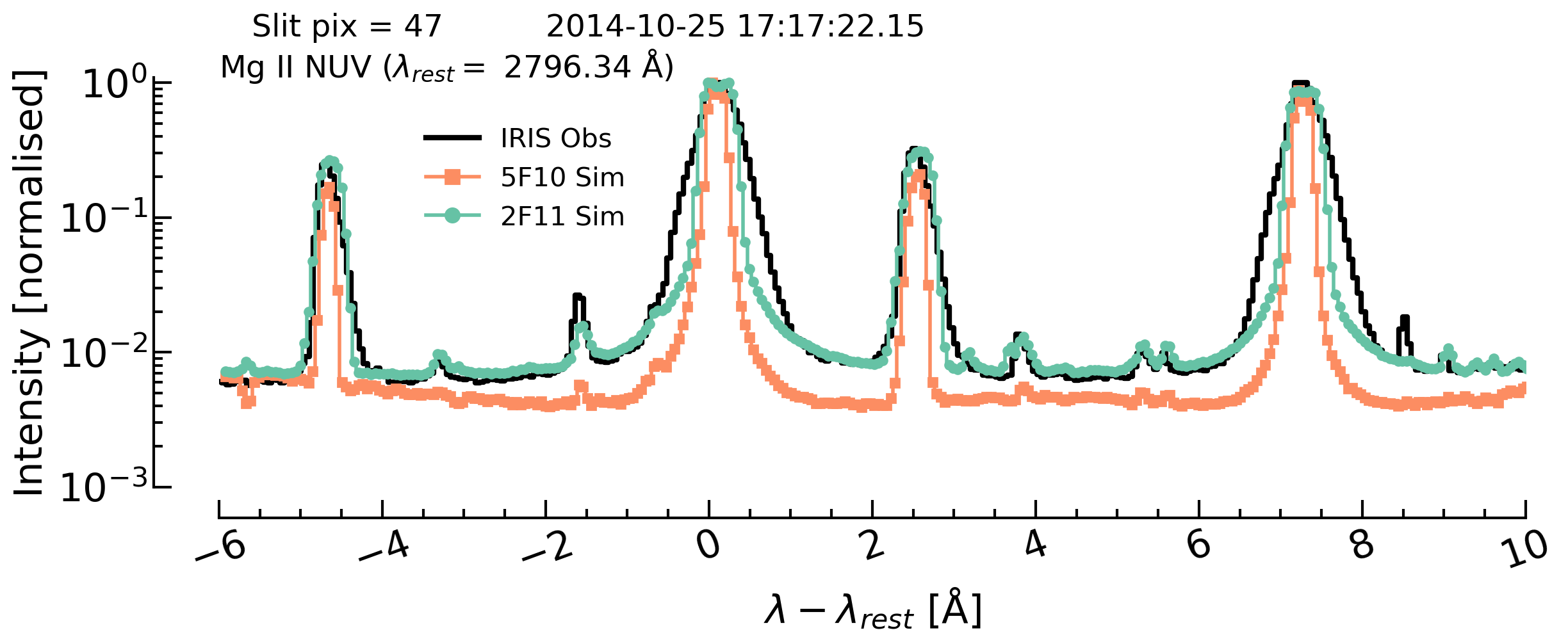}}	
	}
	}
	\vbox{
	\hspace{0.25in} 
	\hbox{
	\subfloat{\includegraphics[width = 0.6\textwidth, clip = true, trim = 0.cm 0.cm 0.cm 0.cm]{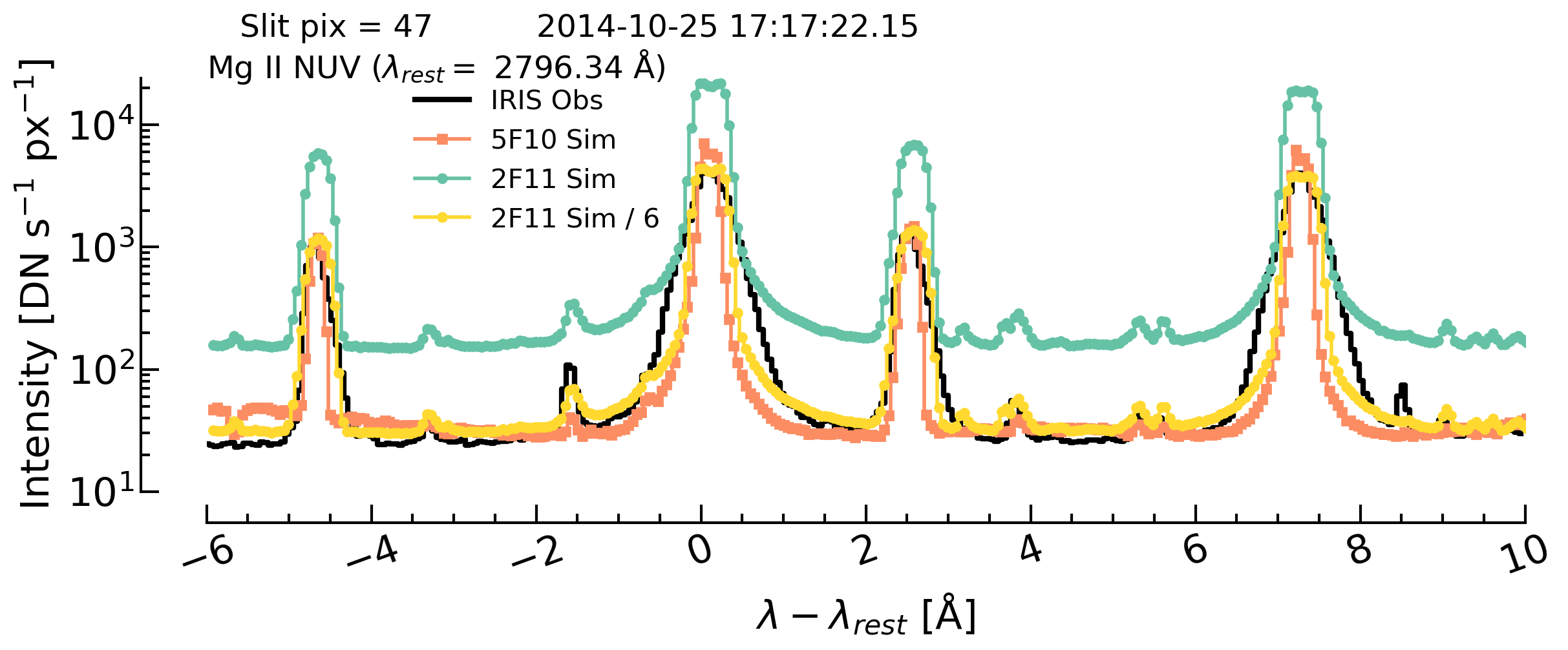}}	
	\subfloat{\includegraphics[width = 0.225\textwidth, clip = true, trim = 0.cm 0.cm 0.cm 0.cm]{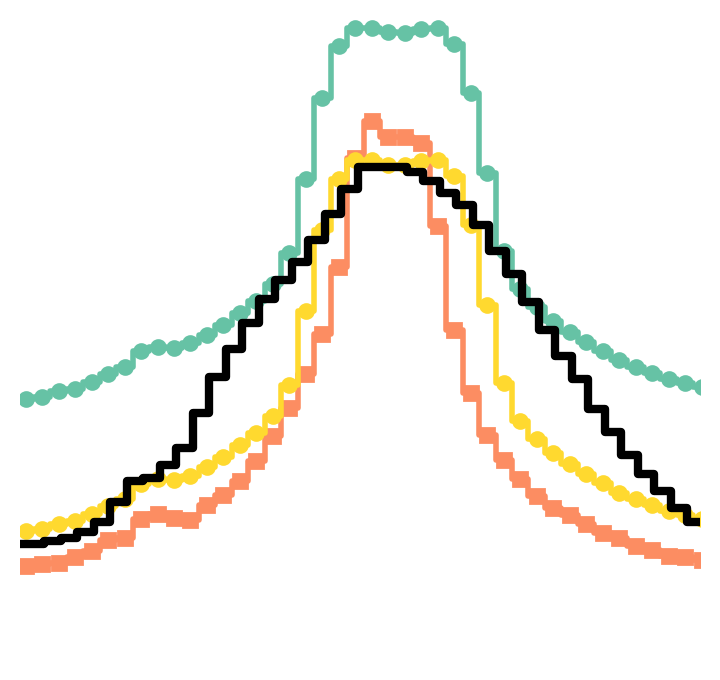}}	
	}
	}
		\caption{\textsl{Model-data comparisons of IRIS spectral lines. In each panel black lines are IRIS observations (pixel position and time are indicated), orange lines are from the 5F10 simulation, and green lines are the 2F11 simulation, both at $t=32.4$~s and both degraded using the IRIS response. Additionally, yellow lines show the NUV spectra (\ion{Mg}{ii} and \ion{Fe}{ii}) from the 2F11 simulation scaled down by a factor of 6. The first row shows the \ion{O}{i}, and \ion{Fe}{ii} lines, where the pre-flare intensity  has been subtracted. The second row shows the same, but also with the continuum level subtracted. The third row shows the \ion{Mg}{ii} NUV spectra, where each spectra has been normalised to the maximum intensity in the wavelength window. The fourth row shows the \ion{Mg}{ii} NUV spectra without any normalisation. Alongside this panel is a zoomed in portion focussing on the \ion{Mg}{ii} k line. The main takeaway points here are that \ion{O}{i} has a rather good model-data comparison. Although the 2F11 simulation is $\approx6\times$ too intense, the \ion{Mg}{ii} subordinate lines and local continuum are otherwise reasonably well captured. The \ion{Mg}{ii} resonance lines are, however, still not broad enough on either side of the line core, and the broadening in the line core is somewhat too large.}}
	\label{fig:model_data_comps}
\end{figure*}
Here we we directly compare the synthetic IRIS spectra to the observed IRIS spectra. Selecting a time near the peak of the flare in both the observations and models we overlay the models and data on Figure~\ref{fig:model_data_comps}, where the black lines on each panel are the observations. Orange lines are the 5F10 simulation, the green are the 2F11, and the yellow are 2F11 where the intensities are divided by a factor of six (this factor was chosen for illustrative purposes only). All of the models include spectral summing.

The top two rows of Figure~\ref{fig:model_data_comps} compare the \ion{O}{i} 1355.598~\AA\ and \ion{Fe}{ii} 2814.445~\AA\ lines. The top row is the excess intensity, that is with pre-flare subtracted, to better contrast the \ion{Fe}{ii} synthetic data with the observations. While the 5F10 simulation produces a continuum intensity comparable to the data for both the FUV and the NUV, it underestimates the \ion{O}{i} line core and width, which also does not exhibit a prominent asymmetry, and overestimates the core intensity of the \ion{Fe}{ii} line. Our larger flare simulation, 2F11, exaggerates the continuum intensity significantly, by around a factor of six in the NUV. This simulation does, however, produce broader lines with asymmetries. If we also subtract the continuum level then we obtain the second row of Figure~\ref{fig:model_data_comps}, which illustrates that the 2F11 \ion{O}{i} line is actually a good match, with a slight underestimate of the core intensity. The \ion{Fe}{ii} line core remains stubbornly too intense, even after dividing by a factor of 6.

Turning to the \ion{Mg}{ii} NUV spectra, we see a similar story. The fourth row of Figure~\ref{fig:model_data_comps} is a direct model-data comparison of the intensities, that demonstrates several key takeaway points. 

First, the 5F10 simulation, while matching the line core intensities reasonably well, does not capture the intensity of the line wings, which are too narrow. This is true both in the mid wings, but also in the far wings, such that the merging with nearby quasi-continuum and other wings is not consistent. 

Second, the continuum intensity, and consequently the line core intensities, in the 2F11 simulation is exaggerated. Similar to the \ion{Fe}{ii} line, if we divide by a factor of 6 we actually capture the continuum rather well, in a rough sense. In that case the merging of the resonance line wings with the nearby quasi-continuum and lines looks good and the line core intensity is also remarkably consistent with the data. If we normalise the synthetic \ion{Mg}{ii} spectra by the maximum value (third row of Figure~\ref{fig:model_data_comps}) then the fact that the general behaviours (aside from line widths) are captured in the 2F11 simulation, but not in the 5F10 simulation, is clear. That is, the core-to-continuum ratio is well-captured.

Third, even with this dilution by an ad-hoc factor of six to match the observed intensities, the \ion{Mg}{ii} h \& k line wings are too narrow, with a non-negligible underestimate. A zoomed-in view of the \ion{Mg}{ii} k line is included to demonstrate this clearly. We remind the reader that these synthetic spectra include what we consider to be the upper limit for microturbulence, based on (a) the observations of \ion{O}{i} 1355.598~\AA, (b) that modelling suggests the formation temperature of \ion{O}{i} 1355.598~\AA\ may be higher than we assumed in the observational section (i.e. the nonthermal width would decrease somewhat), and (c) that since the cores of the \ion{O}{i}, \ion{Cl}{i}, and \ion{Fe}{ii} lines do not exhibit notable broadening, the microturbulence does not suddenly increase with depth to values required by prior studies to explain the wing widths \citep[that is, up to  30-50~km~s$^{-1}$;][]{2019ApJ...879...19Z,2017ApJ...842...82R,2019ApJ...878L..15H}. Additionally, as noted by \cite{2019ApJ...879...19Z}, including such large values at depths below the resonance line core formation heights also results in the nearby continuum and subordinate lines taking on the wrong shape relative to observations, whereas we capture those aspects well.

Fourth, the value of microturbulence included within the formation region of \ion{Mg}{ii} also seems to be at the limit of what is permitted to avoid exaggerating the line core width. Re-running the 2F11 simulation with condensation $V_\mathrm{turb} = [7.5, 14]$~km~s$^{-1}$ demonstrates that smaller values of $V_\mathrm{turb}$ are more consistent with the core of the line, but that larger $V_\mathrm{turb}$ results in a core that is too wide. Perhaps the microturbulence is located primarily along the leading edge of the condensation, such that it decreases with altitude towards the  \ion{Mg}{ii} core formation height which is slightly higher than the \ion{O}{i} 1355.598~\AA\ line, though it must then increase again as relatively large values of $w_{\mathrm{Nthm}} > 30$~km~s$^{-1}$ are often measured for the \ion{Si}{iv} resonance lines \citep[e.g.][]{Jeffreyeaav2794,2022ApJ...936...85S}, and even larger values for hotter transition region and coronal lines are typical \citep[e.g. those analysed from a flare footpoint by][using \textsl{Hinode}/EIS observations]{2011ApJ...740...70M,2022ApJ...936...85S}.

\begin{figure}
	\centering 
	\vbox{
	\hbox{
	\subfloat{\includegraphics[width = 0.375\textwidth, clip = true, trim = 0.cm 0.cm 0.cm 0.cm]{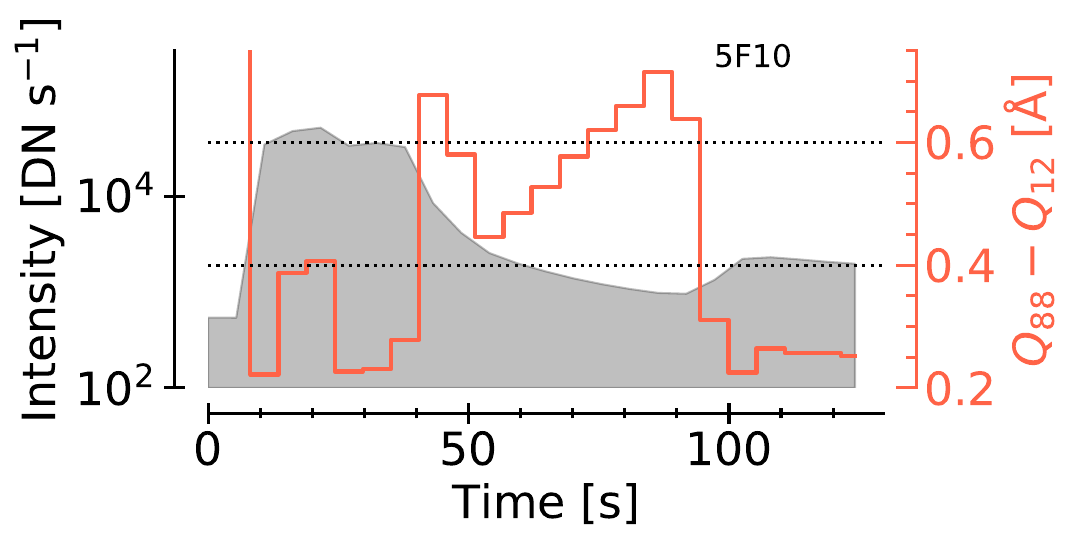}}
	}
	}
	\vbox{
	\hbox{
	\subfloat{\includegraphics[width = 0.375\textwidth, clip = true, trim = 0.cm 0.cm 0.cm 0.cm]{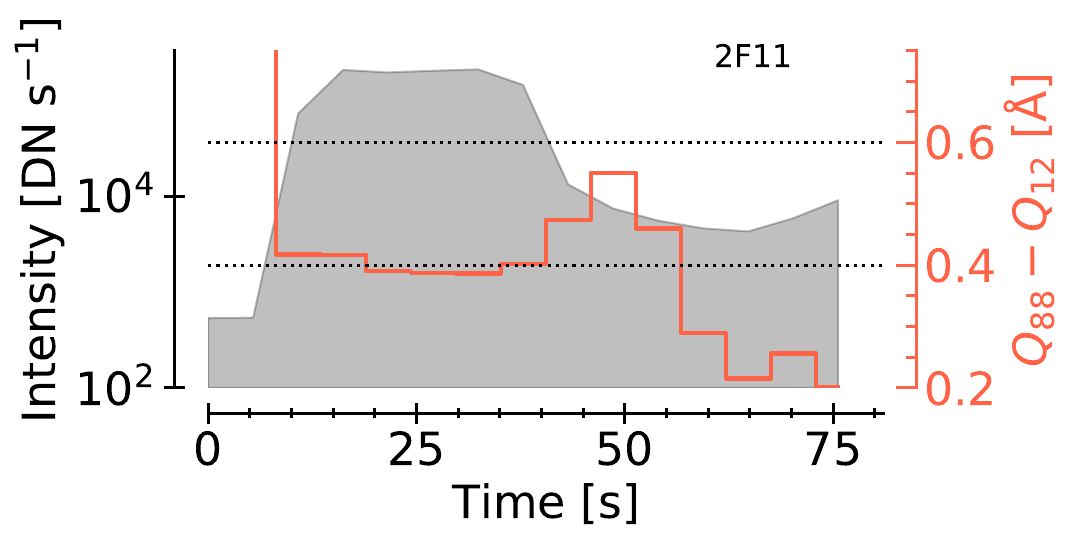}}
	}
	}
	\caption{\textsl{Evolution of the widths (red lines) and integrated intensities (grey shaded areas) of the \ion{Mg}{ii} k line in the 5F10 simulation (top panel) and 2F11 (bottom panel) simulations. Both were measured using the synthetic IRIS data, with spectral summing included. The widths were measured using the quantile method. On each panel the dotted horizontal lines are the typical range of observed values of \ion{Mg}{ii} k line width during the peak of the flare (compare to Figure~\ref{fig:iris_widths_sg_overview2}).}}
	\label{fig:model_data_comps_widths}
\end{figure}
Using the same definition of \ion{Mg}{ii} k line width as we used for the IRIS observations (that is, the quantile-width, $W_{Q}$, c.f. Figure~\ref{fig:iris_widths_sg_overview2}), the under-prediction of the width is quantified in Figure~\ref{fig:model_data_comps_widths} as a function of time. On that figure the dotted horizontal lines indicate the range of typical values calculated in the observations at flare peak. Note that the pre-flare profiles exhibit a core-to-wing intensity that is too low compared to observations of the quiet Sun, so the line width is artificially large in the pre-flare period. At times when the lightcurve of intensity is peaking the width of the line is  lower than the range of the observations. It is only after the intensity starts to decrease that the line width approaches the observed largest widths. Over the peak intensity, the line width in the observations is at least $W_{Q} \sim 0.5-0.6$~\AA, but only around $0.4$~\AA\ in the models. An interesting feature here is that when the line begins initially to decrease in intensity, the width increases for a short while, which is what we see in the observations also. Inspecting the line formation properties suggests this is due to rapid cooling of the upper chromosphere where the line core forms, reducing the gradient between between core-to-line source functions. However, we are missing some amount of width during the peak of the intensity, which presents a problem for our models. 

It is worthwhile to note that the \ion{Mg}{ii} h \& k line widths in this flare are not even amongst the broadest examples observed in solar flares, meaning that the discrepancy between models and data are even more stark for other events. For example see the \ion{Mg}{ii} spectra observed in the X-class 2014-March-29th flare \citep[e.g.][]{2015SoPh..290.3525L,2017ApJ...842...82R,2019ApJ...879...19Z}, and the comprehensive characterisation of \ion{Mg}{ii} profiles by \cite{2018ApJ...861...62P}.

\subsection{Discussion}
Smearing of profiles by unresolved large ($>150-200$~km~s$^{-1}$) macroscopic flows, or including rather large values of microturbulence throughout the chromosphere have been suggested as a resolution to the question of the broadening of \ion{Mg}{ii} resonance lines during solar flares. Neither is compatible with our observations or modelling. Our flare models are clearly missing certain ingredients. We briefly speculate as what those may be, which are not mutually exclusive nor exhaustive.

One missing ingredient may be additional heating at great depths. Temperature gradients between the core-to-wing formation heights may be too steep in current modelling, such that the core is too intense relative to the wings. Easing that gradient, either by increased density to raise the wing formation height closer to the line core formation height, or by additional heating of the lowest parts of the atmosphere that is not currently taking place in our electron-beam models. Heating the chromosphere to a greater depth would enhance the line wing's source function at those depths (the core forms in the upper chromosphere), naturally increasing the emergent intensity (in the Eddington-Barbier approximation, the emergent intensity at some wavelength is reflective of the source function at the height at which $\tau_\lambda = 1$). Broadening due to the stratification of the line source function is often referred to as `opacity broadening' \citep[see e.g.][]{2015ApJ...811...80R}, and increased opacity broadening due to mass-loading of the quiet chromosphere has recently been explored as an explanation for the width of quiescent \ion{Mg}{ii} profiles in MHD models \citep[][]{2023ApJ...944..131H}. Further evidence that heating to greater depth is required in flares includes the outstanding problem of where and how the NUV and optical continua are formed (i.e. the question of white light flares), and the requirement to heat the temperature minimum region or upper photosphere, inferred from spectral inversions \citep[e.g.][]{1990ApJ...350..463M,1990ApJ...365..391M} and spectropolarimetry \citep[e.g.][]{2018A&A...620A.183J}. Certain observations point to a mid-upper chromospheric origin of the optical continuum due to the presence of a Balmer jump, whereas other sources are consistent with heating of the upper photosphere or temperature minimum region. Indeed, some semi-empirical modelling suggests that to explain both the observed NUV and optical data we need heating of the full extent of the photosphere through transition region \citep[e.g.][]{2016ApJ...816...88K}. For more discussion see the introduction to \cite{2014ApJ...783...98K} and Section 3 of \cite{2023FrASS...960862K}. Investigation into the continuum-to-line ratio for \ion{Fe}{ii} \citep[e.g.][]{2019ApJ...878..135K} in this flare could help guide the models, which currently predict ratios too small compared to the observations. 

Another potential missing ingredient is broadening due to interactions between ambient \ion{Mg}{ii} ions and the nonthermal particles within the electron or proton beams. \radyn\ includes nonthermal collisions for H and He, but we do not model the effects on \ion{Mg}{ii}. \cite{2007PASP..119...67H} speculated that the discrepancy between their stellar flare observations and models of \ion{Mg}{ii} resonance lines may be in some part due to the addition of nonthermal collisions to the total collisional rate, since the PRD implementation uses the depopulation rate of the h \& k lines when calculating the coherency rates in the PRD solution. Enhancing the collisional rates may result in increased redistribution of core photons to the wings. As well as nonthermal collisions, the energetic electrons in the beam could produce broadening due to turbulent electric fields from beam-induced plasma waves, that result in an additional Stark-type effect \citep[][]{1988Ap.....29..455A,1999AIPC..467...14G,1990JQSRT..44..171T}, or during charge-exchange between nonthermal particles and ambient plasma \citep[][]{2022JPhB...55c4002G}. The effect of these latter processes on \ion{Mg}{ii} is unknown, but certainly present interesting avenues to explore. 
 

\section{Summary \& Conclusions}\label{sec:conc}

We have performed an observational study of optically thin emission, primarily \ion{O}{i} 1355.598~\AA\ emission during a solar flare, which subsequently informed numerical modelling in an attempt to determine if increased microturbulence within a chromospheric condensation could satisfactorily broaden the \ion{Mg}{ii} lines.

Red-wing asymmetries were identified in lines of \ion{O}{i} 1355.598~\AA\ (optically thin), \ion{Cl}{i} 1351.66~\AA\ (potentially optically thin inside the condensation), and \ion{Fe}{ii} 2814.445~\AA\ (potentially optically thin inside the condensation), as well as in numerous other strong IRIS lines. The metrics obtained from double-component Gaussian fits to each line were similar for all three lines, lending credence to the assumption that the shifted components of \ion{Fe}{ii} and \ion{Cl}{i} could be treated as mostly optically thin.

Assuming that these nonthermal widths were caused by microturbulence, we inserted a value of $V_{\mathrm{turb}} = 10$~km~s$^{-1}$ at heights corresponding to a chromospheric condensation in a radiation hydrodynamics flare simulation. The flare atmospheres were post-processed to synthesise the \ion{Mg}{ii} NUV spectra, the \ion{O}{i} 1355.598~\AA\ line, and the \ion{Fe}{ii} 2814.445~\AA\ line. Despite increasing the \ion{Mg}{ii} line width, this value of microturbulence was insufficient to broaden the \ion{Mg}{ii} spectra enough to match the observations. Some key conclusions, and summary points are:
 
\begin{itemize}
\item The \ion{O}{i}, \ion{Cl}{i}, and \ion{Fe}{ii} lines exhibited line intensities that had a two-stage decay. The rise, peak, and initial decay of intensity was co-temporal with the presence of a red wing asymmetry lasting 30-90~s, followed by a more gradual decay that took a further several minutes to approach background levels.
\item Fitting two-component Gaussians functions to these lines found a stationary component with slightly redshifted components. Those redshifted components had nonthermal widths $w_{\mathrm{Nthm}}\sim5-12$~km~s$^{-1}$, representing a broadening relative to the background values. Based on modelling results, this may indeed be an upper limit, since the formation temperature could be higher than the assumed $T = 10$~kK. These values are consistent with recent models of MHD turbulence in flares \citep{2023ApJ...947...67R}. The stationary component had small blue or redshifts around the rest wavelength, without strong increases in line width. 
\item In our data-guided modelling, \ion{O}{i} 1355.598~\AA\ was confirmed to be optically thin throughout the flare, with the red wing forming alongside the \ion{Mg}{ii} lines and the line core (stationary component in the observations) formed somewhat deeper but still close in height.
\item The \ion{Fe}{ii} line stationary component formed over a large extent of the chromosphere, with opacity playing a smaller role during the flare ($\tau_{\lambda}\sim0.05-0.1$). The red wing formed within the condensation, nearby the \ion{Mg}{ii} spectra, and was mostly optically thin. Thus, both \ion{O}{i} 1355.598~\AA\ and \ion{Fe}{ii} 2814.445~\AA\ are good measures of both the development of the condensation, and of the magnitude of microturbulence near the \ion{Mg}{ii} lines. 
\item Apart from the elevated continuum intensity, the synthetic \ion{O}{i} 1355.598~\AA\ compared well with observations, with similar widths, red-wing asymmetries and line intensities. 
\item The synthetic \ion{Mg}{ii} NUV spectra was around a factor 6 too intense in the stronger flare, but captured the relative shape and intensity of the resonance lines, subordinate lines, and nearby continua rather well. The weaker simulation was not as consistent with regard to the nearby continuum between the resonance and subordinate lines, and the subordinate lines were too weak relative to the resonance lines.
\item The \ion{Mg}{ii} resonances lines were still too narrow, suggesting that the answer to the question of why the observed lines are so broad is not solely due to microturbulence. Increasing microturbulence with depth below the core formation height would not be consistent with the \ion{O}{i} observations, which exhibit narrow stationary components. 
\item Using a range of values of $V_\mathrm{turb} = [7.5, 10, 14]$~km~s$^{-1}$ in the condensation indicates that 7.5~km~s$^{-1}$ or  10~km~s$^{-1}$ captures the width of the core adequately, whereas 14~km~s$^{-1}$ produces a core that is too wide.
\item We speculate two avenues for future efforts to rectify the model-data discrepancy of broad profiles: (1) the \ion{Mg}{ii} lines may be broadened by decreasing the gradient in source function between the line and core, for example by increasing the temperature in the lower chromosphere via additional flare energy deposition at those depths (i.e. opacity broadening), or (2) that interactions between the particle beam itself and ambient \ion{Mg}{ii} may act to broaden the lines nonthermally, but not via the typical Doppler microturbulence. 
\end{itemize}
 
Spectral inversions of chromospheric lines offers another route to estimating an atmosphere that may represent the conditions that produced an observed spectral line \citep[e.g. see][and references therein]{2019A&A...623A..74D}. \cite{2021A&A...649A.106Y} inverted ground based \ion{Ca}{ii} 8542~\AA\ and \ion{Ca}{ii} K line flare observations, finding values of microturbulence ranging a few km~s$^{-1}$ in the lower atmosphere and up to $\sim5-10$~km~s$^{-1}$ in the upper chromosphere ($\log\tau_{500} \sim 10^{-4}$), consistent with our results.  Recent efforts  \citep{2022arXiv221105459S} used spectral inversions to explore the origin of two classes of broad \ion{Mg}{ii} line profiles observed in the 2014-March-29th X class flare: both had Lorenztian wings, one that appeared rather pointed without reversed core, and one with a central reversal and red-peak asymmetry.  \cite{2022arXiv221105459S} produced, via those inversions, model atmospheres that had $V_{\mathrm{turb}} \sim 5-15$~km~s$^{-1}$ in the chromosphere. Whilst these values are consistent with our findings based on \ion{O}{i} 1355.598~\AA, the atmospheres on first glance appear quite different that those produced by electron beam driven radiation hydrodynamic simulations. \cite{2022arXiv221105459S} also state that it is the presence of a strong, divergent velocity field in the upper chromosphere of their inverted atmospheres that produces the `pointy', very broad \ion{Mg}{ii} profiles \citep[though not as extreme as that required by][]{2017ApJ...842...82R}. However, it is not clear then why RHD models that calculate velocity gradients and mass advection self-consistently,  and which often contain strong upflow and downflow gradients, do not similarly reproduce these profiles. We look forward to the opportunity to comprehensively compare our RHD model atmospheres with the results of spectral inversions, including to explore the role that the corrugated pre-flare atmosphere may play compared to our idealised pre-flare conditions. 


\section*{Acknowledgements}
GSK and AFK acknowledge the financial support from a NASA Early Career Investigator Program award (Grant\# 80NSSC21K0460).  JCA acknowledges funding through NASA's IRIS mission and the Heliophysics Innovation Fund. MRK acknowledges the NASA OSTEM Internship Program. This manuscript benefited from discussions held at a meeting of International Space Science Institute team: ``Interrogating Field-Aligned Solar Flare Models: Comparing, Contrasting and Improving,'' led by Dr. G. S. Kerr and Dr. V. Polito. IRIS is a NASA small explorer mission developed and operated by LMSAL with mission operations executed at NASA Ames Research Center and major contributions to downlink communications funded by the Norwegian Space Center (NSC, Norway) through an ESA PRODEX contract.

\section*{Data Availability}
Science-ready (level-2) data from the Interface Region Imaging Spectrograph (IRIS) is publicly available: \url{https://iris.lmsal.com/data.html}. The specific dataset used in this study is available here:  \url{https://www.lmsal.com/hek/hcr?cmd=view-event&event-id=ivo%3A%2F%2Fsot.lmsal.com%2FVOEvent%23VOEvent_IRIS_20141025_145828_3880106953_2014-10-25T14%3A58%3A282014-10-25T14%3A58%3A28.xml}.

\fpc\ is an open source code available for download: \url{https://github.com/solarFP/FP}.

\radyn\ code output is available upon reasonable request to the corresponding author. 

\rh\ is an open source code available for download: \url{https://github.com/ITA-Solar/rh}.

\bibliographystyle{mnras}
\bibliography{Kerr_2023_OI_Flares} 

\begin{thebibliography}{}
\makeatletter
\relax
\def\mn@urlcharsother{\let\do\@makeother \do\$\do\&\do\#\do\^\do\_\do\%\do\~}
\def\mn@doi{\begingroup\mn@urlcharsother \@ifnextchar [ {\mn@doi@}
  {\mn@doi@[]}}
\def\mn@doi@[#1]#2{\def\@tempa{#1}\ifx\@tempa\@empty \href
  {http://dx.doi.org/#2} {doi:#2}\else \href {http://dx.doi.org/#2} {#1}\fi
  \endgroup}
\def\mn@eprint#1#2{\mn@eprint@#1:#2::\@nil}
\def\mn@eprint@arXiv#1{\href {http://arxiv.org/abs/#1} {{\tt arXiv:#1}}}
\def\mn@eprint@dblp#1{\href {http://dblp.uni-trier.de/rec/bibtex/#1.xml}
  {dblp:#1}}
\def\mn@eprint@#1:#2:#3:#4\@nil{\def\@tempa {#1}\def\@tempb {#2}\def\@tempc
  {#3}\ifx \@tempc \@empty \let \@tempc \@tempb \let \@tempb \@tempa \fi \ifx
  \@tempb \@empty \def\@tempb {arXiv}\fi \@ifundefined
  {mn@eprint@\@tempb}{\@tempb:\@tempc}{\expandafter \expandafter \csname
  mn@eprint@\@tempb\endcsname \expandafter{\@tempc}}}

\bibitem[\protect\citeauthoryear{{Abbett} \& {Hawley}}{{Abbett} \&
  {Hawley}}{1999}]{1999ApJ...521..906A}
{Abbett} W.~P.,  {Hawley} S.~L.,  1999, \mn@doi [\apj] {10.1086/307576}, \href
  {http://adsabs.harvard.edu/abs/1999ApJ...521..906A} {521, 906}

\bibitem[\protect\citeauthoryear{{Airapetyan} \& {Nikogosyan}}{{Airapetyan} \&
  {Nikogosyan}}{1988}]{1988Ap.....29..455A}
{Airapetyan} V.~S.,  {Nikogosyan} A.~G.,  1988, \mn@doi [Astrophysics]
  {10.1007/BF01005862}, \href
  {https://ui.adsabs.harvard.edu/abs/1988Ap.....29..455A} {29, 455}

\bibitem[\protect\citeauthoryear{{Allred}, {Hawley}, {Abbett}  \&
  {Carlsson}}{{Allred} et~al.}{2005}]{2005ApJ...630..573A}
{Allred} J.~C.,  {Hawley} S.~L.,  {Abbett} W.~P.,   {Carlsson} M.,  2005,
  \mn@doi [\apj] {10.1086/431751}, \href
  {http://adsabs.harvard.edu/abs/2005ApJ...630..573A} {630, 573}

\bibitem[\protect\citeauthoryear{{Allred}, {Kowalski}  \& {Carlsson}}{{Allred}
  et~al.}{2015}]{2015ApJ...809..104A}
{Allred} J.~C.,  {Kowalski} A.~F.,   {Carlsson} M.,  2015, \mn@doi [\apj]
  {10.1088/0004-637X/809/1/104}, \href
  {http://adsabs.harvard.edu/abs/2015ApJ...809..104A} {809, 104}

\bibitem[\protect\citeauthoryear{{Allred}, {Alaoui}, {Kowalski}  \&
  {Kerr}}{{Allred} et~al.}{2020}]{2020ApJ...902...16A}
{Allred} J.~C.,  {Alaoui} M.,  {Kowalski} A.~F.,   {Kerr} G.~S.,  2020, \mn@doi
  [\apj] {10.3847/1538-4357/abb239}, \href
  {https://ui.adsabs.harvard.edu/abs/2020ApJ...902...16A} {902, 16}

\bibitem[\protect\citeauthoryear{{Allred}, {Kerr}  \& {Gordon Emslie}}{{Allred}
  et~al.}{2022}]{2022ApJ...931...60A}
{Allred} J.~C.,  {Kerr} G.~S.,   {Gordon Emslie} A.,  2022, \mn@doi [\apj]
  {10.3847/1538-4357/ac69e8}, \href
  {https://ui.adsabs.harvard.edu/abs/2022ApJ...931...60A} {931, 60}

\bibitem[\protect\citeauthoryear{{Antiochos} \& {Sturrock}}{{Antiochos} \&
  {Sturrock}}{1978}]{Antiochos1978}
{Antiochos} S.~K.,  {Sturrock} P.~A.,  1978, \mn@doi [\apj] {10.1086/155999},
  \href {https://ui.adsabs.harvard.edu/abs/1978ApJ...220.1137A} {220, 1137}

\bibitem[\protect\citeauthoryear{{Ashfield}, {Longcope}, {Zhu}  \&
  {Qiu}}{{Ashfield} et~al.}{2022}]{2022ApJ...926..164A}
{Ashfield} William~H. I.,  {Longcope} D.~W.,  {Zhu} C.,   {Qiu} J.,  2022,
  \mn@doi [\apj] {10.3847/1538-4357/ac402d}, \href
  {https://ui.adsabs.harvard.edu/abs/2022ApJ...926..164A} {926, 164}

\bibitem[\protect\citeauthoryear{{Bamba}, {Inoue}, {Kusano}  \&
  {Shiota}}{{Bamba} et~al.}{2017}]{2017ApJ...838..134B}
{Bamba} Y.,  {Inoue} S.,  {Kusano} K.,   {Shiota} D.,  2017, \mn@doi [\apj]
  {10.3847/1538-4357/aa6682}, \href
  {https://ui.adsabs.harvard.edu/abs/2017ApJ...838..134B} {838, 134}

\bibitem[\protect\citeauthoryear{{Brown}}{{Brown}}{1971}]{1971SoPh...18..489B}
{Brown} J.~C.,  1971, \mn@doi [\solphys] {10.1007/BF00149070}, \href
  {http://adsabs.harvard.edu/abs/1971SoPh...18..489B} {18, 489}

\bibitem[\protect\citeauthoryear{{Carlsson}}{{Carlsson}}{1998}]{1998LNP...507..163C}
{Carlsson} M.,  1998, in {Vial} J.~C.,  {Bocchialini} K.,   {Boumier} P.,  eds,
   Lecture Notes in Physics, Berlin Springer Verlag Vol. 507, Space Solar
  Physics: Theoretical and Observational Issues in the Context of the SOHO
  Mission. p.~163, \mn@doi{10.1007/BFb0106920}

\bibitem[\protect\citeauthoryear{{Carlsson} \& {Stein}}{{Carlsson} \&
  {Stein}}{1995}]{1995ApJ...440L..29C}
{Carlsson} M.,  {Stein} R.~F.,  1995, \mn@doi [\apjl] {10.1086/187753}, \href
  {http://adsabs.harvard.edu/abs/1995ApJ...440L..29C} {440, L29}

\bibitem[\protect\citeauthoryear{{Carlsson} \& {Stein}}{{Carlsson} \&
  {Stein}}{1997}]{1997ApJ...481..500C}
{Carlsson} M.,  {Stein} R.~F.,  1997, \apjl, \href
  {http://adsabs.harvard.edu/abs/1997ApJ...481..500C} {481, 500}

\bibitem[\protect\citeauthoryear{{Carlsson}, {Leenaarts}  \& {De
  Pontieu}}{{Carlsson} et~al.}{2015}]{2015ApJ...809L..30C}
{Carlsson} M.,  {Leenaarts} J.,   {De Pontieu} B.,  2015, \mn@doi [\apjl]
  {10.1088/2041-8205/809/2/L30}, \href
  {http://adsabs.harvard.edu/abs/2015ApJ...809L..30C} {809, L30}

\bibitem[\protect\citeauthoryear{{Carlsson} et~al.,}{{Carlsson}
  et~al.}{2023}]{2023arXiv230402618C}
{Carlsson} M.,  et~al., 2023, \mn@doi [arXiv e-prints]
  {10.48550/arXiv.2304.02618}, \href
  {https://ui.adsabs.harvard.edu/abs/2023arXiv230402618C} {p. arXiv:2304.02618}

\bibitem[\protect\citeauthoryear{{Cheng}, {Feldman}  \& {Doschek}}{{Cheng}
  et~al.}{1980}]{1980A&A....86..377C}
{Cheng} C.~C.,  {Feldman} U.,   {Doschek} G.~A.,  1980, \aap, \href
  {https://ui.adsabs.harvard.edu/abs/1980A&A....86..377C} {86, 377}

\bibitem[\protect\citeauthoryear{{Cheng}, {Oran}, {Doschek}, {Boris}  \&
  {Mariska}}{{Cheng} et~al.}{1983}]{1983ApJ...265.1090C}
{Cheng} C.~C.,  {Oran} E.~S.,  {Doschek} G.~A.,  {Boris} J.~P.,   {Mariska}
  J.~T.,  1983, \mn@doi [\apj] {10.1086/160751}, \href
  {https://ui.adsabs.harvard.edu/abs/1983ApJ...265.1090C} {265, 1090}

\bibitem[\protect\citeauthoryear{{De Pontieu} et~al.,}{{De Pontieu}
  et~al.}{2014}]{2014SoPh..289.2733D}
{De Pontieu} B.,  et~al., 2014, \mn@doi [\solphys] {10.1007/s11207-014-0485-y},
  \href {http://adsabs.harvard.edu/abs/2014SoPh..289.2733D} {289, 2733}

\bibitem[\protect\citeauthoryear{{De Pontieu} et~al.,}{{De Pontieu}
  et~al.}{2021}]{2021SoPh..296...84D}
{De Pontieu} B.,  et~al., 2021, \mn@doi [\solphys]
  {10.1007/s11207-021-01826-0}, \href
  {https://ui.adsabs.harvard.edu/abs/2021SoPh..296...84D} {296, 84}

\bibitem[\protect\citeauthoryear{{Del Zanna} \& {Mason}}{{Del Zanna} \&
  {Mason}}{2018}]{2018LRSP...15....5D}
{Del Zanna} G.,  {Mason} H.~E.,  2018, \mn@doi [Living Reviews in Solar
  Physics] {10.1007/s41116-018-0015-3}, \href
  {https://ui.adsabs.harvard.edu/abs/2018LRSP...15....5D} {15, 5}

\bibitem[\protect\citeauthoryear{{Emslie} \& {Sturrock}}{{Emslie} \&
  {Sturrock}}{1982}]{1982SoPh...80...99E}
{Emslie} A.~G.,  {Sturrock} P.~A.,  1982, \mn@doi [\solphys]
  {10.1007/BF00153426}, \href
  {https://ui.adsabs.harvard.edu/abs/1982SoPh...80...99E} {80, 99}

\bibitem[\protect\citeauthoryear{{Emslie} et~al.,}{{Emslie}
  et~al.}{2012}]{2012ApJ...759...71E}
{Emslie} A.~G.,  et~al., 2012, \mn@doi [ApJ] {10.1088/0004-637X/759/1/71},
  \href {http://adsabs.harvard.edu/abs/2012ApJ...759...71E} {759, 71}

\bibitem[\protect\citeauthoryear{{Fisher}, {Canfield}  \& {McClymont}}{{Fisher}
  et~al.}{1985a}]{1985ApJ...289..414F}
{Fisher} G.~H.,  {Canfield} R.~C.,   {McClymont} A.~N.,  1985a, \mn@doi [\apj]
  {10.1086/162901}, \href {http://adsabs.harvard.edu/abs/1985ApJ...289..414F}
  {289, 414}

\bibitem[\protect\citeauthoryear{{Fisher}, {Canfield}  \& {McClymont}}{{Fisher}
  et~al.}{1985b}]{1985ApJ...289..425F}
{Fisher} G.~H.,  {Canfield} R.~C.,   {McClymont} A.~N.,  1985b, \mn@doi [\apj]
  {10.1086/162902}, \href {http://adsabs.harvard.edu/abs/1985ApJ...289..425F}
  {289, 425}

\bibitem[\protect\citeauthoryear{{Fisher}, {Canfield}  \& {McClymont}}{{Fisher}
  et~al.}{1985c}]{1985ApJ...289..434F}
{Fisher} G.~H.,  {Canfield} R.~C.,   {McClymont} A.~N.,  1985c, \mn@doi [\apj]
  {10.1086/162903}, \href {http://adsabs.harvard.edu/abs/1985ApJ...289..434F}
  {289, 434}

\bibitem[\protect\citeauthoryear{{Fletcher} \& {Hudson}}{{Fletcher} \&
  {Hudson}}{2008}]{2008ApJ...675.1645F}
{Fletcher} L.,  {Hudson} H.~S.,  2008, \mn@doi [\apj] {10.1086/527044}, \href
  {http://adsabs.harvard.edu/abs/2008ApJ...675.1645F} {675, 1645}

\bibitem[\protect\citeauthoryear{{Fletcher} et~al.,}{{Fletcher}
  et~al.}{2011}]{2011SSRv..159...19F}
{Fletcher} L.,  et~al., 2011, \mn@doi [\ssr] {10.1007/s11214-010-9701-8}, \href
  {http://adsabs.harvard.edu/abs/2011SSRv..159...19F} {159, 19}

\bibitem[\protect\citeauthoryear{{Gavrilenko}}{{Gavrilenko}}{1999}]{1999AIPC..467...14G}
{Gavrilenko} V.~P.,  1999, in {Herman} R.~M.,  ed.,  American Institute of
  Physics Conference Series Vol. 467, Spectral Line Shapes. pp 14--26,
  \mn@doi{10.1063/1.58305}

\bibitem[\protect\citeauthoryear{{Gomez}, {Nagayama}, {Cho}, {Kilcrease},
  {Fontes}  \& {Zammit}}{{Gomez} et~al.}{2022}]{2022JPhB...55c4002G}
{Gomez} T.~A.,  {Nagayama} T.,  {Cho} P.~B.,  {Kilcrease} D.~P.,  {Fontes}
  C.~J.,   {Zammit} M.~C.,  2022, \mn@doi [Journal of Physics B Atomic
  Molecular Physics] {10.1088/1361-6455/ac4f31}, \href
  {https://ui.adsabs.harvard.edu/abs/2022JPhB...55c4002G} {55, 034002}

\bibitem[\protect\citeauthoryear{{Graham} \& {Cauzzi}}{{Graham} \&
  {Cauzzi}}{2015}]{2015ApJ...807L..22G}
{Graham} D.~R.,  {Cauzzi} G.,  2015, \mn@doi [\apjl]
  {10.1088/2041-8205/807/2/L22}, \href
  {http://adsabs.harvard.edu/abs/2015ApJ...807L..22G} {807, L22}

\bibitem[\protect\citeauthoryear{{Graham}, {Cauzzi}, {Zangrilli}, {Kowalski},
  {Sim{\~o}es}  \& {Allred}}{{Graham} et~al.}{2020}]{2020ApJ...895....6G}
{Graham} D.~R.,  {Cauzzi} G.,  {Zangrilli} L.,  {Kowalski} A.,  {Sim{\~o}es}
  P.,   {Allred} J.,  2020, \mn@doi [\apj] {10.3847/1538-4357/ab88ad}, \href
  {https://ui.adsabs.harvard.edu/abs/2020ApJ...895....6G} {895, 6}

\bibitem[\protect\citeauthoryear{{Hansteen}, {Martinez-Sykora}, {Carlsson}, {De
  Pontieu}, {Go{\v{s}}i{\'c}}  \& {Bose}}{{Hansteen}
  et~al.}{2023}]{2023ApJ...944..131H}
{Hansteen} V.~H.,  {Martinez-Sykora} J.,  {Carlsson} M.,  {De Pontieu} B.,
  {Go{\v{s}}i{\'c}} M.,   {Bose} S.,  2023, \mn@doi [\apj]
  {10.3847/1538-4357/acb33c}, \href
  {https://ui.adsabs.harvard.edu/abs/2023ApJ...944..131H} {944, 131}

\bibitem[\protect\citeauthoryear{{Hawley}, {Walkowicz}, {Allred}  \&
  {Valenti}}{{Hawley} et~al.}{2007}]{2007PASP..119...67H}
{Hawley} S.~L.,  {Walkowicz} L.~M.,  {Allred} J.~C.,   {Valenti} J.~A.,  2007,
  \mn@doi [\pasp] {10.1086/510561}, \href
  {https://ui.adsabs.harvard.edu/abs/2007PASP..119...67H} {119, 67}

\bibitem[\protect\citeauthoryear{{Holman} et~al.,}{{Holman}
  et~al.}{2011}]{2011SSRv..159..107H}
{Holman} G.~D.,  et~al., 2011, \mn@doi [\ssr] {10.1007/s11214-010-9680-9},
  \href {http://adsabs.harvard.edu/abs/2011SSRv..159..107H} {159, 107}

\bibitem[\protect\citeauthoryear{{Huang}, {Xu}, {Sadykov}, {Jing}  \&
  {Wang}}{{Huang} et~al.}{2019}]{2019ApJ...878L..15H}
{Huang} N.,  {Xu} Y.,  {Sadykov} V.~M.,  {Jing} J.,   {Wang} H.,  2019, \mn@doi
  [\apjl] {10.3847/2041-8213/ab2330}, \href
  {https://ui.adsabs.harvard.edu/abs/2019ApJ...878L..15H} {878, L15}

\bibitem[\protect\citeauthoryear{Jeffrey, Fletcher, Labrosse  \&
  Sim{\~o}es}{Jeffrey et~al.}{2018}]{Jeffreyeaav2794}
Jeffrey N. L.~S.,  Fletcher L.,  Labrosse N.,   Sim{\~o}es P. J.~A.,  2018,
  \mn@doi [Science Advances] {10.1126/sciadv.aav2794}, 4

\bibitem[\protect\citeauthoryear{{Jur{\v{c}}{\'a}k}, {Ka{\v{s}}parov{\'a}},
  {{\v{S}}vanda}  \& {Kleint}}{{Jur{\v{c}}{\'a}k}
  et~al.}{2018}]{2018A&A...620A.183J}
{Jur{\v{c}}{\'a}k} J.,  {Ka{\v{s}}parov{\'a}} J.,  {{\v{S}}vanda} M.,
  {Kleint} L.,  2018, \mn@doi [\aap] {10.1051/0004-6361/201833946}, \href
  {https://ui.adsabs.harvard.edu/abs/2018A&A...620A.183J} {620, A183}

\bibitem[\protect\citeauthoryear{{Kerr}}{{Kerr}}{2022}]{2022FrASS...960856K}
{Kerr} G.~S.,  2022, \mn@doi [Frontiers in Astronomy and Space Sciences]
  {10.3389/fspas.2022.1060856}, \href
  {https://ui.adsabs.harvard.edu/abs/2022FrASS...960856K} {9, 1060856}

\bibitem[\protect\citeauthoryear{{Kerr}}{{Kerr}}{2023}]{2023FrASS...960862K}
{Kerr} G.~S.,  2023, \mn@doi [Frontiers in Astronomy and Space Sciences]
  {10.3389/fspas.2022.1060862}, \href
  {https://ui.adsabs.harvard.edu/abs/2023FrASS...960862K} {9, 425}

\bibitem[\protect\citeauthoryear{{Kerr} \& {Fletcher}}{{Kerr} \&
  {Fletcher}}{2014}]{2014ApJ...783...98K}
{Kerr} G.~S.,  {Fletcher} L.,  2014, \mn@doi [\apj]
  {10.1088/0004-637X/783/2/98}, \href
  {https://ui.adsabs.harvard.edu/abs/2014ApJ...783...98K} {783, 98}

\bibitem[\protect\citeauthoryear{{Kerr}, {Sim{\~o}es}, {Qiu}  \&
  {Fletcher}}{{Kerr} et~al.}{2015}]{2015A&A...582A..50K}
{Kerr} G.~S.,  {Sim{\~o}es} P.~J.~A.,  {Qiu} J.,   {Fletcher} L.,  2015,
  \mn@doi [\aap] {10.1051/0004-6361/201526128}, \href
  {http://adsabs.harvard.edu/abs/2015A%26A...582A..50K} {582, A50}

\bibitem[\protect\citeauthoryear{{Kerr}, {Fletcher}, {Russell}  \&
  {Allred}}{{Kerr} et~al.}{2016}]{2016ApJ...827..101K}
{Kerr} G.~S.,  {Fletcher} L.,  {Russell} A.~J.~B.,   {Allred} J.~C.,  2016,
  \mn@doi [\apj] {10.3847/0004-637X/827/2/101}, \href
  {http://adsabs.harvard.edu/abs/2016ApJ...827..101K} {827, 101}

\bibitem[\protect\citeauthoryear{{Kerr}, {Carlsson}, {Allred}, {Young}  \&
  {Daw}}{{Kerr} et~al.}{2019a}]{2019ApJ...871...23K}
{Kerr} G.~S.,  {Carlsson} M.,  {Allred} J.~C.,  {Young} P.~R.,   {Daw} A.~N.,
  2019a, \mn@doi [\apj] {10.3847/1538-4357/aaf46e}, \href
  {https://ui.adsabs.harvard.edu/abs/2019ApJ...871...23K} {871, 23}

\bibitem[\protect\citeauthoryear{{Kerr}, {Allred}  \& {Carlsson}}{{Kerr}
  et~al.}{2019b}]{2019ApJ...883...57K}
{Kerr} G.~S.,  {Allred} J.~C.,   {Carlsson} M.,  2019b, \mn@doi [\apj]
  {10.3847/1538-4357/ab3c24}, \href
  {https://ui.adsabs.harvard.edu/abs/2019ApJ...883...57K} {883, 57}

\bibitem[\protect\citeauthoryear{{Kerr}, {Carlsson}  \& {Allred}}{{Kerr}
  et~al.}{2019c}]{2019ApJ...885..119K}
{Kerr} G.~S.,  {Carlsson} M.,   {Allred} J.~C.,  2019c, \mn@doi [\apj]
  {10.3847/1538-4357/ab48ea}, \href
  {https://ui.adsabs.harvard.edu/abs/2019ApJ...885..119K} {885, 119}

\bibitem[\protect\citeauthoryear{{Kerr}, {Allred}, {Kowalski}, {Milligan},
  {Hudson}, {Prado}, {Kucera}  \& {Brosius}}{{Kerr}
  et~al.}{2023}]{2023ApJ...945..118K}
{Kerr} G.~S.,  {Allred} J.~C.,  {Kowalski} A.~F.,  {Milligan} R.~O.,  {Hudson}
  H.~S.,  {Prado} N.~Z.,  {Kucera} T.~A.,   {Brosius} J.~W.,  2023, \mn@doi
  [\apj] {10.3847/1538-4357/acb92a}, \href
  {https://ui.adsabs.harvard.edu/abs/2023ApJ...945..118K} {945, 118}

\bibitem[\protect\citeauthoryear{{Kleint}, {Heinzel}, {Judge}  \&
  {Krucker}}{{Kleint} et~al.}{2016}]{2016ApJ...816...88K}
{Kleint} L.,  {Heinzel} P.,  {Judge} P.,   {Krucker} S.,  2016, \mn@doi [\apj]
  {10.3847/0004-637X/816/2/88}, \href
  {https://ui.adsabs.harvard.edu/abs/2016ApJ...816...88K} {816, 88}

\bibitem[\protect\citeauthoryear{{Kleint}, {Heinzel}  \& {Krucker}}{{Kleint}
  et~al.}{2017}]{2017ApJ...837..160K}
{Kleint} L.,  {Heinzel} P.,   {Krucker} S.,  2017, \mn@doi [\apj]
  {10.3847/1538-4357/aa62fe}, \href
  {https://ui.adsabs.harvard.edu/abs/2017ApJ...837..160K} {837, 160}

\bibitem[\protect\citeauthoryear{{Klimchuk}, {Patsourakos}  \&
  {Tripathi}}{{Klimchuk} et~al.}{2016}]{2016SoPh..291...55K}
{Klimchuk} J.~A.,  {Patsourakos} S.,   {Tripathi} D.,  2016, \mn@doi [\solphys]
  {10.1007/s11207-015-0827-4}, \href
  {https://ui.adsabs.harvard.edu/abs/2016SoPh..291...55K} {291, 55}

\bibitem[\protect\citeauthoryear{{Kontar} et~al.,}{{Kontar}
  et~al.}{2011}]{2011SSRv..159..301K}
{Kontar} E.~P.,  et~al., 2011, \mn@doi [\ssr] {10.1007/s11214-011-9804-x},
  \href {https://ui.adsabs.harvard.edu/abs/2011SSRv..159..301K} {159, 301}

\bibitem[\protect\citeauthoryear{{Kowalski}, {Allred}, {Daw}, {Cauzzi}  \&
  {Carlsson}}{{Kowalski} et~al.}{2017}]{2017ApJ...836...12K}
{Kowalski} A.~F.,  {Allred} J.~C.,  {Daw} A.,  {Cauzzi} G.,   {Carlsson} M.,
  2017, \mn@doi [\apj] {10.3847/1538-4357/836/1/12}, \href
  {https://ui.adsabs.harvard.edu/abs/2017ApJ...836...12K} {836, 12}

\bibitem[\protect\citeauthoryear{{Kowalski}, {Butler}, {Daw}, {Fletcher},
  {Allred}, {De Pontieu}, {Kerr}  \& {Cauzzi}}{{Kowalski}
  et~al.}{2019}]{2019ApJ...878..135K}
{Kowalski} A.~F.,  {Butler} E.,  {Daw} A.~N.,  {Fletcher} L.,  {Allred} J.~C.,
  {De Pontieu} B.,  {Kerr} G.~S.,   {Cauzzi} G.,  2019, \mn@doi [\apj]
  {10.3847/1538-4357/ab1f8b}, \href
  {https://ui.adsabs.harvard.edu/abs/2019ApJ...878..135K} {878, 135}

\bibitem[\protect\citeauthoryear{{Kowalski}, {Allred}, {Carlsson}, {Kerr},
  {Tremblay}, {Namekata}, {Kuridze}  \& {Uitenbroek}}{{Kowalski}
  et~al.}{2022}]{2022ApJ...928..190K}
{Kowalski} A.~F.,  {Allred} J.~C.,  {Carlsson} M.,  {Kerr} G.~S.,  {Tremblay}
  P.-E.,  {Namekata} K.,  {Kuridze} D.,   {Uitenbroek} H.,  2022, \mn@doi
  [\apj] {10.3847/1538-4357/ac5174}, \href
  {https://ui.adsabs.harvard.edu/abs/2022ApJ...928..190K} {928, 190}

\bibitem[\protect\citeauthoryear{{Leenaarts}, {Carlsson}  \& {Rouppe van der
  Voort}}{{Leenaarts} et~al.}{2012}]{2012ApJ...749..136L}
{Leenaarts} J.,  {Carlsson} M.,   {Rouppe van der Voort} L.,  2012, \mn@doi
  [\apj] {10.1088/0004-637X/749/2/136}, \href
  {https://ui.adsabs.harvard.edu/abs/2012ApJ...749..136L} {749, 136}

\bibitem[\protect\citeauthoryear{{Leenaarts}, {Pereira}, {Carlsson},
  {Uitenbroek}  \& {De Pontieu}}{{Leenaarts}
  et~al.}{2013}]{2013ApJ...772...90L}
{Leenaarts} J.,  {Pereira} T.~M.~D.,  {Carlsson} M.,  {Uitenbroek} H.,   {De
  Pontieu} B.,  2013, \mn@doi [\apj] {10.1088/0004-637X/772/2/90}, \href
  {https://ui.adsabs.harvard.edu/abs/2013ApJ...772...90L} {772, 90}

\bibitem[\protect\citeauthoryear{{Li}, {Zhang}, {Yang}  \& {Hou}}{{Li}
  et~al.}{2019}]{2019PASJ...71...14L}
{Li} X.,  {Zhang} J.,  {Yang} S.,   {Hou} Y.,  2019, \mn@doi [\pasj]
  {10.1093/pasj/psy128}, \href
  {https://ui.adsabs.harvard.edu/abs/2019PASJ...71...14L} {71, 14}

\bibitem[\protect\citeauthoryear{{Lin} \& {Carlsson}}{{Lin} \&
  {Carlsson}}{2015}]{2015ApJ...813...34L}
{Lin} H.-H.,  {Carlsson} M.,  2015, \mn@doi [\apj]
  {10.1088/0004-637X/813/1/34}, \href
  {https://ui.adsabs.harvard.edu/abs/2015ApJ...813...34L} {813, 34}

\bibitem[\protect\citeauthoryear{{Lin} et~al.,}{{Lin}
  et~al.}{2002}]{2002SoPh..210....3L}
{Lin} R.~P.,  et~al., 2002, \mn@doi [\solphys] {10.1023/A:1022428818870}, \href
  {http://adsabs.harvard.edu/abs/2002SoPh..210....3L} {210, 3}

\bibitem[\protect\citeauthoryear{{Lin}, {Carlsson}  \& {Leenaarts}}{{Lin}
  et~al.}{2017}]{2017ApJ...846...40L}
{Lin} H.-H.,  {Carlsson} M.,   {Leenaarts} J.,  2017, \mn@doi [\apj]
  {10.3847/1538-4357/aa8458}, \href
  {https://ui.adsabs.harvard.edu/abs/2017ApJ...846...40L} {846, 40}

\bibitem[\protect\citeauthoryear{{Liu}, {Heinzel}, {Kleint}  \& {Ka{\v
  s}parov{\'a}}}{{Liu} et~al.}{2015}]{2015SoPh..290.3525L}
{Liu} W.,  {Heinzel} P.,  {Kleint} L.,   {Ka{\v s}parov{\'a}} J.,  2015,
  \mn@doi [\solphys] {10.1007/s11207-015-0814-9}, \href
  {http://adsabs.harvard.edu/abs/2015SoPh..290.3525L} {290, 3525}

\bibitem[\protect\citeauthoryear{{MacNeice}}{{MacNeice}}{1986}]{1986SoPh..103...47M}
{MacNeice} P.,  1986, \mn@doi [\solphys] {10.1007/BF00154858}, \href
  {https://ui.adsabs.harvard.edu/abs/1986SoPh..103...47M} {103, 47}

\bibitem[\protect\citeauthoryear{{Magain}}{{Magain}}{1986}]{1986A&A...163..135M}
{Magain} P.,  1986, \aap, \href
  {http://adsabs.harvard.edu/abs/1986A%26A...163..135M} {163, 135}

\bibitem[\protect\citeauthoryear{{McLaughlin}, {Milligan}, {Kerr}, {Monson},
  {Sim{\~o}es}  \& {Mathioudakis}}{{McLaughlin}
  et~al.}{2023}]{2023ApJ...944..186M}
{McLaughlin} S.~A.,  {Milligan} R.~O.,  {Kerr} G.~S.,  {Monson} A.~J.,
  {Sim{\~o}es} P. J.~A.,   {Mathioudakis} M.,  2023, \mn@doi [\apj]
  {10.3847/1538-4357/acaf66}, \href
  {https://ui.adsabs.harvard.edu/abs/2023ApJ...944..186M} {944, 186}

\bibitem[\protect\citeauthoryear{{Meegan} et~al.,}{{Meegan}
  et~al.}{2009}]{2009ApJ...702..791M}
{Meegan} C.,  et~al., 2009, \mn@doi [\apj] {10.1088/0004-637X/702/1/791}, \href
  {https://ui.adsabs.harvard.edu/abs/2009ApJ...702..791M} {702, 791}

\bibitem[\protect\citeauthoryear{{Metcalf}, {Canfield}, {Avrett}  \&
  {Metcalf}}{{Metcalf} et~al.}{1990a}]{1990ApJ...350..463M}
{Metcalf} T.~R.,  {Canfield} R.~C.,  {Avrett} E.~H.,   {Metcalf} F.~T.,  1990a,
  \mn@doi [\apj] {10.1086/168400}, \href
  {https://ui.adsabs.harvard.edu/abs/1990ApJ...350..463M} {350, 463}

\bibitem[\protect\citeauthoryear{{Metcalf}, {Canfield}  \& {Saba}}{{Metcalf}
  et~al.}{1990b}]{1990ApJ...365..391M}
{Metcalf} T.~R.,  {Canfield} R.~C.,   {Saba} J. L.~R.,  1990b, \mn@doi [\apj]
  {10.1086/169494}, \href
  {https://ui.adsabs.harvard.edu/abs/1990ApJ...365..391M} {365, 391}

\bibitem[\protect\citeauthoryear{{Milligan}}{{Milligan}}{2011}]{2011ApJ...740...70M}
{Milligan} R.~O.,  2011, \mn@doi [\apj] {10.1088/0004-637X/740/2/70}, \href
  {https://ui.adsabs.harvard.edu/abs/2011ApJ...740...70M} {740, 70}

\bibitem[\protect\citeauthoryear{{Milligan}}{{Milligan}}{2015}]{2015SoPh..290.3399M}
{Milligan} R.~O.,  2015, \mn@doi [\solphys] {10.1007/s11207-015-0748-2}, \href
  {https://ui.adsabs.harvard.edu/abs/2015SoPh..290.3399M} {290, 3399}

\bibitem[\protect\citeauthoryear{{Milligan} \& {Dennis}}{{Milligan} \&
  {Dennis}}{2009}]{2009ApJ...699..968M}
{Milligan} R.~O.,  {Dennis} B.~R.,  2009, \mn@doi [\apj]
  {10.1088/0004-637X/699/2/968}, \href
  {https://ui.adsabs.harvard.edu/abs/2009ApJ...699..968M} {699, 968}

\bibitem[\protect\citeauthoryear{{Milligan}, {Gallagher}, {Mathioudakis},
  {Bloomfield}, {Keenan}  \& {Schwartz}}{{Milligan}
  et~al.}{2006}]{2006ApJ...638L.117M}
{Milligan} R.~O.,  {Gallagher} P.~T.,  {Mathioudakis} M.,  {Bloomfield} D.~S.,
  {Keenan} F.~P.,   {Schwartz} R.~A.,  2006, \mn@doi [\apjl] {10.1086/500555},
  \href {https://ui.adsabs.harvard.edu/abs/2006ApJ...638L.117M} {638, L117}

\bibitem[\protect\citeauthoryear{{Panos}, {Kleint}, {Huwyler}, {Krucker},
  {Melchior}, {Ullmann}  \& {Voloshynovskiy}}{{Panos}
  et~al.}{2018}]{2018ApJ...861...62P}
{Panos} B.,  {Kleint} L.,  {Huwyler} C.,  {Krucker} S.,  {Melchior} M.,
  {Ullmann} D.,   {Voloshynovskiy} S.,  2018, \mn@doi [\apj]
  {10.3847/1538-4357/aac779}, \href
  {https://ui.adsabs.harvard.edu/abs/2018ApJ...861...62P} {861, 62}

\bibitem[\protect\citeauthoryear{{Peat}, {Labrosse}, {Schmieder}  \&
  {Barczynski}}{{Peat} et~al.}{2021}]{2021A&A...653A...5P}
{Peat} A.~W.,  {Labrosse} N.,  {Schmieder} B.,   {Barczynski} K.,  2021,
  \mn@doi [\aap] {10.1051/0004-6361/202140907}, \href
  {https://ui.adsabs.harvard.edu/abs/2021A&A...653A...5P} {653, A5}

\bibitem[\protect\citeauthoryear{{Pereira} \& {Uitenbroek}}{{Pereira} \&
  {Uitenbroek}}{2015}]{2015A&A...574A...3P}
{Pereira} T. M.~D.,  {Uitenbroek} H.,  2015, \mn@doi [\aap]
  {10.1051/0004-6361/201424785}, \href
  {https://ui.adsabs.harvard.edu/abs/2015A&A...574A...3P} {574, A3}

\bibitem[\protect\citeauthoryear{{Pereira}, {Carlsson}, {De Pontieu}  \&
  {Hansteen}}{{Pereira} et~al.}{2015}]{2015ApJ...806...14P}
{Pereira} T. M.~D.,  {Carlsson} M.,  {De Pontieu} B.,   {Hansteen} V.,  2015,
  \mn@doi [\apj] {10.1088/0004-637X/806/1/14}, \href
  {https://ui.adsabs.harvard.edu/abs/2015ApJ...806...14P} {806, 14}

\bibitem[\protect\citeauthoryear{{Polito}, {Testa}  \& {De Pontieu}}{{Polito}
  et~al.}{2019}]{2019ApJ...879L..17P}
{Polito} V.,  {Testa} P.,   {De Pontieu} B.,  2019, \mn@doi [\apjl]
  {10.3847/2041-8213/ab290b}, \href
  {https://ui.adsabs.harvard.edu/abs/2019ApJ...879L..17P} {879, L17}

\bibitem[\protect\citeauthoryear{{Polito}, {Kerr}, {Xu}, {Sadykov}  \&
  {Lorincik}}{{Polito} et~al.}{2023}]{2023ApJ...944..104P}
{Polito} V.,  {Kerr} G.~S.,  {Xu} Y.,  {Sadykov} V.~M.,   {Lorincik} J.,  2023,
  \mn@doi [\apj] {10.3847/1538-4357/acaf7c}, \href
  {https://ui.adsabs.harvard.edu/abs/2023ApJ...944..104P} {944, 104}

\bibitem[\protect\citeauthoryear{{Ramaty} \& {Mandzhavidze}}{{Ramaty} \&
  {Mandzhavidze}}{2000}]{2000AIPC..522..401R}
{Ramaty} R.,  {Mandzhavidze} N.,  2000, in Amer. Inst. of Phy. Conf.Series. pp
  401--410, \mn@doi{10.1063/1.1291742}

\bibitem[\protect\citeauthoryear{{Rathore} \& {Carlsson}}{{Rathore} \&
  {Carlsson}}{2015}]{2015ApJ...811...80R}
{Rathore} B.,  {Carlsson} M.,  2015, \mn@doi [\apj]
  {10.1088/0004-637X/811/2/80}, \href
  {http://adsabs.harvard.edu/abs/2015ApJ...811...80R} {811, 80}

\bibitem[\protect\citeauthoryear{{Reep} \& {Russell}}{{Reep} \&
  {Russell}}{2016}]{2016ApJ...818L..20R}
{Reep} J.~W.,  {Russell} A.~J.~B.,  2016, \mn@doi [ApJL]
  {10.3847/2041-8205/818/1/L20}, \href
  {http://adsabs.harvard.edu/abs/2016ApJ...818L..20R} {818, L20}

\bibitem[\protect\citeauthoryear{{Reep}, {Russell}, {Tarr}  \& {Leake}}{{Reep}
  et~al.}{2018}]{2018ApJ...853..101R}
{Reep} J.~W.,  {Russell} A.~J.~B.,  {Tarr} L.~A.,   {Leake} J.~E.,  2018,
  \mn@doi [ApJ] {10.3847/1538-4357/aaa2fe}, \href
  {http://adsabs.harvard.edu/abs/2018ApJ...853..101R} {853, 101}

\bibitem[\protect\citeauthoryear{{Ruan}, {Schmieder}, {Mein}, {Mein},
  {Labrosse}, {Gun{\'a}r}  \& {Chen}}{{Ruan}
  et~al.}{2018}]{2018ApJ...865..123R}
{Ruan} G.,  {Schmieder} B.,  {Mein} P.,  {Mein} N.,  {Labrosse} N.,
  {Gun{\'a}r} S.,   {Chen} Y.,  2018, \mn@doi [\apj]
  {10.3847/1538-4357/aada08}, \href
  {https://ui.adsabs.harvard.edu/abs/2018ApJ...865..123R} {865, 123}

\bibitem[\protect\citeauthoryear{{Ruan}, {Yan}  \& {Keppens}}{{Ruan}
  et~al.}{2023}]{2023ApJ...947...67R}
{Ruan} W.,  {Yan} L.,   {Keppens} R.,  2023, \mn@doi [\apj]
  {10.3847/1538-4357/ac9b4e}, \href
  {https://ui.adsabs.harvard.edu/abs/2023ApJ...947...67R} {947, 67}

\bibitem[\protect\citeauthoryear{{Rubio da Costa} \& {Kleint}}{{Rubio da Costa}
  \& {Kleint}}{2017}]{2017ApJ...842...82R}
{Rubio da Costa} F.,  {Kleint} L.,  2017, \mn@doi [ApJ]
  {10.3847/1538-4357/aa6eaf}, \href
  {http://adsabs.harvard.edu/abs/2017ApJ...842...82R} {842, 82}

\bibitem[\protect\citeauthoryear{{Rubio da Costa}, {Kleint}, {Petrosian}, {Liu}
   \& {Allred}}{{Rubio da Costa} et~al.}{2016}]{2016ApJ...827...38R}
{Rubio da Costa} F.,  {Kleint} L.,  {Petrosian} V.,  {Liu} W.,   {Allred}
  J.~C.,  2016, \mn@doi [\apj] {10.3847/0004-637X/827/1/38}, \href
  {https://ui.adsabs.harvard.edu/abs/2016ApJ...827...38R} {827, 38}

\bibitem[\protect\citeauthoryear{{Russell} \& {Stackhouse}}{{Russell} \&
  {Stackhouse}}{2013}]{2013A&A...558A..76R}
{Russell} A.~J.~B.,  {Stackhouse} D.~J.,  2013, \mn@doi [\aap]
  {10.1051/0004-6361/201321916}, \href
  {https://ui.adsabs.harvard.edu/abs/2013A&A...558A..76R} {558, A76}

\bibitem[\protect\citeauthoryear{{Sainz Dalda} \& {De Pontieu}}{{Sainz Dalda}
  \& {De Pontieu}}{2022}]{2022arXiv221105459S}
{Sainz Dalda} A.,  {De Pontieu} B.,  2022, \mn@doi [arXiv e-prints]
  {10.48550/arXiv.2211.05459}, \href
  {https://ui.adsabs.harvard.edu/abs/2022arXiv221105459S} {p. arXiv:2211.05459}

\bibitem[\protect\citeauthoryear{{Sellers}, {Milligan}  \& {McAteer}}{{Sellers}
  et~al.}{2022}]{2022ApJ...936...85S}
{Sellers} S.~G.,  {Milligan} R.~O.,   {McAteer} R.~T.~J.,  2022, \mn@doi [\apj]
  {10.3847/1538-4357/ac87a9}, \href
  {https://ui.adsabs.harvard.edu/abs/2022ApJ...936...85S} {936, 85}

\bibitem[\protect\citeauthoryear{{Shih}, {Lin}  \& {Smith}}{{Shih}
  et~al.}{2009}]{2009ApJ...698L.152S}
{Shih} A.~Y.,  {Lin} R.~P.,   {Smith} D.~M.,  2009, \mn@doi [ApJL]
  {10.1088/0004-637X/698/2/L152}, \href
  {https://ui.adsabs.harvard.edu/abs/2009ApJ...698L.152S} {698, L152}

\bibitem[\protect\citeauthoryear{{Shine}}{{Shine}}{1983}]{1983ApJ...266..882S}
{Shine} R.~A.,  1983, \mn@doi [\apj] {10.1086/160835}, \href
  {https://ui.adsabs.harvard.edu/abs/1983ApJ...266..882S} {266, 882}

\bibitem[\protect\citeauthoryear{{Tian}, {Hong}, {Li}  \& {Ding}}{{Tian}
  et~al.}{2022}]{2022A&A...668A..96T}
{Tian} J.,  {Hong} J.,  {Li} Y.,   {Ding} M.~D.,  2022, \mn@doi [\aap]
  {10.1051/0004-6361/202244615}, \href
  {https://ui.adsabs.harvard.edu/abs/2022A&A...668A..96T} {668, A96}

\bibitem[\protect\citeauthoryear{{Tomozov}}{{Tomozov}}{1990}]{1990JQSRT..44..171T}
{Tomozov} V.~M.,  1990, \mn@doi [\jqsrt] {10.1016/0022-4073(90)90098-Q}, \href
  {https://ui.adsabs.harvard.edu/abs/1990JQSRT..44..171T} {44, 171}

\bibitem[\protect\citeauthoryear{{Uitenbroek}}{{Uitenbroek}}{2001}]{2001ApJ...557..389U}
{Uitenbroek} H.,  2001, \mn@doi [\apj] {10.1086/321659}, \href
  {http://adsabs.harvard.edu/abs/2001ApJ...557..389U} {557, 389}

\bibitem[\protect\citeauthoryear{{Vernazza}, {Avrett}  \& {Loeser}}{{Vernazza}
  et~al.}{1981}]{1981ApJS...45..635V}
{Vernazza} J.~E.,  {Avrett} E.~H.,   {Loeser} R.,  1981, \mn@doi [\apjs]
  {10.1086/190731}, \href
  {https://ui.adsabs.harvard.edu/abs/1981ApJS...45..635V} {45, 635}

\bibitem[\protect\citeauthoryear{{Warren}, {Reep}, {Crump}  \&
  {Sim{\~o}es}}{{Warren} et~al.}{2016}]{2016ApJ...829...35W}
{Warren} H.~P.,  {Reep} J.~W.,  {Crump} N.~A.,   {Sim{\~o}es} P. J.~A.,  2016,
  \mn@doi [\apj] {10.3847/0004-637X/829/1/35}, \href
  {https://ui.adsabs.harvard.edu/abs/2016ApJ...829...35W} {829, 35}

\bibitem[\protect\citeauthoryear{{Yadav}, {D{\'\i}az Baso}, {de la Cruz
  Rodr{\'\i}guez}, {Calvo}  \& {Morosin}}{{Yadav}
  et~al.}{2021}]{2021A&A...649A.106Y}
{Yadav} R.,  {D{\'\i}az Baso} C.~J.,  {de la Cruz Rodr{\'\i}guez} J.,  {Calvo}
  F.,   {Morosin} R.,  2021, \mn@doi [\aap] {10.1051/0004-6361/202039857},
  \href {https://ui.adsabs.harvard.edu/abs/2021A&A...649A.106Y} {649, A106}

\bibitem[\protect\citeauthoryear{{Zhu}, {Kowalski}, {Tian}, {Uitenbroek},
  {Carlsson}  \& {Allred}}{{Zhu} et~al.}{2019}]{2019ApJ...879...19Z}
{Zhu} Y.,  {Kowalski} A.~F.,  {Tian} H.,  {Uitenbroek} H.,  {Carlsson} M.,
  {Allred} J.~C.,  2019, \mn@doi [\apj] {10.3847/1538-4357/ab2238}, \href
  {https://ui.adsabs.harvard.edu/abs/2019ApJ...879...19Z} {879, 19}

\bibitem[\protect\citeauthoryear{{de la Cruz Rodr{\'\i}guez}, {Leenaarts},
  {Danilovic}  \& {Uitenbroek}}{{de la Cruz Rodr{\'\i}guez}
  et~al.}{2019}]{2019A&A...623A..74D}
{de la Cruz Rodr{\'\i}guez} J.,  {Leenaarts} J.,  {Danilovic} S.,
  {Uitenbroek} H.,  2019, \mn@doi [\aap] {10.1051/0004-6361/201834464}, \href
  {https://ui.adsabs.harvard.edu/abs/2019A&A...623A..74D} {623, A74}

\makeatother
\end{thebibliography}


\appendix

\section{Description of spectral line evolution}\label{sec:figure4disc}
What follows is a general description of the evolution of each line, as shown in Figures~\ref{fig:iris_spectra_overview1} \& \ref{fig:iris_spectra_overview2}. 

\textbf{\ion{Cl}{i} 1351.66~\AA}: The line starts very narrow in the pre-flare, and the peak intensity is largest before the line broadens. Though still more intense than the pre-flare, the peak line intensity decreases somewhat before displaying clear asymmetries in the red wing. When the line is at its narrowest, the on-board spectral summing means that it is only 3-4 pixels across. The line is at its brightest for approximately 30~s, though remains enhanced for several minutes. The asymmetry is present prominently for around 30-45~s.

\textbf{\ion{O}{i} 1355.598~\AA}: The line brightens in response to the flare, but remains fairly steady in intensity throughout. It broadens, and exhibits a red wing asymmetry that, owing to the lower contrast between the core-to-wing intensity, is easier to see in the initial development than \ion{Cl}{i}. The nearby \ion{C}{i} 1355.844~\AA\ line is considerably brighter, and broadens by a greater extent. The intensity ratio changes in response to the flare, with \ion{C}{i} outshining \ion{O}{i} during flare periods but the opposite is true during quiescent times. Such behaviour has been known for some time \citep[e.g.][]{1980A&A....86..377C,2017ApJ...846...40L}, and a study of their relative formation properties during flares would be interesting, but is outside the scope of this manuscript. Evidently, however, there is some small level of flare heating for more than a minute prior to the strongest enhancements, since the \ion{C}{i} line is brighter than the \ion{O}{i} line. 

\textbf{\ion{Si}{iv} 1402.77~\AA}: The behaviour of the \ion{Si}{iv} resonance line similar to prior observations of this line during flares. That is, the intensity becomes very large, and the line exhibits very broad wings. The red-wing asymmetries in the flare studied here may be quite tame compared to other events. At certain times the detector is saturated.

\textbf{\ion{Fe}{ii} 2814.445~\AA}: This line, and its \ion{Fe}{i} neighbour at 2814.115~\AA\ are initially very weak and comparable to the pre-flare continuum nearby the line. Over time two clear emission lines form, with a clear red wing asymmetry that persists for a similar length of time as the \ion{Cl}{i} and \ion{O}{i} asymmetries. The \ion{Fe}{i} line does not show a very obvious asymmetry, but its formation depth may be lower and less affected by the condensation that produces red wing features in the lines forming at greater altitude. It is also less likely to be optically thin, thus harder to interpret and so we focus on the \ion{Fe}{ii} line. 

\textbf{\ion{Mg}{ii} 2791.60~\AA}: This \ion{Mg}{ii} subordinate line displays a red wing extending out 0.5~\AA\ or more from the rest wavelength. The nearby quasi-continuum intensity is more than an order of magnitude increased from the pre-flare, and the line is fully in emission with no self-absorption features. Though forming at large column depth in the quiescent periods \citep[e.g.][]{2015ApJ...806...14P}, the subordinate lines form close to the \ion{Mg}{ii} resonance lines in the flaring atmosphere \citep[][]{2019ApJ...871...23K,2019ApJ...883...57K,2019ApJ...879...19Z}, so this asymmetry is expected to be caused by the same condensation that produces the asymmetry in the other lines. Note that the asymmetry is clearly present a lot earlier than in the optically thin lines, though again we note that those lines are weaker, which possibly hides the development of an asymmetry. 

 \textbf{\ion{Mg}{ii} k}: Much like the \ion{Mg}{ii} subordinate line, the resonance lines increase in intensity by a significant degree, and both wings broaden in addition to the red wing exhibiting an asymmetry. The k line is single peaked both in the flare and the pre-flare, since this source is in the penumbra/umbra. After the peak of the flare the wings remain broad, and Lorenztian in shape, similar to the profiles discussed by \cite{2017ApJ...842...82R} and \cite{2019ApJ...879...19Z}, though the event we report here is less broad than in those other studies.

\section{Effect of Varying $T_{\mathrm{form}}$ and $W_{\mathrm{I}}$}\label{sec:varyformt}
\begin{figure*}
	\centering 
	\vbox{
	\hbox{
	\subfloat{\includegraphics[width = 0.5\textwidth, clip = true, trim = 0.cm 0.cm 0.cm 0.cm]{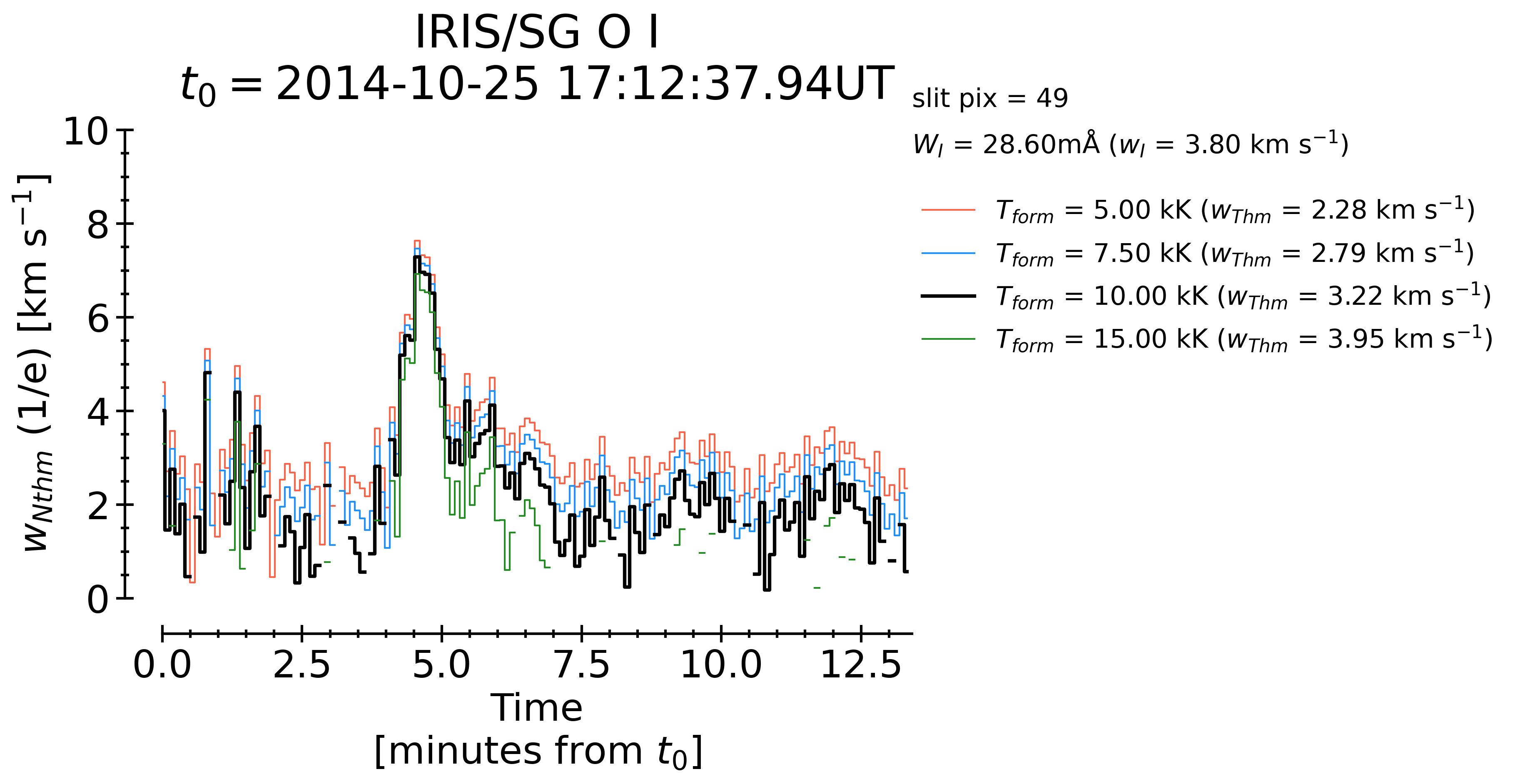}}
	\subfloat{\includegraphics[width = 0.5\textwidth, clip = true, trim = 0.cm 0.cm 0.cm 0.cm]{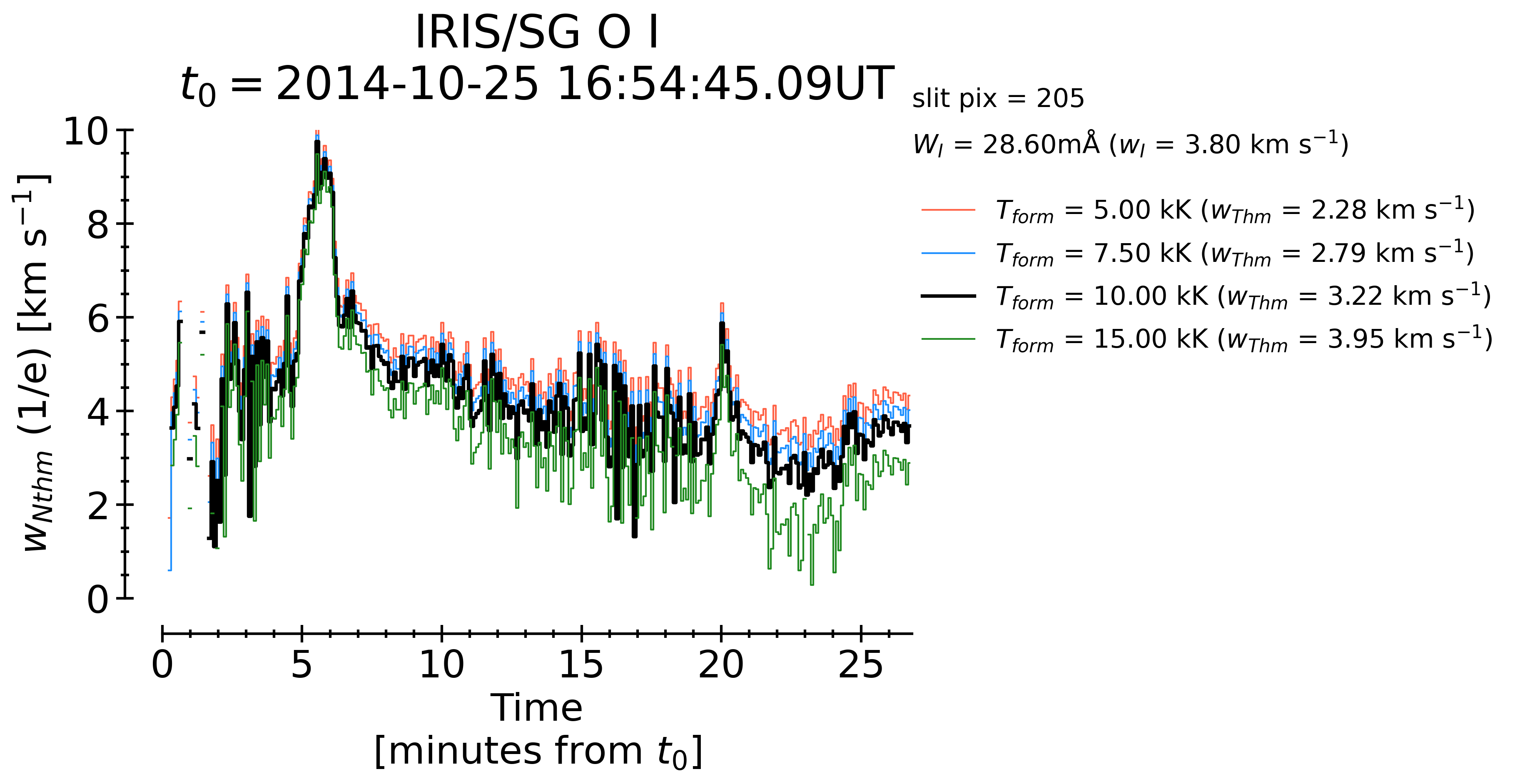}}	
	}
	}
	\vbox{
	\hbox{
	\subfloat{\includegraphics[width = 0.5\textwidth, clip = true, trim = 0.cm 0.cm 0.cm 0.cm]{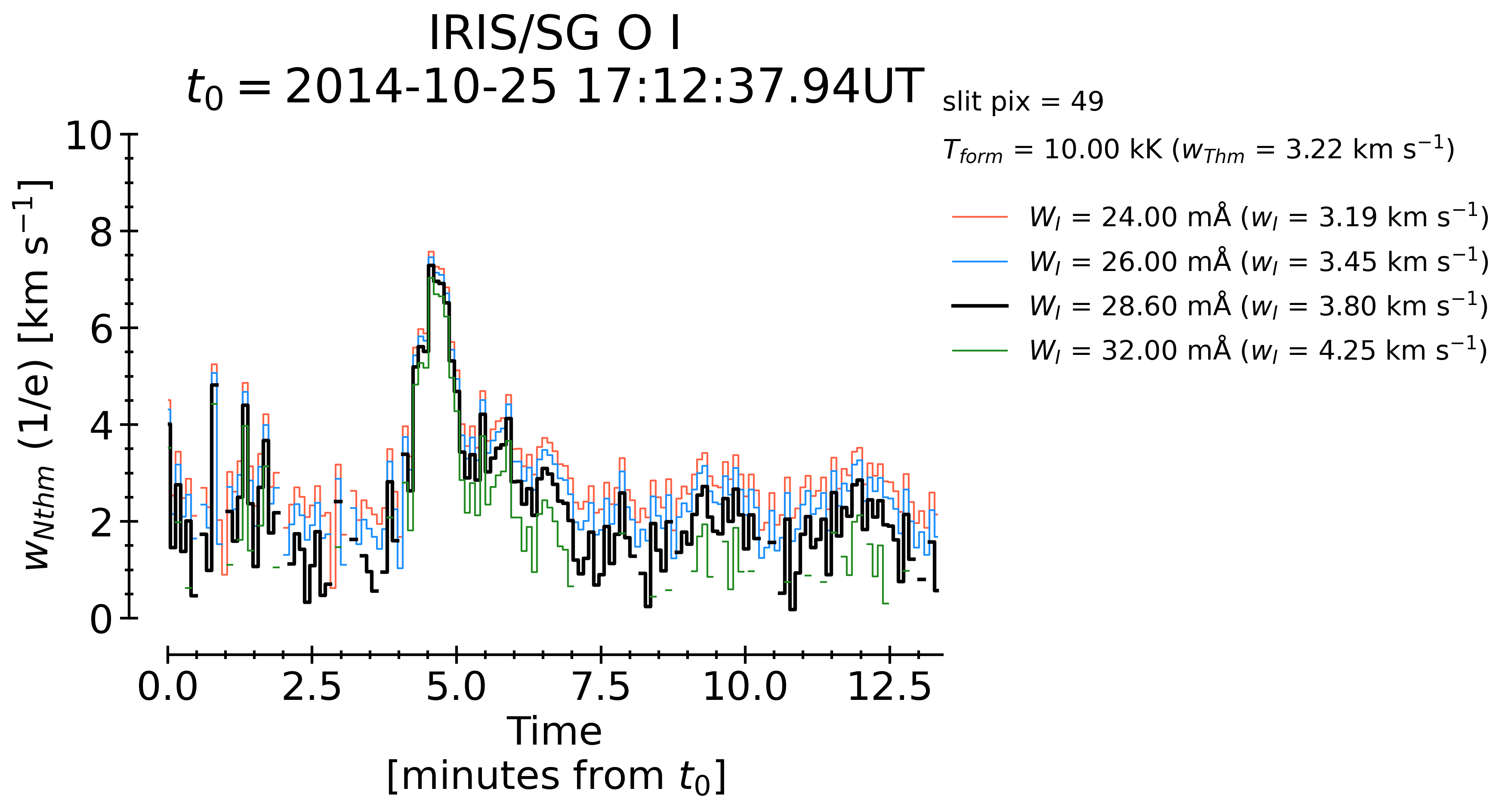}}
	\subfloat{\includegraphics[width = 0.5\textwidth, clip = true, trim = 0.cm 0.cm 0.cm 0.cm]{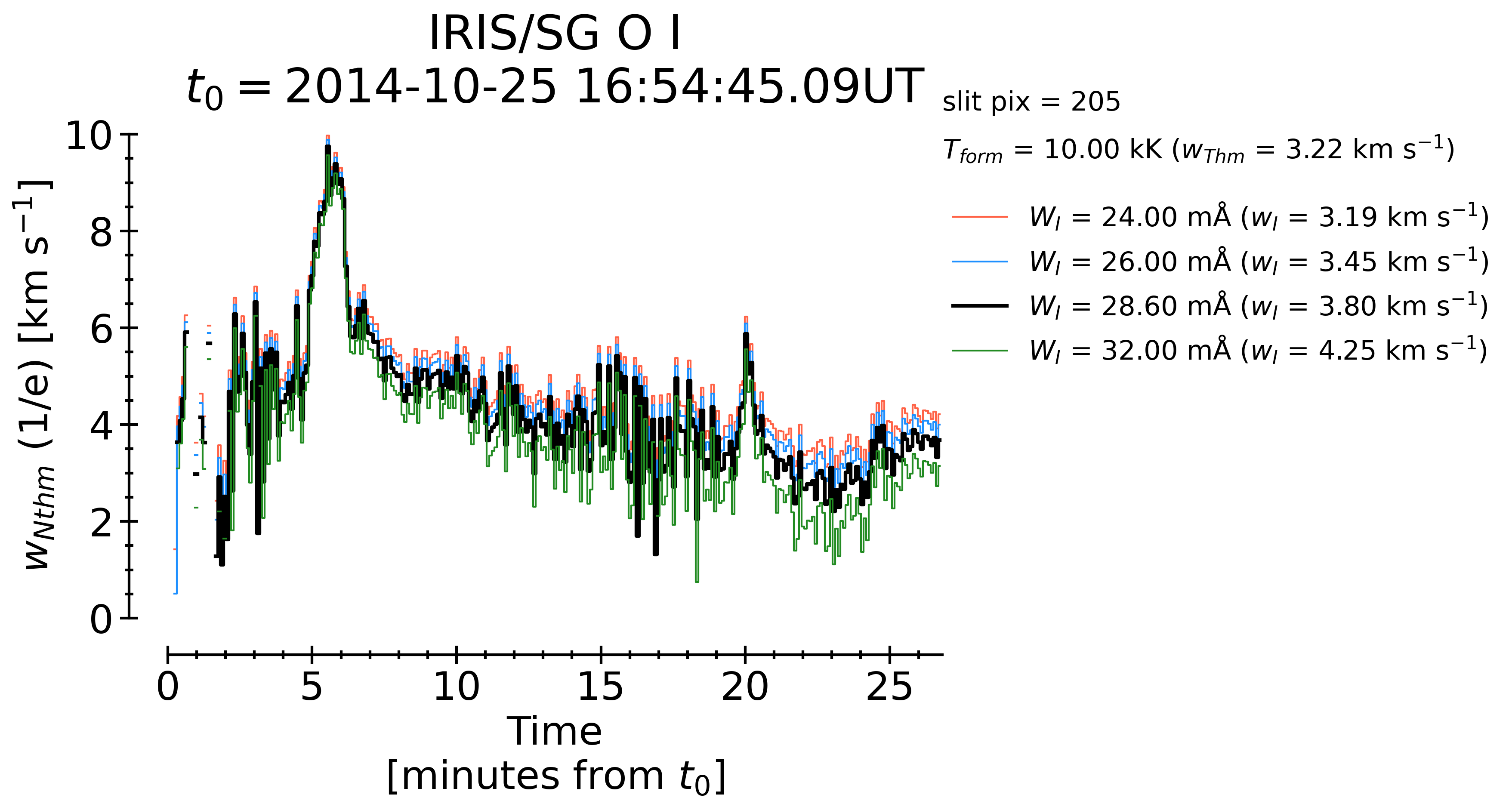}}	
	}
	}
	\caption{\textsl{Effect on calculation of $w_{\mathrm{Nthm}}$ of varying $T_{\mathrm{form}}$ (top row) or $W_{\mathrm{I}}$ (bottom row). For each, two sources are shown, one from the main region of interest (slit pixel 49; first column) and the other from a portion of flare ribbon located further west (slit pixel 205; second colum). In the top row $T_{\mathrm{form}}$ is varied with a fixed  $W_{\mathrm{I}}$. In the bottom row $W_{\mathrm{I}}$ is varied for a fixed $T_{\mathrm{form}}$. Missing segments indicate where the observed total width is smaller than the thermal width + instrumental width (i.e. one of those values is overestimated). This typically happens in the umbral/penumbral source (first column) which has a somewhat narrower profile. Generally, though, varying each of these properties through reasonable values does not have a terribly large impact on the calculated value of $w_{\mathrm{Nthm}}$, particularly during the flare where $w_{\mathrm{Nthm}}$ is a more significant contributor to the total width. Here a single Gaussian component was fit to the \ion{O}{i} 1355.598~\AA\ line. The instrumental widths expressed as (1/e) half-width values are included in parenthesis. }}
	\label{fig:vary_tform_winstr}
\end{figure*}

When calculating $w_{\mathrm{Nthm}}$ it is necessary to make some assumption about the plasma temperature in order to calculate the thermal width $w_{\mathrm{thm}} = \frac{\lambda}{c} \sqrt{\frac{2k_{b}T}{m_{i}}}$. For \ion{O}{i} we assumed $T=10$~kK during the flare, and $T=6$~kK when measuring the ambient non-flaring $w_{\mathrm{Nthm}}$. As discussed in the main text, the flare temperatures may be somewhat larger, and the non-flaring temperatures may take a range of values, up to $T=10$~kK \citep[see][]{2015ApJ...813...34L}. We illustrate the effect on $w_{\mathrm{Nthm}}$ of varying $T$, with fixed values of the instrumental width, in Figure~\ref{fig:vary_tform_winstr} (top row), for two pixels. One source is in the primary region of interest (first column) , the other from the western ribbon (second column). The range of $T$ does cause some variation in the pre-flare $w_{\mathrm{Nthm}}$ values, but during the flare the spread of $w_{\mathrm{Nthm}}$ is quite small and on the order of 1~km~s$^{-1}$ or so from smallest to largest $T$. Given the larger mass of Cl and Fe, the variations with increasing temperature would be smaller.

Similarly, an assumption must be made for the instrumental width, which is particularly important for these narrow lines where $W_{\mathrm{I}}$ can be a non-negligible component of the total observed line width. To demonstrate the impact of our assumption of $W_{\mathrm{I}} = 28.6$~m\AA\ the bottom row of Figure~\ref{fig:vary_tform_winstr} shows $w_{\mathrm{Nthm}}$ calculated with a fixed temperature $T = 10$~kK, and variable $W_{\mathrm{I}}$. That range represents spectral point spread functions equal to $\sim1.85 - 2.46$ IRIS SG pixels in the FUVS channel.  The flare values of $w_{\mathrm{Nthm}}$ vary by less than  $\sim1$~km~s$^{-1}$ over that range of $W_{\mathrm{I}}$. In that figure the  instrumental widths expressed a (1/e) half-width are also listed in parenthesis, in velocity units.

\section{Formation Properties}\label{sec:formprops}
\begin{figure}
	\centering 
	\vbox{
	\subfloat{\includegraphics[width = 0.4\textwidth, clip = true, trim = 0.cm 0.cm 0.cm 0.cm]{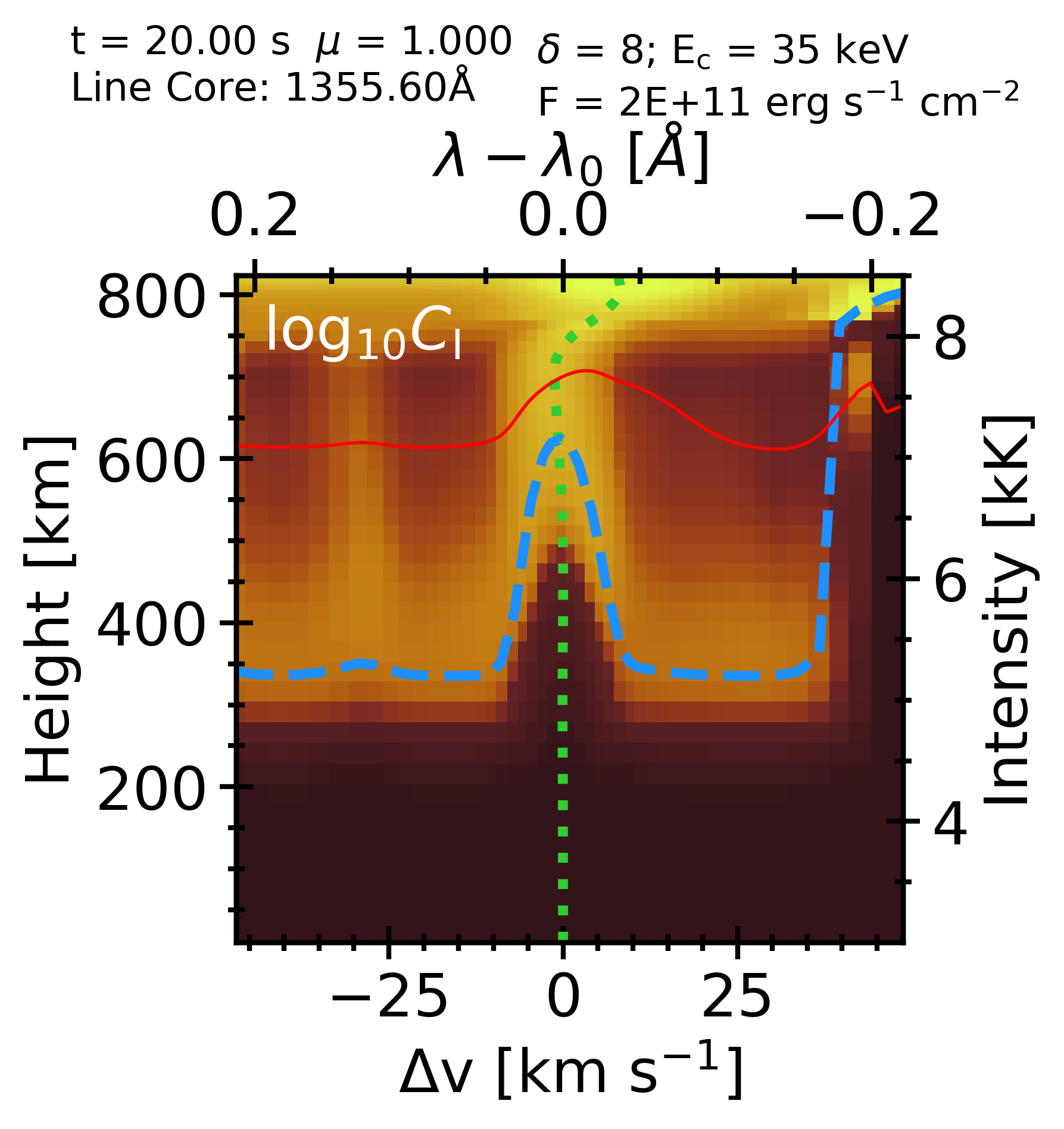}}
	}
	\vbox{
	\subfloat{\includegraphics[width = 0.40\textwidth, clip = true, trim = 0.cm 0.cm 0.cm 0.cm]{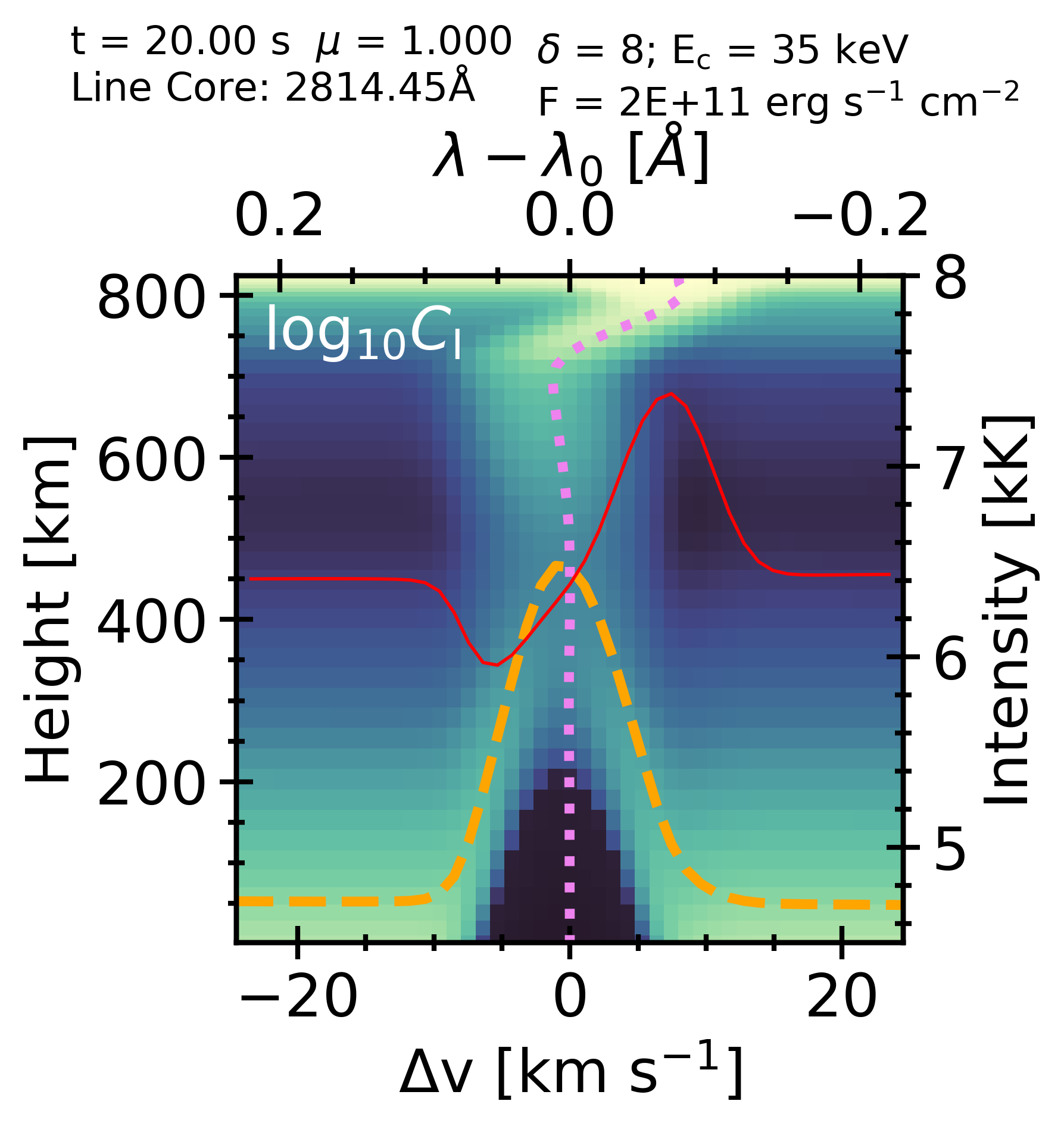}}
	}
	\caption{\textsl{Formation of the \ion{O}{i} 1355.598~\AA\ line (top panel) and \ion{Fe}{ii} 2814.45~\AA\ line (bottom panel) at $t=20$~s in the 2F11 simulation. The background image on each panel is the contribution function to the emergent intensity, $C_{\lambda\mu}(z)$, on a logarithmic scale equalised along each wavelength. The time, viewing angle $\mu = \cos\theta$, and line core wavelengths for each line are printed on the panels. These show locations in the atmosphere where line intensity originates. The dashed line on each panel is the $\tau_{\lambda} = 1$ layer. The dotted lines show the atmospheric bulk velocity, where  positive is a downflow. Finally, the thin solid lines on each panel are the emergent line profile, in units of radiation temperature.}}
	\label{fig:rh_contribs_overview}
\end{figure}
Taken as a whole, our two simulations produced spectral lines that exhibited many of the same characteristics as the observations, especially when degraded to IRIS resolution with appropriate exposure times and spectral summing. It seems that the condensation produced in our higher energy model was too fast compared to the observation, and similarly the condensation was too slow in the lower energy model. Still, given that both produced broadening, these simulations can be informative regarding our three main questions: (1) can the \ion{O}{i} or \ion{Fe}{ii} lines guide us as to the magnitude of nonthermal broadening in the chromosphere?; (2) where do they form in relation to the \ion{Mg}{ii} lines?; (3) can the magnitude of microturbulent broadening within the condensation explain \ion{Mg}{ii} line widths?

To answer (1) \& (2) we study the formation properties of the lines, specifically the plasma properties in the regions where they form. The specific intensity in a plane-parallel semi-infinite atmosphere can be defined as

\begin{equation}
I_{\lambda\mu} = \int \frac{1}{\mu} S_\lambda(z) e^{-\tau_{\lambda}(z)/\mu}\chi_{\lambda}(z)\mathrm{d}z = \int C_{\lambda\mu}(z)\mathrm{d}z,
\end{equation}

\noindent where $\mu = \cos \theta$ ($\theta$ is the viewing angle between the line of sight and the normal), $S_\lambda$ is the wavelength-independent source function, $\tau_\lambda$ is the optical depth, $\chi_{\lambda}$ is the monochromatic opacity, and $C_{\lambda\mu}$ is referred to as the contribution function to the emergent intensity \citep[e.g.][]{1986A&A...163..135M,1998LNP...507..163C}. The contribution function, which varies with height, can in effect show us where in the atmosphere the line forms. If the line forms sufficiently far above the height at which $\tau_{\lambda} = 1$ then the line can be said to form under optically thin conditions, with opacity effects playing little to no role in line formation. If the line forms close to the $\tau_{\lambda} = 1$ layer, though, then opacity effects will play a role in line formation and the line is said to be optically thick. The dividing line between optically thick or thin conditions is not strictly defined in $\tau_{\lambda}$ space, but in this work we assume that if a line component forms at heights above $\tau_{\lambda} \sim 0.1$ then it is likely safe to assume optically thin formation. In this case, physically meaningful results can be extracted from Gaussian fits. 

Example contribution functions, $C_{\lambda\mu}(z)$, for the \ion{O}{i} and \ion{Fe}{ii} lines are shown Figure~\ref{fig:rh_contribs_overview}, for a snapshots from the 2F11 flare simulations, showing the red wing components caused by the condensation.

From study of the $C_{\lambda\mu}(z)$, we found that the \ion{O}{i} 1355.6~\AA\ line is optically thin throughout. Though some contribution does form around the $\tau_{\lambda} = 1$ layer, the bulk forms over several hundred kilometres. Though the line wing of \ion{Fe}{ii} is fairly optically thin once the condensation produces line emission there, the line core does have non-negligible emission close to the $\tau_{\lambda} = 1$ layer, such that the line centre optical depth is $0.1<\tau_{\lambda}<1$. Towards the end of the flare simulations, when the atmosphere has pushed the transition region to greater depth, the \ion{Fe}{ii} line does form closer to the $\tau_{\lambda} = 1$ layer and optical depth effects likely play a role. At these times the shifted component has merged with the stationary component, so there is not a notable red wing asymmetry present. Thus, our speculation that the observed \ion{Fe}{ii} shifted component is optically thin appears valid through most of the simulation.

Following a similar approach to \cite{2017ApJ...836...12K}, \cite{2019ApJ...879...19Z}, and \cite{2023ApJ...944..186M} we can measure the average plasma properties in the region of the atmosphere in which either the line cores or the red wings form. The normalised cumulative distribution function of $C_{\lambda\mu}(z)$ was computed for each line, $C_\mathrm{NCDF}$. The heights in the atmosphere corresponding to where the bulk of the line forms were identified as where $C_\mathrm{NCDF} =  0.95$ and $C_\mathrm{NCDF}  = 0.1$. Then, the average of various properties were measured, weighted by the contribution function itself. For example, for temperature:

\begin{equation}
T_{\mathrm{form}} = \frac{\int^{z(C_\mathrm{NCDF} = 0.95)}_{z(C_\mathrm{NCDF} = 0.10)}~C_{\lambda\mu}(z) T(z) \mathrm{d}z}{\int^{z(C_\mathrm{NCDF} = 0.95)}_{z(C_\mathrm{NCDF} = 0.10)}~C_{\lambda\mu}(z) \mathrm{d}z}.
\end{equation}

\begin{figure*}
	\centering 
	\subfloat{\includegraphics[width = 0.75\textwidth, clip = true, trim = 0.cm 0.cm 0.cm 0.cm]{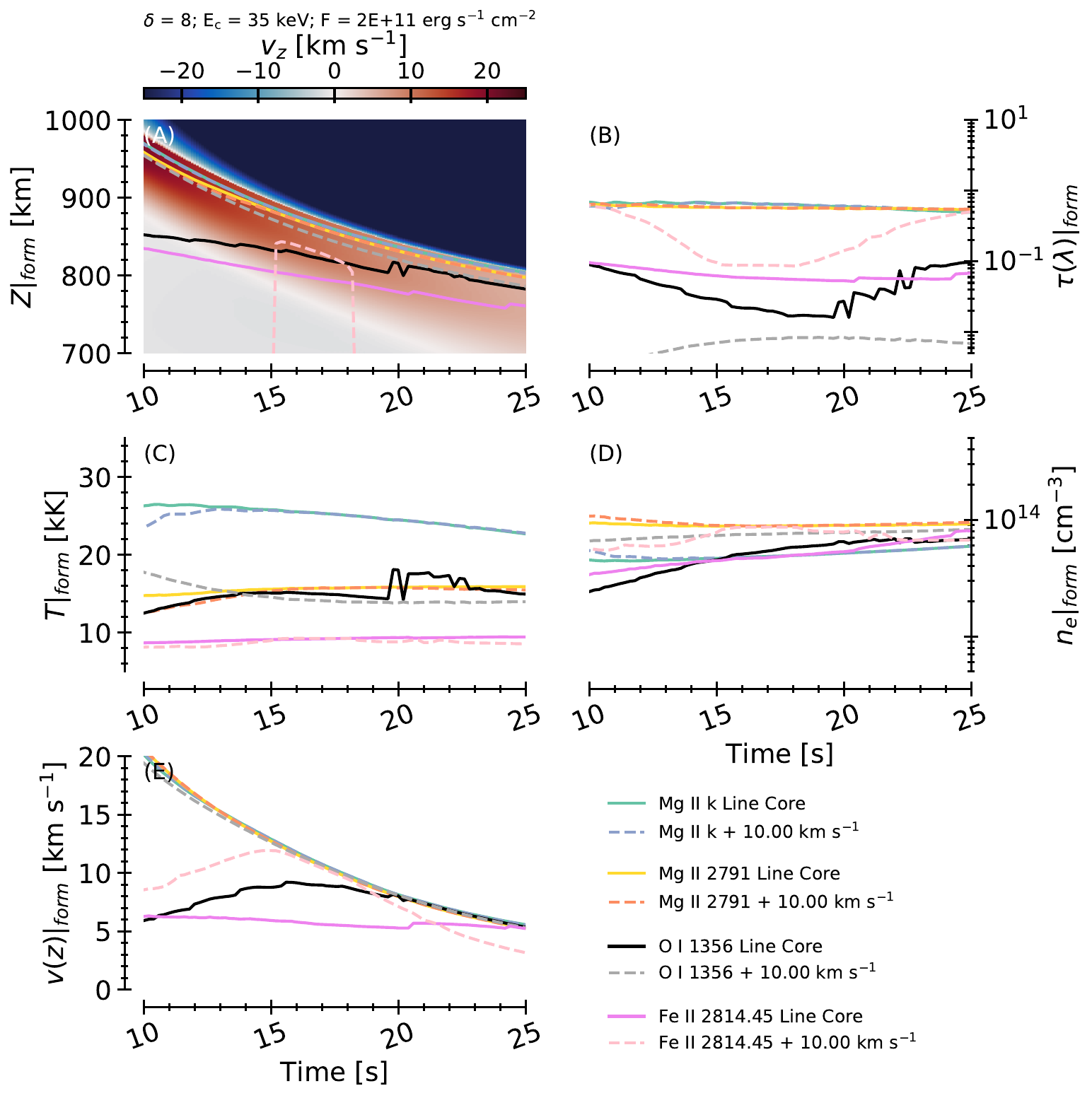}}	
	\caption{\textsl{Mean plasma properties in the formation region of the \ion{Mg}{ii} k, \ion{Mg}{ii} 2791~\AA, \ion{Fe}{ii} 2814.445~\AA, and \ion{O}{1} 1355.6~\AA\ lines for $t = 0-5$~s of the 2F11 simulation, for both the line core (solid lines) and at 10~km~s$^{-1}$ from each line's rest wavelength (dashed lines). Shown are (A) the formation heights, (B) optical depth $\tau_{\lambda}$, (C) temperature in the formation region, (D) electron density in the formation region, and (E) bulk velocity in the formation region (positive is downflow).  In panel (A) the bulk velocity is shown for context (positive is downflow, and the colour table is saturated).}}
	\label{fig:rh_plasmaprops2}
\end{figure*}

\noindent In this manner the mean formation height was measured, as was the mean temperature, electron density, bulk velocity, and $\tau_{\lambda}$ at that height, for each line as a function of time. These results are shown for the 2F11 simulation in \ref{fig:rh_plasmaprops2}, during a period of time that the dense condensation produces a red wing component in each line. The solid lines are the line cores (defined as the wavelength for which the height of $\tau_{\lambda} = 1$ is maximum over the line), and the dashed lines 10~km~s$^{-1}$ into the red wings (defined from the rest wavelength). At times when the condensation does not result in line emission in the wings, the plasma properties are from where the nearby continuum is formed. It is only after $t\sim10$~s or so that wing emission \ion{O}{i} or \ion{Fe}{ii} is strong (which is roughly when the density in the condensation becomes strongest, c.f. Figure~\ref{fig:radyn_flare_atmos}). 

Initially the \ion{Fe}{ii} 2814.445~\AA, \ion{O}{i} 1355.6~\AA, and \ion{Mg}{ii} 2791.6~\AA\ lines cores form several hundred kilometres below the \ion{Mg}{ii} k line. Fairly rapidly the chromosphere is compressed and the line cores form very close to each other, within a few 10s of kilometres or smaller for the two \ion{Mg}{ii} lines, which are together around 100-150~km higher in altitude than the core of \ion{O}{i} 1355.6~\AA. Later when the condensation has accrued density and line emission appears in the extended red wings, the wing emission forms even closer to the \ion{Mg}{ii} lines, within 60~km or so. At these times the lines sample reasonably similar plasma conditions, though there is clearly a strong gradient through the condensation such that the \ion{O}{i} and \ion{Fe}{ii} lines form in a slightly cooler plasma \ion{Mg}{ii} k line, but very close to the conditions of the \ion{Mg}{ii} 2791.6~\AA\ line. That is, a temperature in the range $T\sim8$~kK for \ion{Fe}{ii}, $T\sim12-18$~kK for \ion{O}{i} and \ion{Mg}{ii} 2791.6~\AA, but $T\sim25$~kK for \ion{Mg}{ii} k. The electron density in the formation region of each line increases by more than 2 orders of magnitude, ultimately forming between $n_{e}\sim0.6-1\times10^{14}$~cm$^{-3}$. These densities are below the values that have previously been found to produce single peaked \ion{Mg}{ii} profiles, but the shallow reversals when degraded to IRIS resolution do appear quite flat-topped or single peaked when spectral summing is included.

As expected, the \ion{Mg}{ii} lines form under optically thick conditions throughout, with only a small drop in the mean optical depth when some minor contributions are present above the $\tau_{\lambda} = 1$ layer. The \ion{O}{i} 1355.6~\AA\ emission in the core is generally below $\tau_{\lambda} < 0.1$, particularly during the flare, and the component in the red wing (mostly present at later times) has an optical depth $\tau_{\lambda} < 0.01$. The \ion{Fe}{ii} 2814.445~\AA\ core does initially have an optical depth  $\tau_{\lambda} \sim 0.6$, which drops during the flare to $\tau_{\lambda} \sim 0.05$ when the upper atmosphere contributes more to line formation. When the condensation produces strong emission at 10~km~s$^{-1}$ into the line wing (e.g. 15-19~s) the \ion{Fe}{ii} line wing is optically thin also, tracking the leading edge of the condensation. 

\section{Spectra at native resolution}\label{sec:spectranative}
Below are examples of the \ion{O}{i} 1355.598~\AA and the \ion{Mg}{ii} k spectra from the 2F11 simulation, without any degradation to mimic IRIS observations. Intensities are shown in units of radiation temperature.

\begin{figure}
	\centering 
	\vbox{
	\subfloat{\includegraphics[width = 0.45\textwidth, clip = true, trim = 0.cm 0.cm 0.cm 0.cm]{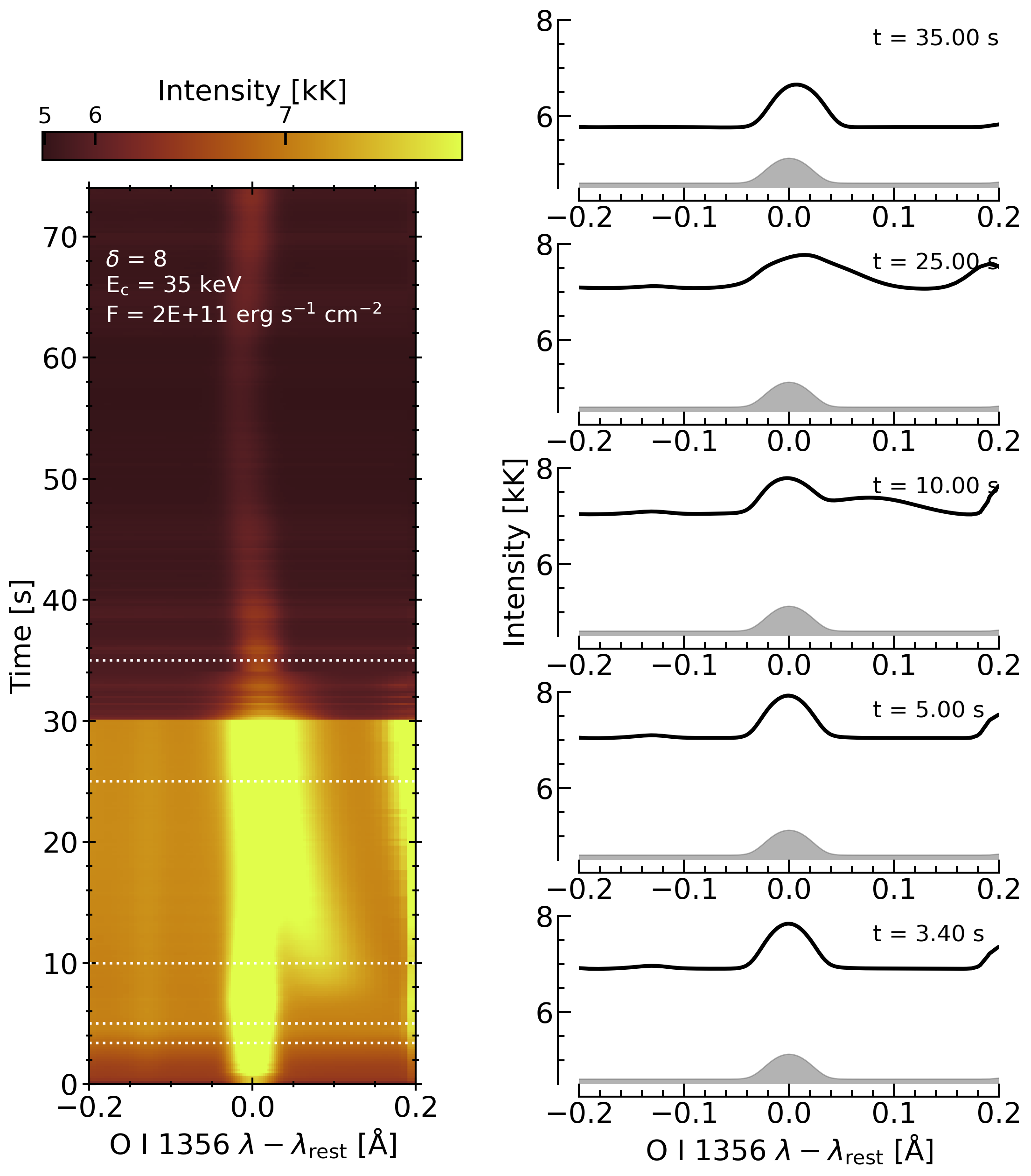}}
	}
	\vbox{
	\subfloat{\includegraphics[width = 0.45\textwidth, clip = true, trim = 0.cm 0.cm 0.cm 0.cm]{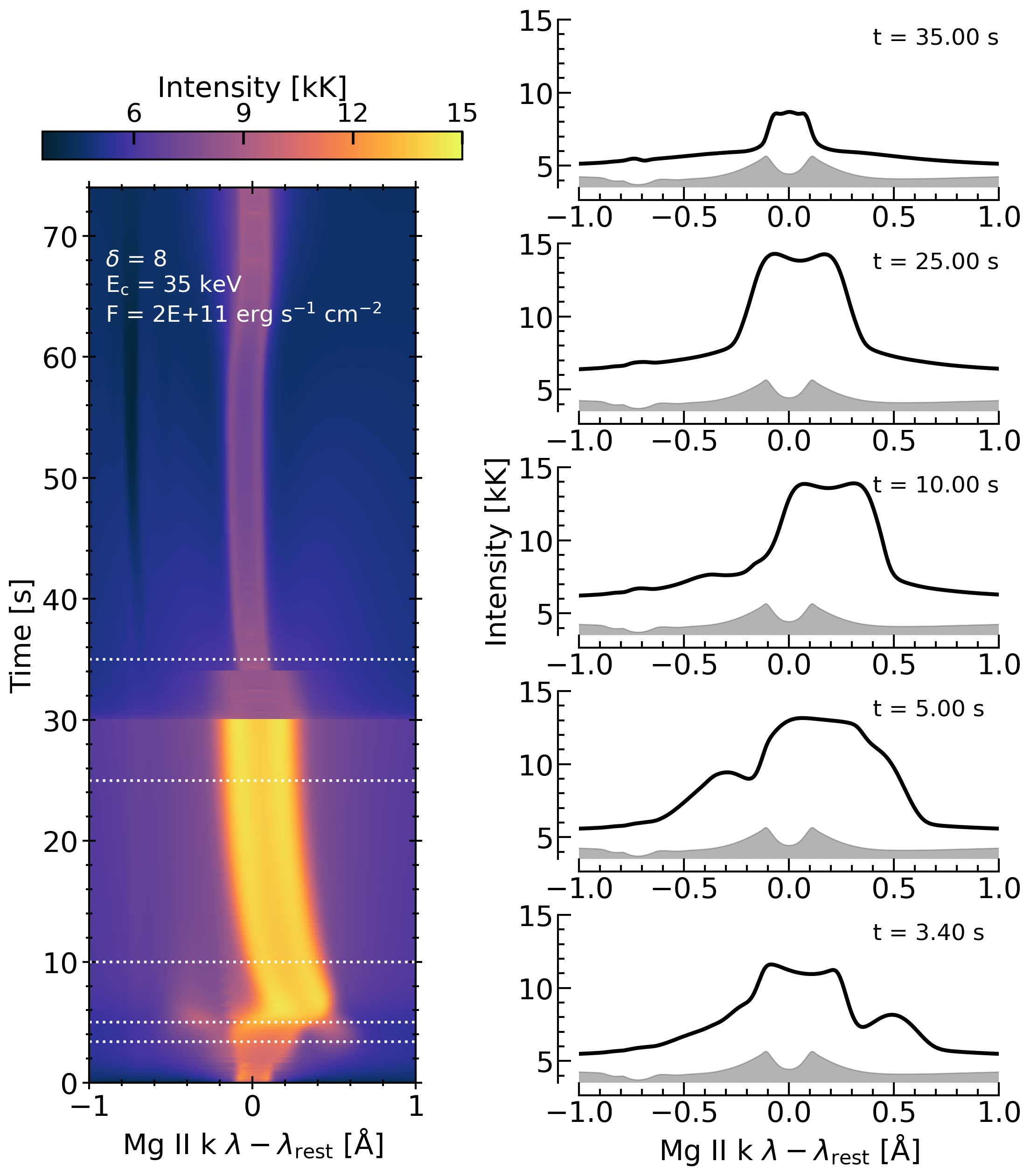}}
	}
	\caption{\textsl{Spectra from the 2F11 simulation, direct from the \rhpar\ simulation, without any degradation to IRIS quality. Shown are \ion{O}{i} 1355.598~\AA\ (top) and \ion{Mg}{ii} k (bottom). The grey shaded area shows the pre-flare spectra.}}
	\label{fig:rh_spectra_native}
\end{figure}

\bsp	
\label{lastpage}
\end{document}